\documentclass[a4paper,11pt]{book}
\usepackage{ae}
\usepackage[T1]{fontenc}
\usepackage[ansinew]{inputenc}
\usepackage{amsmath}
\usepackage{amssymb}
\usepackage{color}
\usepackage{graphicx}
\usepackage{longtable}

\begin{document}

\baselineskip=15pt

\begin{titlepage}
\begin{center}
 \Large{UNIVERSIT\`A DEGLI STUDI DI ROMA 3\\}
\vspace{0.5cm} \Large{ Facolt\`a di Scienze Matematiche Fisiche e
Naturali\\} \vspace{0.3cm} \Large{ Dipartimento di Fisica\\}
\vspace{1cm}
 \large{\textbf{5D DIFFERENTIAL CALCULUS AND NOETHER ANALYSIS}}\\
\vspace{0.3cm}
 \large{\textbf{OF TRANSLATION SYMMETRIES}}\\
\vspace{0.3cm}
 \large{\textbf{IN $\kappa$-MINKOWSKI NONCOMMUTATIVE SPACETIME}}\\
 \vspace{1cm}\large{Laurea (Master) Thesis\\}\vspace{0.3cm}
 \large{Academic Year 2006/2007\\} \vspace{1cm}
Author\\
Daniele Pranzetti\\ \vspace{1cm} \textbf{Abstract}\vspace{-0.1cm}
\end{center}
We perform a Noether analysis for a description of translation
transformations in 4D $\kappa$-Minkowski noncommutative spacetime
which is based on the structure of a 5D differential calculus.
Taking properly into account the properties of the differential
calculus we arrive at an explicit formula for the conserved charges.
We also propose a choice of basis for the 5D calculus which leads to
an intuitive description of time derivatives.
\begin{center}
\vspace{4cm}Internal Supervisor  ~~~~~~~~~~~~~~~~~~~~~~~~~~~~External Supervisor\\
~~~~~~Prof. Orlando Ragnisco ~~~~~~~~~~~~~~~~~~Prof. Giovanni
Amelino-Camelia \\
\end{center}
\end{titlepage}

\clearpage
 \setcounter{page}{1}\pagenumbering{Roman}  \oddsidemargin=50pt  \evensidemargin=10pt
\large\tableofcontents

\chapter*{Introduction}\label{par:Intro}
\addcontentsline{toc}{chapter}{Introduction}

\setcounter{page}{1} \pagenumbering{arabic}
 \evensidemargin=10pt
\oddsidemargin=45pt
 Various arguments suggest that our current
description of particle phy\-sics would require a profound revision
in order to describe processes at the Planck scale $E_p$, defined
as\\
\begin{equation*}
E_p=\sqrt{\frac{\hbar c}{G}}\simeq 10^{19}GeV \,.
\end{equation*}\\
\noindent At such high energy  both quantum and gravitational
effects are important, and the Standard Model of particle physics
appears to be incomplete since it neglects gravity.

A large research effort has been devoted to the search for a
``Quantum Gravity", i.e. a theory giving a unified description of
Quantum Mechanics and General Relativity, the two theories that
respectively govern quantum and gravitational phenomena (see, e.g.,
\cite{Stachel}).

Quantum Mechanics reigns supremely in low energy ($E<<E_p$)
processes where gravity is negligible. In particular, Quantum Field
Theory, following the unification of Special Relativity with Quantum
Mechanics, successfully describes all experimental data up to
energies currently achievable in the laboratory which are in the TeV
range. Several characteristic predictions of the Standard Model of
strong, electromagnetic and weak interactions have been very
successful as in the case of the discovery of the \textit{W} and
\textit{Z} gauge bosons.

On the other hand, Einstein's General Relativity successfully
describes the motion of macroscopic bodies where quantum effects are
negligible.

However, a unified description of these two theories is necessary in
order to produce predictions for some interesting situations in
which both are required, for example the ``Big Bang" - the first
moments of the Universe, when gravitational interactions were very
strong and the scales involved were all microscopic.

If one simply attempts to quantize General Relativity, in the same
sense that Quantum Electro Dynamics is a quantization of Maxwell's
theory, the result is an inconsistent theory. This is due to the
fact that Newton's constant is dimensionful and consequently, the
divergences can not be disposed of by the technique of
renormalization. In addition to this ``renormalizability" problem,
great difficulties of a unified description of Quantum Mechanics and
General Relativity originate from their deep incompatibilities. One
of the most evident aspects of this incompatibility regards the way
in which the geometry of space and time is treated. In the Quantum
Mechanics picture spacetime is a fixed arena where quantum
observables (such as position of a particle) are described. But in
General Relativity spacetime can not be treated as a fixed
background since it acquires a geometrodinamical structure.

The lack of reliable data on the spacetime at very small distance
scales (i.e. for very high energy particles) has led to the proposal
of various models for Quantum Gravity, in particular, “String
Theory” \cite{Witten, Polchinski} and “Loop Quantum Gravity”
\cite{Rovelli, Ashtekar, Smolin, Thiemann}. These models are
sometimes very different in the way they approach the technical and
conceptual problems emerging from a Quantum-Gravity theory; however,
they lead to a common Quantum-Gravity intuition: from any approach
to the unification of General Relativity and Quantum Mechanics
emerges the idea of a limitation to the localization  of the
spacetime point. Different arguments can be produced to identify the
Planck scale, here intended as the length scale
$L_p=\sqrt{\frac{\hbar G}{c^3}}\simeq 1.6\cdot 10^{-35}m$, the
inverse of the Planck (energy) scale $E_p$, as the special scale at
which quantum and gravitational effect are equally important and the
description of spacetime so far adopted has to be radically reviewed
to accommodate the limitation on localization \cite{mead, padma,
ng1994, gacmpla, garay}. For example, in the case of spatial
interval one would expect an uncertainty principle of the type \\
\begin{equation*}
\delta x\gtrsim L_p\,.
\end{equation*}\\
\noindent Intuitively, it is easy to realize how such a limitation
could derive from considerations of Quantum Mechanics and General
Relativity. From basilar equations of Quantum Mechanics follow that
to have a good resolution on small distances it is necessary to use
probe particles of high energy to do the measure. But a very
energetic particle generates an intense gravitational field that
modifies the metric, introducing so a new source of uncertainty on
the measure. Thereby, increasing the probes energy one reduces one
of the contributions to the total measure uncertainty, but
inevitably increases other contributions. Another way to reach the
same conclusion can be based on the observation that no particle can
be localized in a region of linear dimensions inferior to its own
Schwarzschild radius, since the event horizon would then interfere
with the use of a probe. The ``Schwarzschild-radius uncertainty" in
localization, which increases with the particle mass, for a particle
of mass of order $L_p^{-1}$ leads to a localization uncertainty
which is also of order $L_p$. This is due to the fact that if we
combine this gravity-induced ``Schwarzschild-radius uncertainty"
$\delta x\geq r_g\sim GM$, where $G=L_p^2$ (in natural units
$c=\hbar=1$), with the well established ``Compton-wavelength
uncertainty" which decreases with the particle mass, $\delta x\geq
1/M$, for a particle of mass of order $L_p^{-1}$ one cannot do any
better than Planck-length localization.

In this scenario it is conceivable that at Plank-length distance
scales geometry can have a form which is quite different from the
classical one with which we are familiar at large scales.The
description of spacetime as a differentiable manifold might need a
revision and a new description of geometry might lead to a
development of a completely new understanding of physics.

The formalism of noncommutative geometry, which is adopted by this
thesis work, is among the most studied possibilities for such a new
description of spacetime structure. It essentially assumes
(\cite{Amelino1}) that one can describe algebraically Quantum
Gravity corrections replacing the traditional (Minkowski) spacetime
coordinates $x_\mu$ with Hermitian operators $\hat{x}_\mu$ that
satisfies commutation relation of the type:    \\
\begin{equation*}
[\hat{x}_\mu,\hat{x}_\nu]=i\theta_{\mu\nu}(\hat{x})\,.
\end{equation*}\\
A noncommutative spacetime of this type embodies an impossibility to
fully know the short distance structure of spacetime, in the same
way that in the phase space of the ordinary Quantum Mechanics there
is a limit on the localization of a particle. This fact agrees with
the above mentioned intuition of a limitation to localization in the
Quantum-Gravity framework.

There is a wide literature on the simplest, so-called ``canonical",
noncommutativity characterized by commutators of the coordinates of the type\\
\begin{equation*}
[\hat{x}_\mu,\hat{x}_\nu]=i\theta_{\mu\nu}\,,
\end{equation*}\\
\noindent where $\theta_{\mu\nu}$ is a coordinate-independent matrix
of dimensionful parameters.

In this thesis we consider another much studied noncommutative
spacetime, the $\kappa$-Minkowski spacetime, characterized by the
commutation relations:  \\
\begin{equation*}
[\hat{x}_j,\hat{x}_0]=i\lambda\hat{x}_j~~~~~~~~~~~~[\hat{x}_j,\hat{x}_k]=0\,,
\end{equation*}\\
\noindent where $\lambda\footnote{Rather than the lenght scale
$\lambda$ a majority of authors use the energy scale $\kappa$, which
is the inverse of $\lambda$
($\lambda\rightarrow\frac{1}{\kappa}$).}$ has the dimensions of a
length. This type of noncommutativity is an example of
``Lie-Algebra-type" noncommutativity in which commutation relations
among spacetime coordinates exhibit a linear dependence on the
spacetime coordinate themselves.

Recently $\kappa$-Minkowski gained remarkable attention due to the
fact that it provides an example of noncommutative spacetime in
which Lorentz symmetries are preserved as deformed (quantum)
symmetries. The quantum deformation and even a break down of Lorentz
symmetry is not surprising for a quantum spacetime because of the
existence of a minimum spatial length that is not a Lorentz
invariant concept. If ordinary Lorentz invariance was preserved, we
could always perform a boost and squeeze any given length as much as
we want and therefore a minimal length could not exist. If a minimal
length really exists we have to contemplate the possibility that the
Lorentz invariance is lost. The peculiarity of $\kappa$-Minkowski
spacetime is that the symmetry is lost as classical symmetry but
preserved as ``quantum symmetry" in a sense which will be discussed
in detail in this work.

The fact that symmetries are deformed in $\kappa$-Minkowski has
emerged in \cite{Majid1, Majid2} where $\kappa$-Minkowski has been
connected with a dimensionful deformation of the Poincar\'{e}
algebra called $\kappa$-Poincar\'{e}.

The analysis of the physical implications of the deformed
$\kappa$-Poincar\'{e} algebra  have led to interesting  hypotheses
about the possibility that in $\kappa$-Minkowski particles are
submitted to modified dispersion relations \cite{Amelino2}. Since
the growing sensitivity and accuracy of the astrophysical
observations renders experimentally accessible such modified
dispersion relations (see, e.g., \cite{Amelino5} and
\cite{Kowalski2}), there is now strong interest on a systematic
analysis of a field theory in $\kappa$-Minkowski.

In this work we want to investigate the symmetries of
$\kappa$-Minkowski noncommutative spacetime connected with the
translations sector of $\kappa$-Poincar\'{e} algebra for a free
scalar field. Symmetries are introduced directly at the level of the
action, following very strictly commutative field theory in which
the symmetry of a theory is defined as transformation of coordinates
that leaves invariant the action of the theory. Our analysis, in
complete analogy with \cite{Amelino3}, will be based on the
generalization of the Noether theorem within the most studied theory
\cite{Lukierski}, \cite{Kowalski}, \cite{Amelino4} formulated in
$\kappa$-Minkowski spacetime for a scalar field $\Phi(x)$ governed
by the Klein-Gordon-like equation\\
\begin{equation*}
C_\lambda(P_\mu)\Phi=\left[\left(\frac{2}{\lambda} \sinh
{\frac{\lambda}{2} P_0}\right)^2-e^{\lambda
P_0}\vec{P}^2\right]\Phi=m^2\Phi\,.
\end{equation*}\\
In \cite{Amelino3} has been showed that the previous failure to
derive energy-momentum conserved charges associated with the
$\kappa$-Poincar\'{e} translation transformations were due to the
adoption of a rather naive description of translation
transformations, which in particular did not take into account the
properties of the noncommutative $\kappa$-Minkowski differential
calculus. By taking into account the properties of the differential
calculus one encounters no obstruction in following all the steps of
Noether analysis and obtain an explicit formula relating fields and
energy-momentum charges. \cite{Amelino3} used the invariance of the
theory under the four $\kappa$-Poincar\'{e} translation
transformations and a four-dimensional translational invariant
calculus proposed by Majid and Oeckl \cite{Majid3} to find four
energy-momentum conserved charges, showing that Hopf algebra can be
used to describe genuine spacetime symmetries.

The choice of the vector fields generalizing the notion of
derivative in $\kappa$-Minkowski  represents a key point of our line
of analysis. In fact, in the commutative case there is only one
(natural) differential calculus involving the conventional
derivatives, whereas in the $\kappa$-Minkowski case (and in general
in a noncommutative spacetime) the introduction of a differential
calculus is a more complex problem and, in particular, it is not
unique. In our analysis we focus on a possible choice, different
from \cite{Amelino3}, of differential calculus in
$\kappa$-Minkowski: the ``five-dimensional bicovariant calculus"
introduced in \cite{Sitarz}. The vector fields corresponding to this
differential calculus have in fact special covariance properties:
they transform under $\kappa$-Poincar\'{e} in the same way that the
ordinary derivatives (i.e. the vector fields associated to the
differential calculus in the commutative Minkowski space) transform
under Poincar\'{e}. Besides, this is the differential calculus under
which the action of the $\kappa$-Poincar\'{e} group becomes linear.

In Chapter \ref{ch:NoncommutativeGeometry} we introduce the
Hopf-algebras structures which play a fundamental role in the
description of $\kappa$-Minkowski noncommutative spacetime and its
quantum $\kappa$-Poincar\'{e} symmetry group. In analogy with
canonical noncommutative spacetime, where it is used to introduce
fields through the Weyl map \cite{Weyl}, we introduce a field in
$\kappa$-Minkowski through a generalized Weyl map based on the
notion of generalized Weyl system. The Weyl-system description
allows to introduce a field in $\kappa$-Minkowski as a generalized
Fourier transform that establishes a correspondence between
noncommutative positions coordinates (noncommutative coordinate
generators of $\kappa$-Minkowski) and some commutative Fourier
parameters. Thus, such a generalized transform allows us to rewrite
structures living on noncommutative spacetime as structures living
on a classical (commutative) but non-Abelian ``energy-momentum"
space.

However, the interpretation that the Quantum Group language gives to
``momenta" as generators of translations (i.e. the real physical
particle momenta) is based on the notion of quantum group symmetry.
It is puzzling in fact that in the Quantum Group literature it is
stated (see, e.g., \cite{Kowalski}) that the symmetries of
$\kappa$-Minkowski can be described by any one of a large number of
$\kappa$-Poincar\'{e} basis of generators. The nature of this
claimed symmetry-description degeneracy remains obscure from a
physics perspective, in particular we are used to associate
energy-momentum with the translation generators and it is not
conceivable that a given operative definition of energy-momentum
could be equivalently described in terms of different translation
generators. The difference would be easily established by testing,
for example, the different dispersion relations that the different
momenta satisfy (a meaningful physical property, which could, in
particular, have observable consequences in astrophysics
\cite{Amelino5}, \cite{Amelino6} and cosmology \cite{Kowalski2}).

In Chapter \ref{ch:Analysis4D} we present the Noether analysis of
translation symmetry in $\kappa$-Minkowski with a four-dimensional
differential calculus and the four translation generators $P_\mu$ of
the Majid-Ruegg $\kappa$-Poincar\'{e} basis in the definition of
exterior derivative operator \textit{d}, going over the steps of the
analysis reported in \cite{Amelino3} where the conserved charges
associated with the translation sector of the $\kappa$-Poincar\'{e}
symmetry transformation have been obtained. This result confirms
that in $\kappa$-Minkowski there is a non-linear Planck-scale
modification of the energy-momentum relation, but the nonlinearity
intervenes in a way that differs significantly from what had been
conjectured on the basis of some heuristic arguments.

In Chapter \ref{ch:Analysis5D} we perform all the steps of the
Noether analysis for a free scalar field on $\kappa$-Minkowski. In
order to have all the instruments for our Noether analysis of
$\kappa$-Poincar\'{e} translations on $\kappa$-Minkowski, we present
first the five-dimensional differential calculus, introduced by
Sitarz \cite{Sitarz}. We write the exterior derivative operator
\textit{d} of a generic $\kappa$-Minkowski element and the
commutation relations between the one-form generators $d\hat{x}^A$
and the $\kappa$-Minkowski generators $\hat{x}^\mu$. Noether
analysis requires that the exterior derivative operator \textit{d},
defined in terms of five translation generators $\hat{P}_A$,
functions of the four generators $P_\mu$, satisfies the Leibnitz
rule and we show that this is in fact what happens.

Once we have introduced all these needed tools we proceed with our
Noether analysis, relying on direct explicit manipulations of
noncommutative fields, and we investigate explicitly the properties
of the 5 ``would-be currents" that one naturally ends up considering
when working with the 5D differential calculus. To obtain conserved
charges we perform 3D spatial integration of the currents and,
showing how time derivatives are to be formulated in the 5D-calculus
setup, we obtain 5 time-independent charges. In fact, we find that
within the 5D-calculus setup some subtleties must be handled when
trying to establish the time independence of a noncommutative field
and our Noether analysis constructively leads us to identify the
proper time derivative operator in $\kappa$-Minkowski noncommutative
spacetime and to a ``conservation equation" for the currents. This
will motivate a change of basis for the 5D differential calculus
with the introduction of a parameter that can be meaningfully
described as time-translation parameter. The rotation of the
transformation parameters basis does not affect the Noether analysis
in any armful way and leads to a conserved charge associated with
the new time-translation parameter which is a plausible candidate
for the energy observable.

The primary objective of this thesis work is an investigation of the
role that the five-dimensional differential calculus could have in
the description of $\kappa$-Minkowski spacetime symmetries in
alternative to the four-dimensional differential calculus adopted in
\cite{Amelino3}. The results of Chapter \ref{ch:Analysis5D} provides
support to the idea that the five-dimensional differential calculus
can be used for a Noether analysis of the translation sector of
$\kappa$-Poincar\'{e} and all the worries about the presence of five
currents, which produces five charges at the end of the analysis,
vanish. Besides, the fact that the 5D differential calculus is
bicovariant under the action of the full $\kappa$-Poincar\'{e}
algebra and the basis generators $\hat{P}_A$ of translations
transform under $\kappa$-Poincar\'{e} action in the same way as the
operators $P_\mu$ in the commutative case transform under the
standard Poincar\'{e} action, might induce to expect that this
analysis leads to classical results (see, e.g., \cite{Friedel}). The
analysis of Chapter \ref{ch:Analysis5D} shows how the linearity of
$\kappa$-Poincar\'{e} action on the commutation relation of the 5D
differential calculus induces a highly non-trivial structure of the
coalgerba sector of the generators $\hat{P}_A$ and thereby a
non-trivial modification of the quantum symmetry. Recovering the
classical results at the end of the analysis would seem less likely
than expected and the form of the charges obtained shows that this
indeed does not happen.

In Chapter \ref{ch:DispersionRelation} we investigate the
possibility to derive an energy-momentum (dispersion) relation
involving a plausible candidate for the energy observable, by
evaluating the charges obtained in Chapter \ref{ch:Analysis5D} for
some trivial solution of the equation of motion. We see that the
dispersion relation in the massless case is classical, as expected
by \cite{Friedel, Freidel2}, while in the massive case there is a
Plank-scale modification leading to a non special-relativistic
dispersion relation, differently from \cite{Friedel,Freidel2}
prediction. However, it is interesting to notice that this
modification vanishes if one increases arbitrary the intensity of
the fields, i.e. scaling the classical fields by a factor $A$, in
the limit $A\rightarrow{\infty}$, the special-relativistic relation
is reestablished with a mass $m^R=A^2m$.

\clearpage

\chapter{\textbf{Noncommutative Geometry and $\kappa$-Minkowski spacetime}}\label{ch:NoncommutativeGeometry}

In the first part of this chapter we give a brief overview of
Noncommutative Geometry and we introduce the Hopf-algebras
structures which play a fundamental role in the description of
$\kappa$-Minkowski spacetime and its quantum $\kappa$-Poincar\'{e}
symmetry group. In the second part we analyze the symmetries of the
deformed Poincar\'{e} group on $\kappa$-Minkowski and we introduce
fields trough the powerful concept of Weyl maps. Thus, at the end of
the chapter we will be able to write a mass Casimir and a deformed
Klein-Gordon equation for a free scalar field in $\kappa$-Minkowski.

\section{Preliminaries on Noncommutative Geometry}
\label{par:Preliminaries}

The Quantum Mechanics phase space, i.e. the space of microscopic
states of a quantum particle, provides the first example of
noncommutative space. It is defined replacing canonical variables of
position and momentum of a particle $(q_j, p_j)$ with self-adjoint
operators $(\hat{q}_j, \hat{p}_j)$ satisfying Heisenberg's
commutation relations\\
\begin{equation}
[\hat{q}_j, \hat{p}_k]=i\hbar\delta_{jk},~~~~~~~~j,k=1,2,3
\end{equation}\\
\noindent from which follows the Heisenberg uncertainty principle\\
\begin{equation}
\delta\hat{q}_j\delta\hat{p}_k\geq\frac{\hbar\delta_{jk}}{2}.
\end{equation}\\
This principle establishes the existence of an accuracy limitation
for the measurement of the coordinates and the corresponding momenta
of a particle. Consequently, the quantization of phase space can be
viewed as the smearing out of a classical manifold, replacing the
notion of a point with that of a Planck cell. The idealized
classical situation in which one can simultaneously determine the
exact position-momentum measurements is obtained in the limit
$\hbar\rightarrow0$, where the phase space becomes a continuum
manifold.

A very similar idea led to apply noncommutativity to spacetime
itself. The idea of a new structure of spacetime came in the late
40's from Snyder \cite{Snyder} in order to solve the short-distance
singularities of the quantum field theory. Later on, the attention
was focused on a general noncommutative spacetime of Lie-algebra
type with central extension, characterized by the commutation
relations\\
\begin{equation}
[\hat{x}_\mu,
\hat{x}_\nu]=i\theta_{\mu\nu}+i\zeta^\alpha_{\mu\nu}\hat{x}_\alpha
\end{equation}\\
\noindent with coordinate-independent $\theta_{\mu\nu}$ and
$\zeta^\alpha_{\mu\nu}$; in particular, the attention was
concentrated on the canonical noncommutative spacetime,
characterized simply by Heisenberg-like commutation relations\\
\begin{equation}
[\hat{x}_\mu, \hat{x}_\nu]=i\theta_{\mu\nu}.
\end{equation}\\
\noindent As in case of quantum phase space, this spacetime
prescription can be viewed as the smearing out of the classical
manifold losing the notion of the \textit{point}: in fact, a
Heisenberg-type uncertainty principle implies that the notion of the
point is replaced by an analogous of the Planck cell of the quantum
phase space.

In the literature there exist principally two main approaches to
Noncommutative Geometry. In this work we are interested in how it
emerges in the Quantum Groups framework\footnote{In the other widely
spread approach, largely due to Connes \cite{Connes}, in order to be
able to construct field theories on noncommutative spaces in the
same way as on traditional commutative spaces the attention is no
more focused on the spacetime but on the algebra of functions. In
particular, Connes' idea of Noncommutative Geometry is based on the
re-formulation of the manifold geometry in terms of C*-algebras of
functions defined over the manifold, with a generalization of the
corresponding results of differential geometry to the case of a
noncommutative algebra of functions. The characterization of the
Hilbert space and the Dirac operator become the main ingredients and
there is no more an explicit reference neither to the spacetime or
the coordinates, which in general can also do not exist in a
concrete form.} that, for the applications to Quantum Gravity,
reflects more the intuition on the meaning of the Planck length
$L_p$ as the length parameter in which the localization
indetermination of the spacetime points is manifest, due to the
noncommutativity of coordinates. From this point of view it was
Woronowicz \cite{Woronowicz} who initiated a systematic study of the
``noncommutative differential geometry" built on some
``pseudogroups" that are the generalization of the standard Lie
groups related to the commutative differential geometry.

The study of these ``quantum groups" algebras have a fundamental
role in the description of symmetries of noncommutative spacetime.
Our attention in this work is on $\kappa$-Minkowski spacetime,
characterized by the commutation relations:\\
\begin{equation}\label{eq:kappaMinkowski}
[\hat{x}_j,\hat{x}_0]=i\lambda\hat{x}_j~~~~~~~~~~~~[\hat{x}_j,\hat{x}_k]=0\,.
\end{equation}\\
\noindent On the one hand, this example of Lie-algebra-type
noncommutativity which introduces a length deformation parameter
($\lambda$) appears as a natural candidate for a quantized spacetime
with a new limitation on the measurability of geometrical quantity.
In fact the results of \cite{Amelino1}, \cite{Kowalski3} show how
$\kappa$-Minkowski provides a particular realization of the minimal
length concept.

On the other hand, we will show in section
(\ref{par:DeformedPoincare}) how $\kappa$-Minkowski can be seen as
the dual Hopf algebra (the concept of duality will be explained in
detail in (\ref{par:HopfAlgebras}) and Appendix A where
Bicrossproduct Hopf algebras are introduced in detail) of the
momentum sector of the $\kappa$-Poincar\'{e} algebra.

Thus, for a clearer comprehension of the framework we are working
in, we give in section (\ref{par:QuantumGroups}) a brief description
of the Quantum groups language in order to use it in the rest of the
chapter to introduce $\kappa$-Poincar\'{e} and $\kappa$-Minkowski
and investigate their mathematical structure and
symmetries.\\

\section{Quantum Groups and their emergence in Noncommutative Geometry} \label{par:QuantumGroups}

Quantum Groups or Hopf algebras are a generalization of the ordinary
groups (i.e. collections of transformations on a space that are
invertible). They have a rich mathematical structure and numerous
roles in physical situations where ordinary groups are not adequate.
Quantum Groups allows us to generalize many ``classical" physical
ideas in a completely self-consistent way. This generalization is
realized through a ``deformation" induced by the presence of one or
more parameters. The classical case is recovered by setting these
parameters to some fixed values. A very similar case of quantization
is represented by Quantum Mechanics, in which the deformation is
introduced by the Planck constant $\hbar$, and the classical case is
recovered in the limit $\hbar\rightarrow0$. As we will show below,
Quantum Groups have structure, such as the coproduct and the
antipode, that generalize some properties of ordinary groups, such
as the representation on a vector-space tensor product or the
existence of an inverse. This properties are at the basis of the
applications of Quantum Groups in a wide physical domain, from
Statistical Physics to Quantum Gravity.

The mathematical structure of $\kappa$-Minkowski has emerged and has
been described in the context of studies which relate noncommutative
spaces and the world of quantum groups. Just like Lie groups and the
homogeneous spaces associated to them provide a complete description
of the classical differential geometry, so quantum groups and the
homogeneous quantum spaces associated provide a wide class of
examples on which it is possible to built and develop a
noncommutative geometry.

In particular, among the several classes of quantum groups and
algebras, we can consider $\kappa$-Poincar\'{e}, built as a deformed
algebra of the usual relativistic symmetries.

A special class of Quantum Groups (called of ``bicrossproduct" type)
was largely investigated by S. Majid in the approach to Planck-scale
Physics \cite{Majid4}. As we will see below, this line of research
represents an important point for our study of $\kappa$-Minkowski.
In fact, as shown in section (\ref{par:DeformedPoincare}), for a
particular choice of the generators basis, $\kappa$-Poincar\'{e}
algebra has a manifest structure of bicrossproduct Hopf algebra with
the properties of duality that enable to identify $\kappa$-Minkowski
as the space where  $\kappa$-Poincar\'{e} acts on in a covariant
way, i.e the commutation relations that characterize
$\kappa$-Minkowski remain unchanged under the action of
$\kappa$-Poincar\'{e} algebra (in Appendix A we provide the
demonstration of $\kappa$-Poincar\'{e} invariant action on
$\kappa$-Minkowski).

The nature of quantum groups will be clearer at the end of this
section, but we can anticipate here the fact that a quantum group is
a noncommutative noncocommutative Hopf algebra. Let us clarify the
term ``quantum group". The term ``group" refers to the
correspondence between topological groups and commutative Hopf
algebras, since it is always possible to associate a commutative
Hopf algebra to every topological compact group $G$ and all the
properties of the group $G$ can be reformulated in terms of the Hopf
algebra $A=C(G)$, the space of continuum functions on $G$. The term
``quantum" refers to the deformation of the Hopf algebra $A$ into a
certain noncommutative Hopf algebra $A_q$, were $q$ is the
deformation parameter. In practice we do not deform the group $G$,
but its dual object $A=C(G)$. Thereby the quantum groups category
can be considered dual to that of noncommutative noncocommutative
Hopf algebras. In other words a quantum group can be considered as
the geometric object, with noncommutative coordinates, corresponding
to a general Hopf algebra. As we will see, this particular
mathematical object has the property that its dual space turns out
to be again a Hopf algebra. In particular, $\kappa$-Minkowski itself
will be a Hopf algebra with a dual space of momenta that will have
the typical commutative structure and where we will define our field
theory.

In order to introduce the notion of quantum group symmetry that
should preserve a covariant action over the associated homogeneous
space, we provide in this section some notions about the definition
of Hopf Algebra (or Quantum Group), which is relevant for our
description of quantum deformations of Poincar\'{e} group.

\subsection{Hopf Algebras}\label{par:HopfAlgebras}

The central structure of all quantum groups theory is that of Hopf
algebra. The extra structures that characterize a Hopf Algebra with
respect to a Lie algebra turn out to be very useful in order to
translate in the mathematical language some physical properties. In
particular, one finds that it is necessary to introduce some new
mathematical in the rules of composition of representations. Let us
start with the definition of $\mathbb{C}$-algebra (associative
algebra with unity).

\textbf{Definition}. \textit{A vector space $A$ on the complex field
$\mathbb{C}$ endowed with two maps $m$ ($m:A\otimes A\rightarrow A$)
and $\eta$ ($\eta:\mathbb{C}\rightarrow A$) is defined
$\mathbb{C}$-algebra if $m$ and $\eta$ satisfy:}\\
\begin{equation}
m(m\otimes1)=m(1\otimes m)~~~~~~~~~(associativity)
\end{equation}\\
\begin{equation}
m(1\otimes\eta)=m(\eta\otimes 1)=id~~~~~~~~~~~~~~~(unity)
\end{equation}\\
\noindent where $id:A\rightarrow A$ denotes the identity map on
\textit{A}.

A representation of an algebra \textit{A} over a vector space
\textit{V} is a set $(V,\rho)$, where $\rho$ is a linear map from
\textit{A} to the space of linear operator in \textit{V},
\textit{Lin(V)}, satisfying\\
\begin{equation*}
\rho(ab)=\rho(a)\rho(b)~~~~~~a,b\in A\,.
\end{equation*}\\
\noindent If we now take two vector spaces $V_1$ and $V_2$ and we
want to use the representations of the algebra \textit{A} on them,
$(V_1,\rho)$ and $(V_2,\rho)$, to determine the representation of
\textit{A} on the tensor product of the spaces $(V_1\otimes
V_2,\rho)$ we need a new structure in order to satisfy linearity and
homomorphism property, and to reflect the associativity of the
algebra. This structure is the coproduct, defined as a linear map
that splits an algebra element into a sum of elements belonging to
the tensor product of algebras:\\
\begin{equation}
\Delta:A\rightarrow A\otimes A\,.
\end{equation}\\
\noindent In this way the coproduct is a sum of tensor products and
is indicated as $\Delta(a)=\sum_i a_{(1)}^i\otimes a_{(2)}^i$ or, in
the Sweedler notation, $\Delta(a)=a_{(1)}\otimes a_{(2)}$.

Using the coproduct the representation of \textit{A} is given by\\
\begin{equation}
\rho(a)=((\rho_1\otimes \rho_2)\cdot \Delta(a))(v_1\otimes
v_2)~~~a\in A\,.
\end{equation}\\
\noindent To ensure the homomorphism property of $\Delta$ and
associativity of the algebra, $\Delta$ must satisfy these
conditions\\
\begin{equation}
\Delta(ab)=\Delta(a)\Delta(b),
\end{equation}\\
\begin{equation} \label{eq:coassociativity}
(\Delta\otimes
id)\Delta=(id\otimes\Delta)\Delta~~~~~~~~~(coassociativity)\,.
\end{equation}\\
It is then natural to generalize also the unity in the so-called
co-unity, a map $\epsilon$ such that:\\
\begin{equation}
\epsilon:~~~~~~A\rightarrow \mathbb{C}
\end{equation}\\
\begin{equation} \label{eq:counity}
(1\otimes \epsilon)\cdot \Delta=(\epsilon\otimes 1)\cdot
\Delta=id~~~~~~~~~(counity)\,.
\end{equation}\\
In this way we can give the definition of a coalgebra. A
\textit{coalgebra} \textit{C} is a vector space over a field
$\mathbb{C}$ endowed with a linear coproduct $\Delta:C\rightarrow
C\otimes C$ and a linear counit $\epsilon:C\rightarrow \mathbb{C}$,
which satisfies the \textit{coassociativity}
(\ref{eq:coassociativity}) and \textit{counity} (\ref{eq:counity})
properties.

\textbf{Definition}\textit{.A Hopf algebra $(H, m, \eta; \Delta,
\varepsilon, S)$ is a vector space that is both an algebra and a
coalgebra in a compatible way endowed with a linear antipode map
$S:H\rightarrow H$ such that:}\\
\begin{equation}\label{eq:antipode}
m(S\otimes id)\Delta=m(id\otimes S)\Delta=\eta\epsilon\,;
\end{equation}\\
\noindent the compatibility is given by the following homomorphism
properties\\
\begin{equation}\label{eq:Homomorphism}
\Delta(ab)=\Delta(a)\Delta(b)~~~~~\Delta(1)=1\otimes 1
\end{equation}\\
\begin{equation}
\epsilon(ab)=\epsilon(a)\epsilon(b)~~~~~\epsilon(1)=1
\end{equation}\\
\noindent for all $a,b\in H$. By the definition (\ref{eq:antipode})
it follows that the antipode is unique and satisfies:\\
\begin{equation}
S(a\cdot b)=S(a)S(b),~~~S(1)=1~~~(algebra ~~antirepresentation)
\end{equation}\\
\begin{equation}
(S\otimes S)\Delta(a)=\tau\Delta S(a)
\end{equation}\\
\noindent $a,b\in H$ and $\tau$ represent the flip map
$\tau:\tau(a\otimes b)=b\otimes a$. In a Hopf algebra the antipode
plays a role that generalizes the concept of group inversion.

We want to introduce now the notion of \textit{duality}. Two Hopf
algebras $H$ and $H^*$ are said to be \textit{dually paired} if
there exists a non degenerate inner product $<,>$ such that the
following axioms are satisfied\\
\begin{equation}\label{eq:axiom1}
<ab,c>=<a\otimes b,\Delta(c)>
\end{equation}\\
\begin{equation}\label{eq:axiom2}
<1_{H^*},c>=\epsilon(c)
\end{equation}\\
\begin{equation}\label{eq:axiom3}
<\Delta(a),c\otimes d>=<a,cd>
\end{equation}\\
\begin{equation}\label{eq:axiom4}
\epsilon(a)=<a,1_H>
\end{equation}\\
\begin{equation}\label{eq:axiom5}
<S(a),c>=<a,S(c)>
\end{equation}\\
\noindent where $a,b\in H$ and $<a\otimes b, c\otimes
d>=<a,c><b,d>$.

Note that the relations above may be used constructively, i.e. given
a Hopf algebra $H$, one can construct a dually paired Hopf algebra
$H^*$; this method is used to construct the spacetime coordinate
algebra from the Hopf algebra of the translation generators, as we
will show in the following for $\kappa$-Minkowski spacetime,
obtained by duality from the momenta sector of the
$\kappa$-Poincar\'{e} Hopf algebra.

One can show that to each proposition over an algebra corresponds a
dual proposition over the dual structure that is obtained by
substituting each operation over the algebra with the corresponding
operation over the dual structure. In this way one can establish,
for example, that the dual of a \textit{commutative} Hopf algebra is
\textit{co-commutative}, and vice-versa. In fact, from the
commutativity of $H~~(cd=dc~~\forall c,d\in H)$ it follows\\
\begin{eqnarray*}
<a_{(1)},c><a_{(2)},d>&=&<\Delta(a),c\otimes d> =<a,cd>=<a,dc>= \\
\nonumber\\
&=&<a_{(1)},d><a_{(2)},c>=<a_{(2)},c><a_{(1)},d>\,,\nonumber\\
\end{eqnarray*}\\
\noindent comparing the first and the last members we find that
$\tau\Delta=\Delta$.

\section{Deformation of the Poincar\'{e} algebra and $\kappa$-Minkow\-ski
spacetime} \label{par:DeformedPoincare}

In the framework of Quantum Groups, the deformation of the
Poincar\'{e} group has attracted much attention in the early 1990s
mostly for the motivation arising from Quantum Gravity, in which a
loss of the classical Lorentz symmetry would not be surprising due
to the existence of a \textit{minimum length}. Different approaches
have been attempted in this direction, but interesting developments
have been found in looking for a deformation of the algebra rather
than the group. Following the very powerful technique of contraction
procedure introduced in \cite{Celeghini}, which consider the
q-deformation of the anti-De Sitter algebra $SU(2)_Q$, one recovers
a quantum deformation $U_k(P_4)$ of the Poincar\'{e} algebra $P_4$
which depends on a \textit{dimensionful} parameter $\kappa$. In this
way a fundamental length $\lambda=\kappa^{-1}$ enters the theory.
This quantum algebra has been obtained firstly in \cite{Lukierski2}
in the so-called \textit{standard basis}, whose characteristic
commutation relations are:\\
\begin{equation*}
[P_\mu, P_\nu]=0,
\end{equation*}\\
\begin{equation*}
[M_j, P_0]=0,~~~~~[M_j, P_k]=i\epsilon_{jkl}P_l,
\end{equation*}\\
\begin{equation*}
[N_j, P_0]=iP_j,~~~~~[N_j,
P_k]=i\delta_{jk}\lambda^{-1}\sinh{\lambda P_0},
\end{equation*}\\
\begin{equation*}
[M_j, M_k]=i\epsilon_{jkl}M_l,~~~~~[M_j, N_k]=i\epsilon_{jkl}N_l,
\end{equation*}\\
\begin{equation}
[N_j, N_k]=-i\epsilon_{jkl}(M_l\cosh{\lambda
P_0}-\frac{\lambda^2}{4}P_l\vec{P}\cdot\vec{M})\,,
\end{equation}\\
\noindent where $P_\mu$ are the four-momentum generators, $M_j$ are
the spatial rotation generators and $N_j$ are the boost generators.
The algebra obtained in this way contains the subalgebra of the
classical rotations $O(3)$. The cross-relations between the boost
and the rotation generators are instead deformed, and consequently
the full Lorentz sector do not form a sub-algebra. The coalgebra
sector of the $\kappa$-Poincar\'{e} standard basis is given by:\\
\begin{equation}\label{eq:non-cocommutation}
\Delta (P_0) = P_0 \otimes 1 + 1 \otimes P_0,~~~\Delta (P_j)= P_j
\otimes e^{\frac{\lambda P_0}{2}} + e^{-\frac{\lambda P_0}{2}}
\otimes P_j,
\end{equation}\\
\begin{equation}
\Delta (M_j) = M_j\otimes 1 + 1 \otimes M_j,
\end{equation}\\
\begin{equation}
\Delta (N_j)= N_j \otimes e^{\frac{\lambda P_0}{2}} +
e^{-\frac{\lambda P_0}{2}} \otimes
N_j+\frac{\lambda}{2}\varepsilon_{jkl}(P_k\otimes
M_le^{\frac{\lambda P_0}{2}}+e^{-\frac{\lambda P_0}{2}}M_k\otimes
P_l)\,.
\end{equation}\\
\noindent The mass Casimir $C_\lambda(P)$, i.e. the function that
commutes with all the generators of the algebra, is:\\
\begin{equation}
C_\lambda(P)=(\frac{2}{\lambda} \sinh {\frac{\lambda P_0}{2}
})^2-\vec{P}^2~~~\overrightarrow{_{\lambda\rightarrow0}}~~~P_0^2-\vec{P}^2 \,;
\end{equation}\\
\noindent it provides a deformation of the Casimir of the
Poincar\'{e} algebra $C(P)=P_0^2-\vec{P}^2$.

In the quantum Groups language it is said that the pair of a Hopf
algebra and its dual determines a generalized phase space, i.e. the
space of the generalized momenta and the corresponding generalized
coordinates. The quantum algebra $U_k(P_4)$ contains a translation
subalgebra, and it is natural to consider the dual of the enveloping
algebra of translations as a $\kappa$-Minkowski space. This space
must necessarily be noncommutative, because the duality axioms (see
\ref{eq:axiom1}) state that a non-cocommutative  algebra in the
momenta corresponds to a noncommutative algebra in the spacetime
coordinates. So, the non-cocommutative relations
(\ref{eq:non-cocommutation}) imply that the generators of the dual
space (spacetime coordinates) do not commute.

However, one expects that the quantum deformation of a group
symmetry (such as $U_k(P_4)$) represents, in some sense, a ``quantum
symmetry" for the corresponding homogenous space. In our case, for
example, $\kappa$-Poincar\'{e} is expected to act on
$\kappa$-Minkowski spacetime in a covariant way, preserving its
algebra structure. For this reason, a new $\kappa$-Poincar\'{e}
basis has been introduced, in which the ``covariance" of its action
on $\kappa$-Minkowski is clearly manifest. This is the case of the
Majid-Ruegg \textit{bicrossproduct basis} introduced in
\cite{Majid2}.

One has a large freedom in the choice of the generators of the
quantum algebra $U_k(P_4)$. One can define a very large number of
basis through nonlinear combinations of the generators. Thereby the
choice of the generators of $U_k(P_4)$ is not unique: different
choices of the basis generators modify the form of the
$\kappa$-Poincar\'{e} Hopf algebra in the algebra sector (i.e. the
commutation relations among generators) and in the coalgebra sector
(i.e. the form of the coproduct and the counit). It has been found
in \cite{Majid2} that the $\kappa$-deformed Poincar\'{e} algebra, in
a particular choice of generators basis, has a manifest structure if
bicrossproduct Hopf algebra $U(so(1,3))\triangleright\triangleleft
T$ (see Appendix A), i.e. the semidirect product of the classical
Lorentz group $so(1,3)$ acting in a deformed way on the momentum
sector $T$, and in which also the coalgebra is semidirect with a
back-reaction of the momentum sector on the Lorentz rotations. The
following change of variables:\\
\begin{equation}\label{eq:MajidBasis}
\mathcal{P}_0=-P_0,~~~\mathcal{P}_j=-P_je^{\frac{\lambda
P_0}{2}},~~~\mathcal{N}_j=N_je^{\frac{\lambda
P_0}{2}}-\frac{\lambda}{2}\epsilon_{jkl}M_kP_le^{\frac{\lambda
P_0}{2}}
\end{equation}\\
\noindent leads to the $\kappa$-Poincar\'{e} algebra in the
so-called \textit{Majid-Ruegg bicrossproduct basis} in which the
Lorentz sector is not deformed. The deformation occurs only in the
cross-relations between the Lorentz and translational sectors\\
\begin{equation*}
[\mathcal{P}_\mu, \mathcal{P}_\nu]=0
\end{equation*}\\
\begin{equation*}
[M_j, \mathcal{P}_0]=0
\end{equation*}\\
\begin{equation*}
[M_j, \mathcal{P}_k]=i\epsilon_{jkl}\mathcal{P}_l
\end{equation*}\\
\begin{equation*}
[\mathcal{N}_j, \mathcal{P}_0]=i\mathcal{P}_j
\end{equation*}\\
\begin{equation}\label{eq:MajidAlgebra}
[\mathcal{N}_j,\mathcal{P}_k]=i\delta_{jk}\left(\frac{1}{2\lambda}(1-e^{-2\lambda
\mathcal{P}_0})+\frac{\lambda}{2}\mathcal{P}^2\right)-i\lambda
\mathcal{P}_j\mathcal{P}_k
\end{equation}\\
\noindent and the Lorentz subalgebra remains classical\\
\begin{equation*}
[M_j, M_k]=i\epsilon_{jkl}M_l
\end{equation*}\\
\begin{equation*}
[M_j, \mathcal{N}_k]=i\epsilon_{jkl}\mathcal{N}_l
\end{equation*}\\
\begin{equation}
[\mathcal{N}_j, \mathcal{N}_k]=-i\epsilon_{jkl}M_l\,.
\end{equation}\\
The coproducts are given by\\
\begin{equation*}
\Delta (\mathcal{P}_0) = \mathcal{P}_0 \otimes 1 + 1 \otimes
\mathcal{P}_0
\end{equation*}\\
\begin{equation*}
\Delta (\mathcal{P}_j)= \mathcal{P}_j \otimes e^{-\lambda
\mathcal{P}_0} + 1\otimes \mathcal{P}_j
\end{equation*}\\
\begin{equation*}
\Delta (M_j) = M_j\otimes 1 + 1 \otimes M_j
\end{equation*}\\
\begin{equation}\label{eq:MajidCoproducts}
\Delta (\mathcal{N}_j)= \mathcal{N}_j \otimes e^{-\lambda
\mathcal{P}_0} +1 \otimes\mathcal{N}_j
-\lambda\epsilon_{jkl}\mathcal{M}_k\otimes P_l
\end{equation}\\
\noindent and the antipodes are\\
\begin{equation*}
S(\mathcal{P}_j)=-\mathcal{P}_je^{\lambda
\mathcal{P}_0}\,,~~~S(\mathcal{P}_0)=-\mathcal{P}_0\,,~~~S(M_j)=-M_j\,,
\end{equation*}\\
\begin{equation}\label{eq:antipodes}
S(\mathcal{N}_j)=-\mathcal{N}_je^{\lambda
\mathcal{P}_0}-\lambda\epsilon_{jkl}M_k\mathcal{P}_le^{\lambda
\mathcal{P}_0}\,.
\end{equation}\\
The mass Casimir of this algebra, i.e. the function that commute
with all the generators of the algebra, is given by:\\
\begin{equation}\label{eq:DeformedCasimir}
C_\lambda(P)=(\frac{2}{\lambda} \sinh {\frac{\lambda P_0}{2}
})^2-e^{\lambda
P_0}\vec{P}^2~~~\overrightarrow{_{\lambda\rightarrow0}}~~~P_0^2-\vec{P}^2 \,.
\end{equation}\\
\noindent This deformation of the Poincar\'{e} Casimir has led to
many discussion about the phenomenological implications of a
deformed group symmetry. This is essentially due to the connections
of $\kappa$-Poincar\'{e} with $\kappa$-Minkowski spacetime, in which
the relation (\ref{eq:DeformedCasimir}) is considered to have the
interpretation of deformed dispersion relation for particle.

$\kappa$-Minkowski noncommutative spacetime, whose coordinates
satisfy the commutation relations (\ref{eq:kappaMinkowski}), is
shown to be the spacetime associated to $\kappa$-Poincar\'{e}
algebra. In fact, expressing the $\kappa$-Poincar\'{e} generators in
this basis (which from now on we denote by $(P_\mu,M_j,N_j)$), it is
possible to see that $\kappa$-Poincar\'{e} acts covariantly as a
Hopf algebra on $\kappa$-Minkowski spacetime as shown in Appendix A.
In this way the commutation relations (\ref{eq:kappaMinkowski}) that
characterize $\kappa$-Minkowski remain unchanged under the action of
$\kappa$-Poincar\'{e} algebra, this is consistent with the notion of
quantum group symmetry that should preserve a covariant action over
the associated homogeneous space. In the limit $\lambda\rightarrow0$
one recovers the standard Minkowski space, with the ordinary
Poincar\'{e} group.

 In this basis it is very easy to show that the
dual Hopf algebra $T^*$ of the translation sector $T$ of the
$\kappa$-Poincar\'{e} algebra, $T\subset U_k(P_4)$, is the Hopf
algebra of the $\kappa$-Minkowski generators $\hat{x}_\mu$. We can
assume the duality relations\\
\begin{equation}
<\hat{x}_\mu,P_\nu>=-i\eta_{\mu\nu}
\end{equation}\\
\noindent and, applying the duality axioms, we can determine the
Hopf algebra of $\hat{x}_\mu$ if we know the Hopf algebra of
$P_\mu$. For example, using the axiom (\ref{eq:axiom1}) and the
coproduct (\ref{eq:MajidCoproducts}), one finds:\\
\begin{eqnarray*}
<[\hat{x}_j,\hat{x}_0],P_k>&=&<\hat{x}_j\otimes
\hat{x}_0,\Delta(P_k)>-<\hat{x}_0\otimes \hat{x}_j,\Delta(P_k)>=
\\ \nonumber\\
&=&<\hat{x}_j\otimes \hat{x}_0,P_k\otimes e^{-\lambda P_0}+1\otimes
P_k>+\\ \nonumber\\
&-&<\hat{x}_0\otimes \hat{x}_j,P_k\otimes e^{-\lambda
P_0}+1\otimes P_k>=
\\ \nonumber\\
&=&<\hat{x}_j,P_k><\hat{x}_0,e^{-\lambda
P_0}>-<\hat{x}_0,1><\hat{x}_j,P_k>=
\\ \nonumber\\
&=&-\lambda <\hat{x}_j,P_k><\hat{x}_0,P_0>=<i\lambda\hat{x}_j,P_k>\,,\nonumber\\
\end{eqnarray*}\\
\noindent from which it follows:\\
\begin{equation*}
[\hat{x}_j,\hat{x}_0]=i\lambda\hat{x}_j\,;
\end{equation*}\\
\noindent this is the non-zero commutator between space and time
``coordinates" of $\kappa$-Minkowski.\\

\section{Fields in $\kappa$-Minkowski and Weyl
maps}\label{par:Weylmaps}

For the development of a field theory in $\kappa$-Minkowski it is
fundamental to have a convenient characterization of the concept of
field as function of noncommutative variables. In our description
and handling of functions of the noncommuting coordinates (fields in
the noncommutative geometry) an important role will be played by
\textit{Weyl maps}, which allow to introduce structures for the
functions of the noncommuting coordinates in terms of the
corresponding structures that are meaningful for the ordinary
functions.

Weyl maps establish a correspondence between elements of
$\kappa$-Minko-wski and analytic functions of four variables $x_\mu$
that commute. This correspondence is not unique, i.e. there exist
several Weyl maps which can be defined. Thereby a coherence
criterion for the proposed theories is that of Weyl map choice
independence.

Among the several Weyl maps that one can define, it can be useful to
consider two explicit examples, which we denote by $\Omega_R$ and
$\Omega_S$, so that we get some intuition for the differences which
may arise. To characterize the Weyl maps $\Omega_R$ and $\Omega_S$
let us consider a simple function $f(x)=x_jx_0$ of the Minkowski
commutative spacetime. The action on the function $f$ of the
\textit{time-to-the-right} $\Omega_R$  Weyl map and of the
\textit{time-symmetrized} Weyl map $\Omega_S$ are the following:\\
\begin{equation*}
\Omega_R(f)=\hat{x}_j\hat{x}_0~~~~~~~~~~
\Omega_S(f)=\frac{1}{2}(\hat{x}_j\hat{x}_0+\hat{x}_0\hat{x}_j) \,.
\end{equation*}\\
\noindent These two maps are related to two possible orderings that
one can choose for the noncommutative functions of coordinates in
$\kappa$-Minkowski spacetime.

It is sufficient to specify the Weyl map on the complex exponentials
and extend it to the generic function $\Omega_{R,S}(f(x))$, whose
Fourier transform is $\tilde{f}(p)=\frac{1}{(2\pi)^4}\int
f(x)e^{-ipx}d^4x$, by linearity\\
\begin{equation}
\Omega_{R,S}(f(x))=\int \tilde{f}(p)\Omega_{R,S}(e^{ipx})d^4p\,.
\end{equation}\\
The $\Omega_R$ Weyl map is implicitly defined through\\
\begin{equation}
\Omega_R(e^{ipx})=e^{i\vec{p}\hat{\vec{x}}}e^{-ip_0\hat{x}_0} \,,
\end{equation}\\
\noindent while the alternative $\Omega_S$ Weyl map is such that\\
\begin{equation}
\Omega_S(e^{ipx})=e^{-i\frac{p_0\hat{x}_0}{2}}e^{i\vec{p}\hat{\vec{x}}}e^{-i\frac{p_0\hat{x}_0}{2}}  \,,
\end{equation}\\
\noindent where $p_\mu$ are four real commutative parameters. It is
so possible, using the definition of Fourier transform we have in
Minkowski commutative spacetime, to define the function of
$\kappa$-Minkowski ordered through the two maps $\Omega_R$ and
$\Omega_S$ as the Fourier integrals with the $\kappa$-Minkowski
exponentials ordered through the two maps.

Notice that it is possible to go from time-to-the-right to
time-symme-trized ordering through a transformation of the Fourier
parameters\\
\begin{equation}
\Omega_R(e^{ipx})=\Omega_S(e^{i\vec{p}e^{\frac{\lambda
p_0}{2}}\vec{x}-ip_0x_0})\,.
\end{equation}\\
In the development of a field theory, with these fields of
noncommuting spacetime coordinates, the description of products of
fields plays of course a central role. And it is useful to describe
the product of two fields $F$ and $G$ of noncommuting spacetime
coordinates in terms of (correspondingly deformed) rule of product
for the commuting fields $f$ and $g$ through the Weyl map:\\
\begin{equation*}
F=\Omega(f)~~~~~~~~ G=\Omega(g)\,.
\end{equation*}\\
\noindent Of course, as a result of the noncommutativity, the
product $FG$ cannot be described as $\Omega(fg)$. Instead one has
that $FG=\Omega(f\star g)$, where\\
\begin{equation}
(f\star g)=\Omega^{-1}(\Omega (f)\Omega(g))
\end{equation}\\
\noindent is the ``$\star$-product" (often also called Moyal
product).

In the case of the $\Omega_R$ and $\Omega_S$ Weyl maps for
$\kappa$-Minkowski one finds:\\
\begin{equation*}
\Omega_R(e^{ipx})\cdot\Omega_R(e^{iqx})=\Omega_R(e^{ipx}\star_Re^{iqx})=\Omega_R(e^{i(\vec{p}+\vec{q}e^{-\lambda
p_0})\vec{x}-i(p_0+q_0)x_0})  \,,
\end{equation*}\\
\begin{equation*}
\Omega_S(e^{ipx})\cdot\Omega_S(e^{iqx})=\Omega_S(e^{ipx}\star_Se^{iqx})=\Omega_S(e^{i(\vec{p}e^{\frac{\lambda
q_0}{2}}+\vec{q}e^{\frac{-\lambda p_0}{2}})\vec{x}-i(p_0+q_0)x_0})        \,.
\end{equation*}\\
The Weyl map can also be used to introduce a notion of integration
in the noncommutative spacetime. We can assume a rule of integration
that is naturally expressed using the $\Omega_R$ Weyl map\\
\begin{equation}\label{eq:Rintegration}
\int_R \Omega_R(f)=\int f(x)d^4x
\end{equation}\\
\noindent which states that the integral of a right-ordered function
of $\kappa$-Minkowski corresponds exactly to the integral of the
corresponding commutative function. In this way the right integral
of a right-ordered exponential corresponds to the standard delta
function:\\
\begin{equation}
\frac{1}{(2\pi)^4}\int_Re^{i\vec{k}\hat{\vec{x}}}e^{-ik_0\hat{x}_0}=\frac{1}{(2\pi)^4}\int
d^4x\Omega_R^{-1}\Omega_R(e^{ikx})=\delta^4(k)   \,.
\end{equation}\\
\noindent This rule has been largely investigated in literature (see
for example \cite{Amelino7}). Our alternative choice of Weyl map
would naturally invite us to consider the integration rule\\
\begin{equation}\label{eq:Sintegration}
\int_S \Omega_S(f)=\int f(x)d^4x\,.
\end{equation}\\
\noindent Actually these integrals are equivalent, i.e. $\int_R
\Phi=\int_S \Phi$ for each element $\Phi$ of $\kappa$-Minkowski.
This is easily verified by expressing the most general element of
$\kappa$-Minkowski both in its $\Omega_R$-inspired form and its
$\Omega_S$-inspired form\\
\begin{equation*}
\Phi=\int d^4p \tilde{f}(p)\Omega_R(e^{ipx})=\int d^4p
\tilde{f}(p_0,\vec{p}e^{-\frac{\lambda p_0}{2}})e^{-\frac{3\lambda
p_0}{2}}\Omega_S(e^{ipx})
\end{equation*}\\
\noindent and observing that\\
\begin{equation}\label{eq:IntegrationRule}
\int_R \Phi=\int_S \Phi=(2\pi)^4\tilde{f}(0)\,.
\end{equation}\\
\noindent Because of the equivalence we will omit
indices $R$ or $S$ on the integration symbol.\\

\section{Free scalar fields in classical
Minkowski}\label{par:ScalarFieldClassical}

While for the canonical noncommutative spacetimes the naive choice
of action $S(\Phi)=\int d^4x\Phi(\partial^2 -M^2)\Phi$ (for free
scalar fields) is fully satisfactory, in the description of free
scalar fields in $\kappa$-Minkowski a nontrivial choice of action
emerges very naturally. This originates from the desire to work with
a ``maximally symmetric" action, and in the case of
$\kappa$-Minkowski it is possible to introduce an action which is
invariant under the 10 Poincar\'{e}-like symmetries, but this action
has nontrivial form.

In preparation for this $\kappa$-Minkowski analysis we find useful
to devote this section to a description of the simple action
$S(\Phi)=\int d^4x\Phi(\partial^2 -M^2)\Phi$ for a free scalar field
$\Phi$ in commutative Minkowski spacetime
($\partial^2=\partial_\mu\partial^\mu$ is the familiar D'Alambert
operator).

Let us start by introducing some notation and convention for the
description of symmetry transformations. The most general
infinitesimal transformation is of the form $x'_\mu=x_\mu+\epsilon
A_\mu(x)$, with $A_\mu$ four real functions of the coordinates.

A field is scalar if $\Phi'(x')=\Phi(x)$, and in leading order in
$\epsilon$ one finds\\
\begin{equation*}
\Phi'(x')-\Phi(x)={\partial^\mu\Phi(x)}(x_\mu-x'_\mu)=-\epsilon
A_\mu(x)\partial^\mu\Phi(x)\,;
\end{equation*}\\
\noindent in terms of the generator $T$ of the transformation,
$T=iA_\mu(x)\partial^\mu$, one obtains $x'=(1-i\epsilon T)x$ and
$\Phi'=(1+i\epsilon T)\Phi$.

Correspondingly the variation of the action can be written as\\
\begin{equation*}
S(\Phi')-S(\Phi)=i\epsilon\int d^4x \left(T\{\Phi(\partial^2
-M^2)\Phi\}+\Phi[\partial^2,T]\Phi\right)=
\end{equation*}\\
\begin{equation*}
=i\epsilon\int d^4x \left(TL(x)+\Phi[\partial^2,T]\Phi\right)
\end{equation*}\\
\noindent and therefore the action is invariant under $T$-generated
transformations,\\
\begin{equation*}
S(\Phi')-S(\Phi)=0
\end{equation*}\\
\noindent if and only if\\
\begin{equation}\label{eq:SymmetryDefinition}
\int d^4x \left(TL(x)+\Phi[\partial^2,T]\Phi\right)=0\,.
\end{equation}\\
For the action $S(\Phi)=\int d^4x\Phi(\partial^2 -M^2)\Phi$ in
classical Minkowski spacetime it is well established that the
symmetries are described in terms of the classical Poincar\'{e}
algebra, generated by the elements\\
\begin{equation*}
P_\mu=-i\partial_\mu\,,~~~~~M_j=\epsilon_{jkl}x_kP_l\,,~~~~~~N_j=x_jP_0-x_0P_j\,,
\end{equation*}\\
\noindent which satisfy the commutation relations\\
\begin{equation*}
[P_\mu,P_\nu]=0\,,~~~~~~[M_j,P_0]=0\,,~~~~~~[M_j,P_k]=i\epsilon_{jkl}P_l \,,
\end{equation*}\\
\begin{equation*}
[M_j,M_k]=i\epsilon_{jkl}M_l\,,~~~~~~[M_j,N_k]=i\epsilon_{jkl}N_l \,,
\end{equation*}\\
\begin{equation}
[N_j,P_0]=iP_j\,,~~~~~~[N_j,P_k]=i\delta_{jk}P_0\,,~~~~~~[N_j,N_k]=-i\epsilon_{jkl}M_l\,.
\end{equation}\\
\noindent The operator $\partial^2=-P_\mu P^\mu$ is the first
Casimir of the algebra, and of course satisfies $[\partial^2,T]=0$.

For this case of a maximally-symmetric theory in commutative
Mink-owski spacetime it is conventional to describe the symmetries
fully in terms of Poincar\'{e} Lie algebra. For $\kappa$-Minkowski
noncommutative spacetime a description of symmetry in terms of a
Hopf algebra turns out to be necessary. But we must stress that
essentially the difference between symmetries described in terms of
a Lie algebra and symmetries described in terms of a Hopf algebra
resides in the description of the action of symmetry transformations
on products of functions: if for all generators $T_a$ one finds that
$T_a(fg)=[T_a(f)]g+f[T_a(g)]$, one may say that the coproduct is
trivial and a description in terms of a Lie algebra is sufficient,
whereas for the case when the coproduct is nontrivial one speaks of
a Hopf-algebra symmetry.

Once the algebra properties are specified (action of symmetry
transformation on functions of noncommutative coordinates) the
property of the counit, coproduct and antipode can always be
formally derived, but this will not in general satisfy the Hopf
algebra criteria since they may require the introduction of new
operators, not included in the algebra sector. If this does not
occur (if the counit, coproduct and antipode that one obtains on the
basis of the algebra sector can be expressed fully in terms of
operators in the algebra) the Hopf-algebra criteria are
automatically satisfied.\\

\section{Symmetry analysis in $\kappa$-Minkowski
spacetime} \label{par:SymmetryAnalysis}

We want to discuss in this section the form of the action for a free
scalar field in $\kappa$-Minkowski which most naturally replaces the
$S(\Phi)=\int d^4x\Phi(\partial^2 -M^2)\Phi$ action of the
classical-Minkowski case assuming the integration rule
(\ref{eq:IntegrationRule}).

A key point is that it is possible to introduce an action for a free
scalar field in $\kappa$-Minkowski which is invariant under
translations, space-rotations and boosts, in the Hopf-algebra sense.

By straightforward generalization of the result
(\ref{eq:SymmetryDefinition}) reviewed in the previous section, a
symmetry transformation $T$ must be such that\\
\begin{equation}
\int d^4x \left(T\{\Phi(\partial^2_\lambda
-M^2)\Phi\}+\Phi[\partial^2_\lambda,T]\Phi\right)=0\,,
\end{equation}\\
\noindent if the action takes the form\\
\begin{equation*}
S(\Phi)=\int d^4x\Phi(\partial^2_\lambda -M^2)\Phi
\end{equation*}\\
\noindent with $\partial^2_\lambda$ to be determined.

The next step is the description of the Poincar\'{e}-like symmetries
which will be implemented as invariances of the action. One of
course wants to introduce a description of translations,
space-rotations and boosts that follows as closely as possible the
analogy with the well-established descriptions that apply in the
commutative limit $\lambda\rightarrow0$. Since functions in
$\kappa$-Minkowski can be fully described in terms of Weyl map, and
since the Weyl map are fully specified once given on Fourier
exponentials, one can, when convenient, confine the discussion to
the Fourier exponentials.

\subsection{Translations} \label{par:Translations}

Since in classical Minkowski the translation generator acts
according to\\
\begin{equation}
P_\mu(e^{ikx})=k_\mu e^{ikx}
\end{equation}\\
\noindent in an analysis of $\kappa$-Minkowski based on the
time-to-the-right Weyl map it is natural to define translations as
generated by the operators $P_\mu^R$ such that\\
\begin{equation}
P_\mu^R\Omega_R(e^{ikx})=k_\mu \Omega_R(e^{ikx})\,.
\end{equation}\\
Since, as mentioned, the exponentials
$e^{i\vec{k}\vec{\hat{x}}}e^{-ik_0\hat{x}_0}$ form a basis of
$\kappa$-Minkowski, in order to establish the form of the action of
these translation generators on products of functions of the
$\kappa$-Minkowski coordinates, the structure which is codified in
the coproduct $\Delta P_j^R$, one can simply observe that\\
\begin{eqnarray}
P_j^R\Omega_R(e^{ikx})\Omega_R(e^{ipx})\!\!\!&=&\!\!\!-i\Omega_R(\partial_je^{i(k\dot{+}p)x})=\nonumber\\
\!\!&=&\!\!\!-i\Omega_R((k\dot{+}p)_je^{i(k\dot{+}p)x})=\nonumber\\
\!\!\!\!&=&\!\!\!\![P_j^R\Omega_R(e^{ikx})][\Omega_R(e^{ipx})]+[e^{-\lambda
P_0^R}\Omega_R(e^{ikx})][P_j^R\Omega_R(e^{ipx})]\,, \nonumber\\
\end{eqnarray}\\
\noindent where $k\dot{+}p\equiv (k_0+p_0,\vec{k}+e^{-\lambda
k_0}\vec{p})$. This is conventionally described by the symbolic
notation\\
\begin{equation}
\Delta P_j^R=P_j^R\otimes1+e^{-\lambda P_0^R}\otimes P_j^R \,.
\end{equation}\\
\noindent Following an analogous procedure one can derive\\
\begin{equation}
\Delta P_0^R=P_0^R\otimes1+1\otimes P_0^R\,,
\end{equation}\\
\noindent i.e., while for space translations one has a nontrivial
coproduct, for time translations the coproduct is trivial.

Using the full machinery of the mathematics of Hopf algebras one can
verify that the quadruplet of operators $P_\mu^R$ does give rise to
a genuine Hopf algebra of translation-like symmetry transformations.

\subsection{Rotations} \label{par:Rotations}

Following the same idea that allows us to introduce translations in
$\kappa$-Minkowski, we attempt now to obtain a 7-generators Hopf
algebra, describing four transla\-tion-like operators and three
rotation-like generators.

For what concerns the translations we have found that an acceptable
Hopf-algebra description was obtained by straightforward
``quantization"  of the classical translations: the $P_\mu^R$
translations were just obtained from the commutative-spacetime
translations through the $\Omega_R$ Weyl map. Also for rotations
this strategy turns out to be successful:\\
\begin{equation}
M_j^R\Omega_R(f)=\Omega_R(M_jf)=\Omega_R(-i\epsilon_{jkl}x_k\partial_lf)\,.
\end{equation}\\
And, while for the (spatial) translations one finds nontrivial
coproduct, the coproduct of rotations is trivial:\\
\begin{equation}
\Delta M_j=M_j\otimes1+1\otimes M_j\,;
\end{equation}\\
\noindent it is also straightforward to verify that\\
\begin{equation}
[M_j,M_k]=i\epsilon_{jkl}M_l\,.
\end{equation}\\
\noindent Therefore the triplet $M_j$ forms a 3-generator Hopf
algebra that is completely undeformed (classical) both in the
algebra and coalgebra sectors. (Using the intuitive description
introduced earlier this is a trivial rotation Hopf algebra, whose
structure could be equally well captured by the standard Lie algebra
of rotations.)

There is therefore a difference between the translations sector and
the rotations sector. Both translations and rotations can be
realized as straightforward (up to ordering) quantization of their
classical actions, but while for rotations even the coalgebraic
properties are classical (trivial coalgerba) for the translations we
found a nontrivial coalgebra sector.

Our translations and rotations can be put together straightforwardly
to obtain a 7-generator translations-rotations symmetry Hopf
algebra. It is sufficient to observe that\\
\begin{equation}
[M_j,P_\mu^R]\Omega(e^{ikx})=\epsilon_{jkl}\Omega([-x_k\partial_\mu+\partial_\mu
x_k]\partial_le^{ikx})= \delta_{\mu
k}\epsilon_{jkl}\Omega(\partial_le^{ikx})
\end{equation}\\
\noindent from which it follows that\\
\begin{equation}
[M_i,P_j^R]=i\epsilon_{ijk}P_k^R,~~~[M_i,P_0^R]=0\,,
\end{equation}\\
\noindent i.e. the action of rotations on energy-momentum is
undeformed. Accordingly, the generators $M_j$ can be represented as
differential operators over energy-momentum space in the familiar
way: $M_j=-i\epsilon_{jkl}P_k\partial_{P_l}$.

\subsection{Boosts} \label{par:Boosts}

In the analysis of translations and rotations in $\kappa$-Minkowski
we have already encountered two different situations: rotations are
essentially classical in all respects, while translations have a
``classical" action (straightforward $\Omega$-map ``quantization" of
the corresponding classical action) but have nontrivial coalgebraic
properties (nontrivial coproduct). Of course, the fact that some
symmetry transformations in a noncommutative spacetime allow
``classical" description (through the Weyl map) is not to be
expected in general. In general one can only require that the
results should reproduce the familiar ones for commutative Minkowski
in the limit of vanishing noncommutativity parameters
($\lambda\rightarrow0$). As we now intend to include also boosts,
and obtain 10-generator symmetry algebras, we encounter another
possibility: for boosts non only the coalgebra sector is nontrivial
but even the action cannot be obtained by ``quantization" of the
classical action.

The ``classical" boosts $N_j^R$ should have action\\
\begin{equation}
N_j^R\Omega_R(f)=\Omega_R(N_jf)=\Omega_R(i[x_0\partial_j-x_j\partial_0]f)\,.
\end{equation}\\
\noindent And actually it is easy to see (and it is obvious) that
these boosts combine with the rotations $M_j^R$ to close the
(undeformed) Lorentz algebra, and that adding also the translations
$P_\mu^R$ one obtains the undeformed Poincar\'{e} algebra. However,
these algebras cannot be extended (by introducing a suitable
coalgebra sector) to obtain a Hopf algebra of symmetries of theories
in our noncommutative $\kappa$-Minkowski spacetime. In particular,
one finds an inconsistency in the coproduct of these boosts $N_j^R$,
which signals an obstruction originating from an inadequacy in the
description of the action of boosts on (noncommutative) products of
$\kappa$-Minkowski functions. The problem is that $\Delta(N_j^R)$
would not be an element of the algebraic tensor product, i.e. it is
not a function only of the elements $M$, $N$, $P$.

Since the ``classical" choice $N_j^R$ is inadequate there are two
possible outcomes: either there is no 10-generator symmetry-algebra
extension of the 7-generator symmetry algebra $(P_\mu^R,M_j)$ or the
10-generator symmetry-algebra extension exists but requires
nonclassical boosts. The latter is true.

The generators of the needed modified boost action, $\mathcal{N}_j$,
are found through a rather tedious analysis which can be found in
literature \cite{Amelino4} and we do not report in detail here. One
starts by observing that, by imposing that the deformed boost
generator $\mathcal{N}_j$ (although possibly having a nonclassical
action) transform as a vector under spatial rotations, the most
general form of $\mathcal{N}_j$ is\\
\begin{eqnarray*}
\mathcal{N}_j\Omega(\Phi(x))&=&\Omega\{[ix_0A(-i\partial_x)\partial_j+\lambda^{-1}x_jB(-i\partial_x)+\\
\nonumber\\
&-&\lambda
x_lC(-i\partial_x)\partial_l\partial_j-i\epsilon_{jkl}x_kD(-i\partial_x)\partial_l]\Phi(x)\}\,,\nonumber\\
\end{eqnarray*}
\noindent where $A$,$B$,$C$,$D$ are unknown functions of $P_\mu^R$
(in the classical limit $A=i$, $D=0$; moreover, as
$\lambda\rightarrow0$ one obtains the classical limit if $\lambda
C\rightarrow0$ and $B\rightarrow\lambda P_0$).

Imposing that in the formula above\\
\begin{equation*}
\mathcal{N}_j^R[\Omega(e^{ikx})\Omega(e^{ipx})]=[\mathcal{N}_{(1),j}^R\Omega(e^{ikx})]
[\mathcal{N}_{(2),j}^R\Omega(e^{ipx})]
\end{equation*}\\
\noindent it should be possible to write $\mathcal{N}_{(1),j}$ and
$\mathcal{N}_{(2),j}$ in terms of generators of the Hopf algebra,
one clearly obtains some constraints on the function
$A$,$B$,$C$,$D$. The final result is\\
\begin{equation}
\mathcal{N}_j^R\Omega_R(f)=\Omega_R([ix_0\partial_j+
x_j(\frac{1-e^{2i\lambda\partial_0}}{2\lambda}-\frac{\lambda}{2}\nabla^2)-\lambda
x_l\partial_l\partial_j]f)\,.
\end{equation}\\
It is easy to verify that the Hopf algebra
$(P_\mu^R,M_j^R,\mathcal{N}_j^R)$ satisfy all the requirements for a
candidate symmetry algebra for theories in $\kappa$-Minkowski
spacetime.\\

\section{Mass Casimir and deformed Klein-Gordon
e\-quation}\label{par:DeformedKlein-Gordon}

Of course, the fact that one replaces the ``classical" Poincar\'{e}
Lie algebra with the ``quantum" deformed-Poincar\'{e} Hopf algebra
has some striking consequences. For what concerns the search of a
description of scalar fields the key ingredient is to find the
``mass Casimir" in the quantum version, i.e. a differential operator
$\Box_\lambda$ , which in the classical limit reduces to the
D'Alambert operator $\Box=\partial^2$, such that the action\\
\begin{equation}\label{eq:DeformedAction}
S(\Phi)=\int d^4x\Phi(\Box_\lambda -M^2)\Phi
\end{equation}\\
\noindent is invariant under the realization of Hopf-algebra
symmetry we have constructed $(P_\mu^R,M_j^R,\mathcal{N}_j^R)$. We
therefore must verify that, for some choice of
$\Box_\lambda$,$[\Box_\lambda,T]=0$ for every $T$ in the Hopf
algebra.

Guided by the intuition that $\Box_\lambda$ should be a scalar with
respect to $(P_\mu^R,M_j^R,\\\mathcal{N}_j^R)$ transformations, one
is led to the proposal\\
\begin{equation}
\Box_\lambda=\left(\frac{2}{\lambda}\sinh (\frac{\lambda
P_0^R}{2})\right)^2-e^{\lambda P_0^R}(\vec{P}^R)^2\,.
\end{equation}\\
In fact, it is easy to verify that with this choice of
$\Box_\lambda$ the action (\ref{eq:DeformedAction}) is invariant
under the $(P_\mu^R,M_j^R,\mathcal{N}_j^R)$ transformations.
Therefore, we have finally managed to construct an action describing
free scalar fields in $\kappa$-Minkowski that enjoys 10-generator
(Hopf-algebra) symmetries $(P_\mu^R,M_j^R,\mathcal{N}_j^R)$.

Since the Casimir $\Box_\lambda$ is a scalar, we can ask a free
scalar field theory to satisfy the Klein-Gordon-like motion equation
with respect to the deformed D'Alambert operator $\Box_\lambda$:\\
\begin{equation}
(\Box_\lambda-M^2)\Phi(x)=0\,.
\end{equation}\\
\noindent This equation of motion can be obtained from the variation of the
action (\ref{eq:DeformedAction}), as shown in \cite{Amelino3}.

\clearpage

\chapter{\textbf{Noether analysis with four-dimensional differential
calculus}}\label{ch:Analysis4D}

The new result reported in this thesis is a Noether analysis of
translation symmetries in $\kappa$-Minkowski using a
five-dimensional bicovariant differential calculus. In preparation
for that derivation, which is the subject of the next chapter, we
find useful to review briefly the known result for the corresponding
Noether analysis of \cite{Amelino3} with the four-dimensional
differential calculus. We will present the analysis for the massless
case and show at the end of the chapter how, for a complex plane
wave field, the expression for the translation-symmetry conserved
charges obtained gives a non-linear Planck-scale
modification of the dispersion relation.\\

\section{Translation transformation and 4D differential calculus}
\label{par:4DDifferentialCalculus}

Before \cite{Amelino3}, previous attempts to derive
translation-symmetry conserved charges in $\kappa$-Minkowski
noncommutative spacetime failed due to the adoption of a rather
naive description of translation transformation, which in particular
did not take into account the properties of the noncommutative
$\kappa$-Minkowski differential calculus. In \cite{Amelino3} it was
shown that by taking properly into account the properties of the
differential calculus one encounters no obstruction in following all
the steps of the Noether analysis and one obtains an explicit
formula relating fields and energy-momentum charges.

In order to characterize translation transformations, if one
concentrates on the infinitesimal translation parameters, rather
than the generators, and tries to enforce in $\kappa$-Minkowski the
view of infinitesimal translation as a map $x_\mu\rightarrow
x_\mu+\epsilon_\mu$, as customary in the commutative limit, then one
finds that the translation parameters must have nontrivial algebraic
properties\\
\begin{equation}\label{eq:EpsilonCommutation}
[\epsilon_j,x_0]=i\lambda
\epsilon_j\,,~~~[\epsilon_j,x_k]=0\,,~~~[\epsilon_0,x_\mu]=0
\end{equation}\\
\noindent in order to ensure that the ``point" $x+\epsilon$ still
belongs to the $\kappa$-Minkowski spacetime:\\
\begin{equation}
[x_j+\epsilon_j,x_0+\epsilon_0]=i\lambda
(\epsilon_j+x_j)\,,~~~[x_i+\epsilon_i,x_j+\epsilon_j]=0\,.
\end{equation}\\
\noindent These algebraic relations reflect the known properties of
the $\kappa$-Minkowski differential calculus \cite{Majid5} (the
$\epsilon$'s describe the difference between the coordinates of two
spacetimes points and are therefore related to the $dx$'s of the
differential calculus).

In order to perform the Noether analysis it is necessary to describe
the action of translation transformations on the fields $f$, which
will be of the type $f\rightarrow f+df$. The definition of $df$ has
not to be treated as a freedom allowed by the formalism: the
exterior derivative operator $d$ must of course satisfy the Leibnitz
rule\\
\begin{equation}\label{eq:Leibnitz}
d(f\cdot g) =df\cdot g+f\cdot dg\,.
\end{equation}\\
\noindent If we consider the translation transformation in the
commutative case, we have $df=i[P^\mu f(x)]\epsilon_\mu$. If one
tries to extend this definition to $\kappa$-Minkowski just
substituting the commutative translation generators with the
$\kappa$-Poincar\'{e} ones, the $P_\mu$ translation generators of
the Majid-Ruegg $\kappa$-Poincar\'{e} basis (\ref{eq:MajidBasis}),
(\ref{eq:MajidAlgebra}), Leibnitz rule (\ref{eq:Leibnitz}) cannot be
satisfied due to the nontrivial coproduct of $P_\mu$. It is crucial
for the analysis to observe that the form of the generators $P_\mu$
and the properties of the infinitesimal translation parameters
$\epsilon_\mu$ must be combined in the description of the $df$. And
the fact that in the $\kappa$-Minkowski case the transformation
parameters have nontrivial algebraic properties poses an ordering
issue, there is in fact an infinity of different formulations of the
$df$ which all reduce to $df=i[P^\mu f(x)]\epsilon_\mu$ in the
classical-spacetime (commutative) limit.

Taking into account the $\epsilon$'s algebraic properties
(\ref{eq:EpsilonCommutation}) and the coalgebra of
$\kappa$-Poincar\'{e} translation generators, one easily finds that
the requirement (\ref{eq:Leibnitz}) singles out the formula\\
\begin{equation}\label{eq:df4D}
df=i\epsilon_\mu P^\mu f(x)\,.
\end{equation}\\
\noindent It is through this formula, involving both generators and
infinitesimal parameters, that one can truly characterize the
translation transformations. The exclusive knowledge of the
translation generators properties is clearly insufficient.\\

\section{Noether analysis for the massless case}\label{par:4DNoetherAnalysism=0}

It is easy to verify that this improved description of translation
transformations actually allows to complete the Noether analysis,
thereby obtaining the energy-momentum charges.

We perform the Noether analysis for a theory of massless free scalar
fields solutions of the following much studied
\cite{Lukierski,Kowalski,Amelino4}, Klein-Gordon-like equation\\
\begin{equation}\label{eq:MotionEquationm=0}
C_\lambda(P_\mu)\Phi=\left[\left(\frac{2}{\lambda}\sinh
(\frac{\lambda P_0}{2})\right)^2-e^{\lambda
P_0}(\vec{P})^2\right]\Phi=0\,.
\end{equation}\\
\noindent This equation of motion can be derived from the following
action\\
\begin{equation}\label{eq:Actionm=0}
S[\Phi]=\int d^4x \mathcal{L}[\Phi(x)]=-\frac{1}{2}\int d^4x
\tilde{P^\mu} \Phi \tilde{P_\mu} \Phi \,,
\end{equation}\\
\noindent where we introduced the compact notation
$\tilde{P_\mu}$,\\
\begin{equation} \label{eq:Ptilde}
\tilde{P_0} = \frac{2}{\lambda} \sinh {\frac{\lambda}{2} P_0} ~~~~~~~~~
\tilde{P_i} = P_i e^{\frac{\lambda}{2} P_0},
\end{equation}\\
\noindent which also allows to rewrite $C_\lambda(P_\mu)$ as
$\tilde{P_\mu}\tilde{P^\mu}$. The most general solution of eq.
(\ref{eq:MotionEquationm=0}) can be written as\\
\begin{equation} \label{eq:Solutionm=0}
\Phi(x)=\int d^4k \tilde{f}(k_0,\vec{k})e^{i\vec{k} \cdot
\vec{x}}e^{- k_0x_0}\delta(C_\lambda(k_\mu))\,.
\end{equation}\\
If we now consider the variation $\delta\Phi$ applied to the field
$\Phi$\\
\begin{equation}
\Phi~~\rightarrow~~\Phi'=\Phi+\delta\Phi\,,
\end{equation}\\
\noindent the action then varies according to\\
\begin{equation*}
\delta S[\Phi]=\int
d^4x(\mathcal{L}[\Phi'(x')]-\mathcal{L}[\Phi(x)])=
\end{equation*}\\
\begin{equation*}
=\frac{1}{2}\int d^4x \left\{e^{\frac{\lambda
P_0}{2}}\left[[(\tilde{P_\mu}\tilde{P^\mu})\Phi]\delta\Phi\right]+e^{-\frac{\lambda
P_0}{2}}\left[\delta\Phi(\tilde{P_\mu}\tilde{P^\mu})\Phi\right]\right\}+
\end{equation*}\\
\begin{equation} \label{eq:Variation}
+\int d^4x\left\{-\frac{1}{2}\tilde{P^\mu}\left[e^{\frac{\lambda
P_0}{2}}\tilde{P_\mu}\Phi\delta\Phi+\delta\Phi e^{-\frac{\lambda
P_0}{2}}\tilde{P_\mu}\Phi\right]+\mathcal{L}[\Phi(x')]-\mathcal{L}[\Phi(x)]\right\} \,,
\end{equation}\\
\noindent where we also used the observation that\\
\begin{equation}
\tilde{P_\mu}[f(x)g(x)]=[\tilde{P_\mu}f(x)][e^{\frac{\lambda}{2}P_0}g(x)]+
[e^{-\frac{\lambda}{2}P_0}f(x)][\tilde{P_\mu}g(x)]
\end{equation}\\
\noindent for any field $f(x)$ and $g(x)$.

In (\ref{eq:Variation}) there are two separated integrals: the first
integral represents the action variation that gives the equation of
motion, while the second integral gives the border terms in the
action variation from which we obtain the conserved currents, once
imposed the equation of motion. This second integral contains itself
two terms: the first one originates from the variation
$\delta\Phi\equiv \Phi'(x)-\Phi(x)$ of the fields, the second one
from the variation $d\Phi\equiv\Phi(x')-\Phi(x)$ of the field
coordinates. We remind that, by definition of scalar field, holds\\
\begin{equation}
0=\Phi'(x')-\Phi(x)=|\Phi'(x')-\Phi(x')|-|\Phi(x')-\Phi(x)|~~\rightarrow~~\delta\Phi=-d\Phi
\end{equation}\\
\noindent whit the approximation $\delta{\Phi(x')}\equiv
\Phi'(x')-\Phi(x')\simeq \Phi'(x)-\Phi(x)\equiv\delta\Phi$, in order
to consider variations at the first order.

Using the equation of motion (\ref{eq:MotionEquationm=0}) one easily
obtains the following description of the total variation of our
action (\ref{eq:Actionm=0}) under a translation transformation
($x\rightarrow x+dx$ and $f\rightarrow f+df$):\\
\begin{eqnarray}
\delta S[\Phi]&=&-\frac{1}{2}\int d^4x\left\{
\epsilon^\mu\left((\tilde{P_\alpha}e^{-\lambda P_0\delta_{\mu
j}}\Phi)(\tilde{P^\alpha}P_\mu\Phi)+(P_\mu\tilde{P_\alpha}\Phi)\tilde{P^\alpha}\Phi\right)\right\}+
\nonumber\\
\nonumber\\
&-&\int d^4x\{\epsilon^\mu P_\mu\mathcal{L}\}=
-\frac{1}{2}\epsilon^\mu\int d^4x\tilde{P^\alpha}
[(\tilde{P_\alpha}e^{(-\delta_{\mu j}+\frac{1}{2})\lambda
P_0}\Phi)(P_\mu\Phi)+
\nonumber\\
\nonumber\\
&+&(P_\mu\Phi)\tilde{P_\alpha}e^{-\frac{\lambda
P_0}{2}}\Phi]-\epsilon^\mu\int d^4x\{ P_\mu\mathcal{L}\}=
\nonumber\\
\nonumber\\
&=&\int d^4x\{\epsilon^\mu \tilde{P^\nu} J_{\mu\nu}\}\,,
\end{eqnarray}\label{eq:4Dcurrents} \\
\noindent where\\
\begin{equation}
J_{j\mu}=-\frac{1}{2}(\tilde{P_j}e^{(-\delta_{\mu
j}+\frac{1}{2})\lambda
P_0}\Phi)(P_\mu\Phi)+\frac{1}{2}(P_\mu\Phi)\tilde{P_j}e^{-\frac{\lambda
P_0}{2}}\Phi-\delta_{\mu j}P_j\tilde{P_j}^{-1}\mathcal{L} \,,
\end{equation}\\
\begin{equation}
J_{0\mu}=-\frac{1}{2}(\tilde{P_0}e^{(-\delta_{\mu
j}+\frac{1}{2})\lambda
P_0}\Phi)(P_\mu\Phi)+\frac{1}{2}(P_\mu\Phi)\tilde{P_0}e^{-\frac{\lambda
P_0}{2}}\Phi-\delta_{\mu 0}P_0\tilde{P_0}^{-1}\mathcal{L}.
\end{equation}\\
\noindent Performing a 3D spatial integration of the component
$J_{0\mu}$ and evaluating the charges on the solution of the
equation of motion, whose general form is given in
(\ref{eq:Solutionm=0}), one easily finds the following expression
for the charges carried by the solutions of the equation of
motion:\\
\begin{equation}\label{eq:4Dcharges}
Q_\mu\!=\!\int d^3x J_{0\mu}\!=\!\frac{1}{2}\int d^4p \,e^{3\lambda P_0}\, p_\mu\,
\tilde{\Phi}(p_0, \vec{p})\tilde{\Phi}(-\!p_0,\!- e^{\lambda
P_0}\vec{p})\frac{p_0}{|p_0|}\delta(C_\lambda(p_\mu))\,.
\end{equation}\\
\noindent The fact that these energy-momentum charges $Q_\mu$,
computed by 3D spatial integration of the $J_{0\mu}$, are indeed
time independent confirms that the Noether analysis has been
successful.

It is rather clear from the form of (\ref{eq:4Dcharges}) that the
energy-momentum relation is Planck-scale-($\lambda$-)deformed with
respect to the special-relativistic (Poincar\'{e}-Lie-algebra)
limit. Let us consider for example a ``regularized plane wave
solution" whose Fourier transform is\\
\begin{equation}
\tilde{\Phi}(k)=\frac{2\sqrt{|\vec{k}|}\theta(k_0)\delta(\vec{k}-\vec{p})
}{\sqrt{V}}           \,.
\end{equation}\\
\noindent It is easy to see that the field $\Phi(x)$ can be written
as\\
\begin{eqnarray}
\Phi(x)&=&\int d^4k
\frac{2\sqrt{|\vec{k}|}\theta(k_0)\delta(\vec{k}-\vec{p})
}{\sqrt{V}}e^{i\vec{k}\cdot\vec{x}}e^{-ik_0x_0}\delta(C_{\lambda}(k_0,\vec{k})-m^2)=
\nonumber\\
\nonumber\\
&=&\int \frac{d^4k}{2|\vec{k}|}
\frac{2\sqrt{|\vec{k}|}\delta(\vec{k}-\vec{p})
}{\sqrt{V}}e^{i\vec{k}\cdot\vec{x}}e^{-ik_0x_0}\delta(k_0 -k^+_0)=
\nonumber\\
\nonumber\\
&=&\frac{1}{\sqrt{V|\vec{p}|}}e^{i\vec{p}\cdot\vec{x}}e^{-ip^+_0x_0}\,.
\end{eqnarray}\label{eq:Phy}\\
For a complex scalar classic field $\Phi$, solution of
$C_{\lambda}(k)\Phi=0$ on $\kappa$-Minkowski\\
\begin{equation*}
\Phi(x)=\int
d^4k\tilde{\Phi}(k)\delta(C_{\lambda}(k))e^{i\vec{k}\cdot\vec{x}}e^{-ik_0x_0}
\end{equation*}\\
\noindent holds the condition\\
\begin{equation}
\left(\tilde{\Phi}(k_0,\vec{k})\right)^*=\left(\tilde{\Phi}^*(-k_0,-\vec{k}e^{\lambda
k_0})\right)^*e^{3\lambda k_0}
\end{equation}\\
\noindent that allows us to rewrite the charges as\\
\begin{equation}
Q_\mu=\frac{1}{2}\int  d^4k |\tilde{\Phi}(k_0,\vec{k})|^2 k_\mu
\frac{k_0}{|k_0|}\delta(C_{\lambda}(k))\,.
\end{equation}\\
Using these results and the solutions of $\delta(C_{\lambda}(k))$\\
\begin{eqnarray}
\delta(C_{\lambda}(k))&=&\delta\left((\frac{2}{\lambda}\sinh{\frac{\lambda
k_0}{2}})^2-|\vec{k}|^2e^{\lambda k_0}\right)=
\nonumber\\
\nonumber\\
&=&\frac{1}{2|\vec{k}|}(\delta(k_0 -k^+_0)+\delta(k_0 -k^-_0))\,,
\end{eqnarray}\\
\noindent where\\
\begin{equation*}
k^+_0=\frac{1}{\lambda}\ln\left(\frac{1}{1-(\lambda
|\vec{k}|)}\right)
\end{equation*}\\
\begin{equation}
k^-_0=\frac{1}{\lambda}\ln\left(\frac{1}{1+(\lambda
|\vec{k}|)}\right)\,,
\end{equation}\\
\noindent we are now ready to compute the charges\\
\begin{eqnarray}
(Q_0,\vec{Q})&=&\frac{1}{2}\int  d^4k
\Big|\frac{2\sqrt{|\vec{k}|}\theta(k_0)\delta(\vec{k}-\vec{p})
}{\sqrt{V}}\Big|^2\frac{k_0}{|k_0|}
(k_0,\vec{k})\delta(C_{\lambda}(k))=
\nonumber\\
\nonumber\\
&=&\frac{1}{2}\int
d^4k\frac{\big|2\sqrt{|\vec{k}|}\big|^2}{2|\vec{k}|}\frac{k_0}{|k_0|}(k_0,\vec{k})
\delta(k_0 -k^+_0)\delta(\vec{k}-\vec{p})=
\nonumber\\
\nonumber\\
&=&\int d^3k\frac{k_0}{|k_0|}(k^+_0,\vec{k})\delta(\vec{k}-\vec{p})=
(p_0^+,\vec{p})\,
\end{eqnarray}\\
\noindent which are on shell with respect to the
Casimir, i.e.  $C_\lambda(Q_\mu)=0$. Therefore we can write the
dispersion relation of the charges $Q_\mu$ associated to the field
(\ref{eq:Phy}):\\
\begin{equation}\label{eq:4DDispersion}
\left(\frac{2}{\lambda}\right)^2\sinh^2\left(\frac{\lambda
Q_0}{2}\right)-e^{\lambda Q_0}Q_i^2=0\,.
\end{equation}\\
\noindent Of course, in the special-relativistic limit,
$\lambda\rightarrow0$, one recovers the standard energy-momentum
relation $Q_o^2-Q_i^2=0$ (for our massless fields), but in general
some $\lambda$-dependent corrections are present.

\clearpage

\chapter{\textbf{Noether analysis with five-dimensional bicovariant differential
calculus}}\label{ch:Analysis5D}

In the previous chapter we saw how taking into account the
properties of the noncommutative $\kappa$-Minkowski differential
calculus turns out to be a key point to obtain conserved
translation-symmetry charges. While in the commutative case there is
only one (natural) differential calculus involving the conventional
derivatives, in the $\kappa$-Minkowski case (and in general in a
noncommutative spacetime) the introduction of a differential
calculus is a more complex problem and, in particular, it is not
unique. In our analysis of this chapter we focus on a possible
choice of differential calculus in $\kappa$-Minkowski: the
five-dimensional (5D) bicovariant differential calculus introduced
by Sitarz in \cite{Sitarz}. The vector fields corresponding to this
differential calculus has in fact special covariance properties:
they transform under $\kappa$-Poincar\'{e} in the same way that the
ordinary derivatives (i.e. the vector fields associated to the
differential calculus in the commutative Minkowski space) transform
under Poincar\'{e}. The fact that this calculus is bi-covariant
under the action of the full $\kappa$-Poincar\'{e}
algebra\footnote{We remind that the differential calculus used in
the previous chapter and proposed by \cite{Majid3} is a
four-dimensional translational invariant calculus, but is not
covariant, in the sense of Sitarz \cite{Sitarz}, under the action of
the full $\kappa$-Poincar\'{e} algebra.} motivated some authors
(see, e.g., \cite{Friedel, Freidel2}) to argue that the charges
associated to the translation symmetry and the translation-symmetry
relation derived from this calculus should have the same properties
of the corresponding charges in the classical Minkowski spacetime.

Before this thesis work, the possibility to use the properties of
the five-dimensional differential calculus and work exclusively on
the noncommutative $\kappa$-Minkowski spacetime in performing a
Noether analysis had never been considered. An attempt to obtain
charges from the 5D differential calculus was present in the
literature (\cite{Friedel}) but relied on an uncontrolled map
between the noncommutative spacetime theory here of interest and a
commutative spacetime theory (the results reported in this chapter
expose the inadequacy of that proposed correspondence). Working
exclusively on the noncommutative spacetime, one obtains explicit
formulas relating fields and energy-momentum charges, that turn out
to be non-classical functionals of the fields with a non-trivial
$\lambda$ dependence.

In the first section we introduce the five-dimensional differential
calculus, whose construction following the Sitarz procedure is
reported in Appendix B, and the proper translation generators with
their co-algebra sector and the suitable commutation relations
between fields and one-forms derived from the 5D differential
calculus. In the second part of the chapter, equipped with all these
tools, we perform all the steps of the Noether analysis, in analogy
with the previous chapter.\\

\section{Bicovariant differential calculus on
$\kappa$-Minkow\-ski}\label{par:5DDifferentialCalculus}

In the commutative case there is only one ``natural" differential
calculus, which involves the ordinary derivatives. In this case, the
exterior derivative operator $d$ of a commutative function $f(x)$ is
the usual one:\\
\begin{equation}
df(x)=dx^\mu\partial_\mu f(x)=idx^\mu P_\mu f(x)
\end{equation}\\
\noindent where we have expressed the vector fields $\partial_\mu$
in terms of the standard translation generators
$P_\mu=-i\partial_\mu$. In this way it is clear that the
$\partial_\mu$ transform \textit{covariantly} under the standard
Lorentz algebra (generated by $M_j$, $N_j$):\\
\begin{equation*}
[M_j,P_0]=0\,,~~~~~~[M_j,P_k]=i\epsilon_{jkl}P_l\,,
\end{equation*}\\
\begin{equation}
[N_j,P_0]=iP_j\,,~~~~~~[N_j,P_k]=i\delta_{jk}P_0 \,.
\end{equation}\\
In the case of $\kappa$-Minkowski, instead, the choice of a
differential calculus is not unique. We introduce below a possible
choice of differential calculus in $\kappa$-Minkowski, the 5D
differential calculus. In this differential calculus the exterior
derivative operator $d$ of a generic $\kappa$-Minkowski element
$F(\hat{x})=\Omega(f(x))$ can be written in the form\footnote{We use
greek letters for indexes running over ($\alpha,\mu=0,1,2,3$), small
latin letters for indexes running over
($j,k=1,2,3$) and capital latin letters for indexes running over ($A,B=0,1,2,3,4$).}:\\
\begin{equation}\label{eq:df}
dF(\hat{x}) = d \hat{x}^A \hat{P_A}(P) F(\hat{x})\,,~~~A=0,...,4 \,,
\end{equation}\\
\noindent where the operators
$\hat{P}_0,\hat{P}_1,\hat{P}_2,\hat{P}_3$ form a basis for the
translation generators of $\kappa$-Poincar\'{e}, while $\hat{P}_4$
is connected with the Casimir $C_\lambda(P)$. The operators
$\hat{P}_A$ are defined\\
\begin{equation}\label{eq:generatori1}
\hat{P_0} = \frac{1}{\lambda} ( \sinh {\lambda P_0} +
\frac{\lambda^2}{2} P^2 e^{ \lambda P_0} )
\end{equation}\\
\begin{equation}\label{eq:generatori2}
\hat{P_i} = P_i e^{ \lambda P_0}~~~i=1,2,3
\end{equation}\\
\begin{equation}\label{eq:generatori3}
\hat{P_4} = \frac{1}{\lambda} ( \cosh {\lambda P_0} -1 -
\frac{\lambda^2}{2} P^2 e^{ \lambda P_0} )=\frac{\lambda}{2}m^2,
\end{equation}\\
\noindent where $P_\mu$ denotes again (as in chapter
\ref{ch:NoncommutativeGeometry} and \ref{ch:Analysis4D}) the
translation generators of the Majid-Ruegg $\kappa$-Poincar\'{e}
basis, whose action on a right-ordered function of
$\kappa$-Minkowski is
$P_\mu(e^{ik\hat{x}}e^{-ik_0\hat{x}_0})=k_\mu(e^{ik\hat{x}}e^{-ik_0\hat{x}_0})$.
The commutation relations between the one-form generators
$d\hat{x}^A$ and the $\kappa$-Minkowski generators $\hat{x}^\mu$
are:\\
\begin{equation*}
[\hat{x}_0, d\hat{x}_4] = i \lambda d\hat{x}_0\,, ~~~~~~ [\hat{x}_0,
d\hat{x}_0] = i \lambda d\hat{x}_4\,, ~~~~~~ [\hat{x}_0, d\hat{x}_i] = 0\,,
\end{equation*}\\
\begin{equation}\label{eq:5Dcommutators}
[\hat{x}_i, d\hat{x}_4] = [\hat{x}_i, d\hat{x}_0] = -i \lambda
d\hat{x}_i \,,~~~~~~ [\hat{x}_i, d\hat{x}_j] = i \lambda \delta_{ij}
(d\hat{x}_4 - d\hat{x}_0)\,.
\end{equation}\\
\noindent The introduction of such a 5D calculus in our 4D spacetime
may at first appear to be surprising, but it can be naturally
introduced on the basis of the fact that the
$\kappa$-Poincar\'{e}/$\kappa$-Minkowski framework can be obtained
(and was indeed originally obtained \cite{Lukierski2}) by
\``{I}non\"{u}-Wigner contraction of a 5D q-deformed anti-De Sitter
algebra. The fifth one-form generator is here denoted by
``$d\hat{x}^4$", but this is of course only a formal notation, since
there is no fifth $\kappa$-Minkowski coordinate $\hat{x}^4$. And the
peculiar role of $d\hat{x}^4$ in this differential calculus is also
codified in the fact that the last component $\hat{P_4}(P)$ is
essentially the Casimir (\ref{eq:DeformedCasimir}) of
$\kappa$-Poincar\'{e}:\\
\begin{equation}
\hat{P_4}(P)=\frac{\lambda}{2} C_\lambda(P)\,.
\end{equation}\\
This differential calculus is characterized by interesting
transformation properties under the action of the Lorentz sector of
$\kappa$-Poincar\'{e}. In fact taking into account
(\ref{eq:MajidAlgebra}) one finds that:\\
\begin{equation*}
[M_j, \hat{P}_0]=0\,,~~~~~~ [M_j,
\hat{P}_k]=i\epsilon_{jkl}\hat{P}_l\,,~~~~~~[M_j, \hat{P}_4]=0\,,
\end{equation*}\\
\begin{equation}\label{eq:5DAlgebra}
[N_j, \hat{P}_0]=i\hat{P}_j\,,~~~~~~
[N_j,\hat{P}_k]=i\delta_{jk}\hat{P}_0\,,~~~~~~[N_j, \hat{P}_4]=0 \,.
\end{equation}\\
\noindent Thus the operators $\hat{P}_\mu$ transform under
$\kappa$-Poincar\'{e} action in the same way as the $P_\mu$
operators transform under the standard Poincar\'{e} action, while
$\hat{P}_4(P)$ is invariant.

This differential calculus originates in \cite{Sitarz} by the
request that it remains invariant under the action of the
$\kappa$-Poincar\'{e} action, i.e. the commutation relations
(\ref{eq:5Dcommutators}) that characterize it remain
invariant\footnote{The demonstration is reported in detail in
Appendix A} under the action of the $\kappa$-Poincar\'{e}
generators; practically, one seeks some $d\hat{x}^A$ such that their
commutator with the $\kappa$-Minkowski coordinates $\hat{x}^\mu$\\
\begin{equation}
[d\hat{x}^A, \hat{x}^\mu]=\upsilon^{A\mu}_\rho d\hat{x}^\rho\,,
\end{equation}\\
\noindent for some numbers $\upsilon^{A\mu}_\rho$, are invariant in
the sense\\
\begin{equation}
T[d\hat{x}^A, \hat{x}^\mu]=\upsilon^{A\mu}_\rho Td\hat{x}^\rho\,,
\end{equation}\\
\noindent where $T$ denotes any one of the $\kappa$-Poincar\'{e}
generators $(P_\mu, M_j, N_j)$. A differential calculus in which the
commutation relations between the one-form generators and the
$\kappa$-Minkowski generators remain \textit{invariant} under the
action of the symmetry algebra ($\kappa$-Poincar\'{e} in our case),
is called ``bicovariant" differential calculus. In \cite{Gonera} it
was claimed that the 5D differential calculus of Sitarz is the
unique bicovariant one with respect to the left action of
$\kappa$-Poincar\'{e} group.

In order to perform our Noether analysis of the translation sector
of $\kappa$-Poincar\'{e} with the 5D differential calculus it is
convenient to first derive formulas for the coproducts of the
operators $\hat{P}_A$ and their commutation relations with the
time-to-the-right-ordered plane wave basis of $\kappa$-Minkowski.

The form of the coproducts of $\hat{P}_A$ is easily obtained
exploiting the relationship (\ref{eq:generatori1}),
(\ref{eq:generatori2}) and (\ref{eq:generatori3}) between
$\hat{P}_A$ and $P_\mu$ and the fact that we have already provided
formulas, (\ref{eq:MajidCoproducts}), for the coproduct of $P_\mu$.
Taking into account the homomorphism property of the coproduct map
(\ref{eq:Homomorphism}), one finds that\\
\begin{equation}\label{eq:coproduct1}
\Delta (\hat{P_0} ) = \hat{P_0} \otimes e^{\lambda P_0} +
e^{-\lambda P_0} \otimes \hat{P_0} + \lambda P_i \otimes \hat{P_i}
\end{equation}\\
\begin{equation}\label{eq:coproduct2}
\Delta (\hat{P_i} ) = \hat{P_i} \otimes e^{\lambda P_0} + 1 \otimes
\hat{P_i}
\end{equation}\\
\begin{equation}\label{eq:coproduct3}
\Delta (\hat{P_4} ) = \hat{P_4} \otimes e^{\lambda P_0} -
e^{-\lambda P_0} \otimes \hat{P_0} - \lambda P_i \otimes \hat{P_i} +
1\otimes (\frac{e^{\lambda P_0} -1}{\lambda})\,.
\end{equation}\\
For the commutation relations between the time-to-the-right-ordered
plane waves $e^{ik \cdot \hat{x}}e^{- k_0 \hat{x_0}}$ and the $d
\hat{x_A}$ elements of the 5D differential calculus one finds\\
\begin{equation}\label{eq:planewave1}
e^{ik \cdot \hat{x}}e^{- k_0 \hat{x}_0} d \hat{x}_0 = [ (
 \lambda \hat{P}_0+  e^{ -\lambda P_0} ) d \hat{x}_0 + \lambda
 \hat{P_i} d \hat{x}_i +
+(\lambda \hat{P}_4+1-  e^{ -\lambda P_0} )
 d \hat{x}_4 ] e^{ik \cdot \hat{x}}e^{- k_0 \hat{x}_0}
\end{equation}\\
\begin{equation}\label{eq:planewave2}
e^{ik \cdot \hat{x}}e^{- k_0 \hat{x}_0} d \hat{x}_i = [
 \lambda P_i d \hat{x}_0 + d \hat{x}_i - \lambda P_i
 d \hat{x}_4 ] e^{ik \cdot \hat{x}}e^{- k_0 \hat{x}_0}
\end{equation}\\
\begin{equation}\label{eq:planewave3}
e^{ik \cdot \hat{x}}e^{- k_0 \hat{x}_0} d \hat{x}_4 = [\lambda
\hat{P_0} d \hat{x}_0 + \lambda
 \hat{P_i} d \hat{x}_i + (\lambda \hat{P_4}+1)
 d \hat{x}_4 ] e^{ik \cdot \hat{x}}e^{- k_0 \hat{x}_0}\,.
\end{equation}\\

\section{Noether analysis}\label{par:5DNoetherAnalysis}

\subsection{Leibnitz rule}\label{par:LeibnitzRule}

With the tools introduced in the previous section we are now ready
to perform the Noether analysis of $\kappa$-Poincar\'{e} translation
symmetry for $\kappa$-Minkowski.

The exterior derivative operator $d$ of a general element of
$\kappa$-Minkow\-ski, defined in (\ref{eq:df}), must of course
satisfy the Leibnitz rule with respect of the coproducts of the
translation generators $\hat{P}_A$. Let us show this using the
commutation relations (\ref{eq:planewave1})-(\ref{eq:planewave3})
between the one-forms generators and the time-to-the-right-ordered
plane waves
and remembering that\\
\begin{equation*}
\hat{P_0}+\hat{P_4}=\frac{e^{\lambda P_0} -1}{\lambda}\,.
\end{equation*}\\
We have:\\
\begin{eqnarray}
(d \Psi) \Phi+\Psi (d\Phi) &=& ( d\hat{x^0} \hat{P_0} \Psi ) \Phi +
\Psi ( d\hat{x^0} \hat{P_0} \Phi )  + ( d\hat{x^i} \hat{P_i} \Psi )
\Phi +\nonumber\\
&+& \Psi ( d\hat{x^i} \hat{P_i} \Phi )
 +( d\hat{x^4} \hat{P_4} \Psi )
\Phi+ \Psi ( d\hat{x^4} \hat{P_4}\Phi )=\nonumber\\
&=&( d\hat{x^0} \hat{P_0} \Psi )\Phi + [ (
 \lambda \hat{P}_0+  e^{ -\lambda P_0} ) d \hat{x^0} \Psi \hat{P_0} \Phi + \lambda
 \hat{P_i} d \hat{x^i} \Psi \hat{P_0} \Phi +\nonumber\\
 &+& (\lambda \hat{P}_4+1-  e^{ -\lambda P_0})
 d \hat{x^4} \Psi \hat{P_0} \Phi ] + ( d\hat{x^i} \hat{P_i} \Psi )
 \Phi +\nonumber\\
&+&[ \lambda P_i d \hat{x^0} \Psi \hat{P_i} \Phi  + d \hat{x^i} \Psi
\hat{P_i} \Phi - \lambda P_i
 d \hat{x^4} \Psi \hat{P_i} \Phi ]+( d\hat{x^4} \hat{P_4} \Psi )\Phi +\nonumber\\
&+& [ \lambda \hat{P_0} d \hat{x^0} \Psi \hat{P_4} \Phi + \lambda
 \hat{P_i} d \hat{x^i} \Psi \hat{P_4} \Phi + (\lambda \hat{P_4}+1)
 d \hat{x^4} \Psi \hat{P_4} \Phi ]=\nonumber\\
&=&  d\hat{x^0} [\hat{P_0} \Psi \Phi + \lambda \hat{P_0}\Psi
\hat{P_0} \Phi + e^{ -\lambda P_0} \Psi\hat{P_0} \Phi +\lambda P_i
 \Psi \hat{P_i} \Phi+\lambda \hat{P_0} \Psi
\hat{P_4} \Phi]+\nonumber\\
&+& d\hat{x^i} [\hat{P_i} \Psi
 \Phi +  \Psi \hat{P_i} \Phi + \lambda \hat{P_i}
  \Psi \hat{P_0} \Phi +\lambda \hat{P_i}
  \Psi \hat{P_4} \Phi]+\nonumber\\
&+& d\hat{x^4}[(\lambda \hat{P_4}+1)\Psi(\hat{P_0}+\hat{P_4})\Phi
-e^{ -\lambda P_0}\Psi\hat{P_0}\Phi-\lambda
P_i\Psi\hat{P_i}\Phi+\hat{P_4}\Psi\Phi]=\nonumber\\
&=&  d\hat{x^0} [\hat{P_0} \Psi \Phi + \lambda \hat{P_0}\Psi
(\frac{e^{\lambda P_0} -1}{\lambda}) \Phi + e^{ -\lambda P_0}
\Psi\hat{P_0} \Phi +\lambda P_i
 \Psi \hat{P_i} \Phi]+\nonumber\\
&+& d\hat{x^i} [\hat{P_i} \Psi \Phi +  \Psi \hat{P_i} \Phi + \lambda
\hat{P_i} \Psi (\frac{e^{\lambda P_0} -1}{\lambda}) \Phi ]+\nonumber\\
&+& d\hat{x^4}[(\lambda \hat{P_4}+1)\Psi(\frac{e^{\lambda P_0}
-1}{\lambda}) \Phi -e^{ -\lambda P_0}\Psi\hat{P_0}\Phi-\lambda
P_i\Psi\hat{P_i}\Phi+\hat{P_4}\Psi\Phi]=\nonumber\\
&=&  d\hat{x^0} [ \hat{P_0}\Psi e^{\lambda P_0} \Phi + e^{ -\lambda
P_0} \Psi\hat{P_0} \Phi +\lambda P_i
 \Psi \hat{P_i} \Phi]+\nonumber\\
&+& d\hat{x^i} [\hat{P_i} \Psi e^{\lambda P_0}\Phi +  \Psi \hat{P_i}
\Phi ]+\nonumber\\
&+& d\hat{x^4}[\Psi(\frac{e^{\lambda P_0} -1}{\lambda}) \Phi -e^{
-\lambda P_0}\Psi\hat{P_0}\Phi-\lambda
P_i\Psi\hat{P_i}\Phi+\hat{P_4}\Psi e^{\lambda
P_0}\Phi]=\nonumber\\
&=&d(\Psi\Phi) \,,
\end{eqnarray}\\
\noindent where the last equality holds with respect of the
coproducts (\ref{eq:coproduct1}), (\ref{eq:coproduct2}) and
(\ref{eq:coproduct3}). Therefore Leibnitz is satisfied.

\subsection{Currents}\label{par:5DCurrents}

In the following Noether analysis we assume that a massive scalar
field $\Phi(x)$ is governed by one of the most studied equation of
motions in the $\kappa$-Minkowski
literature~\cite{Lukierski,Amelino4}, i.e. the
Klein-Gordon-like equation\\
\begin{equation}\label{eq:5DMotionEquation}
C_\lambda(P_\mu)\,\Phi \equiv \left[\left(\frac{2}{\lambda} \sinh
{\frac{\lambda}{2} P_0}\right)^2-e^{\lambda
P_0}\vec{P}^2\right]\Phi=m^2\Phi ~,
\end{equation}\\
\noindent which can be derived from the following action\\
\begin{equation*}
S[\Phi]=\int d^4x \mathcal{L}[\Phi(x)]
\end{equation*}\\
\begin{equation}\label{eq:5DAction}
\mathcal{L}[\Phi(x)]=\frac{1}{2} \left(\Phi(x)
\,C_\lambda\,\Phi(x) - m^2\Phi(x)\Phi(x)\right)     \,.
\end{equation}\\
\noindent We remind that the operator $C_\lambda(P_\mu)$ is the mass
Casimir of the $\kappa$-Poincar\'{e} Hopf algebra and we find
sometimes useful to also write it as
$C_\lambda=\tilde{P}_\mu\tilde{P}^\mu$ in terms of the operators\\
\begin{equation} \label{eq:5DPtilde}
\tilde{P_0} = \frac{2}{\lambda} \sinh {\frac{\lambda}{2} P_0} ~~~~~~~~~
\tilde{P_i} = P_i e^{\frac{\lambda}{2} P_0}\,,
\end{equation}\\
\noindent whose coproducts are given by\\
\begin{equation}
\tilde{P_\alpha}[f(x)g(x)]=[\tilde{P_\alpha}f(x)][e^{\frac{\lambda}{2}P_0}g(x)]+
[e^{-\frac{\lambda}{2}P_0}f(x)][\tilde{P_\alpha}g(x)]\,.
\end{equation}\\
We can now derive the total variation of our action
(\ref{eq:5DAction}) under a translation transformation
($x\rightarrow x+dx$ and $f\rightarrow f+df$), using eq.
(\ref{eq:df}), (\ref{eq:planewave1})-(\ref{eq:planewave3}) and the
observation that, by definition of a scalar field,\\
\begin{equation}
0=\Phi'(\hat{x}')-\Phi(\hat{x})=[\Phi'(\hat{x}')-\Phi(\hat{x}')]-[\Phi(\hat{x}')-\Phi(\hat{x})]~,
\end{equation}\\
\noindent i.e. $\delta\Phi=-d\Phi =-i\left( \,\hat{\epsilon}^0
\hat{P}_0 + \hat{\epsilon}^j \hat{P}_j +\hat{\epsilon}^4 \hat{P}_4\,
\right)\Phi$ (where we have identified the infinitesimal transformation parameters with the one-forms generators of the 5D differential calculus):\\
\begin{eqnarray}
\delta S&=&\frac{1}{2} \int d^4x \left( \delta\Phi C_\lambda \Phi
+\Phi C_\lambda \delta\Phi -
m^2\delta\Phi\Phi-m^2\Phi\delta\Phi\right)=\nonumber\\
&=& \frac{1}{2} \int d^4x\bigg[e^{\frac{\lambda
P_0}{2}}\tilde{P}^0\left((\frac{2}{\lambda}+\lambda
m^2-\frac{e^{\lambda
P_0}}{\lambda})\Phi\hat{\epsilon}^A\hat{P}_A\Phi-\Phi
\frac{e^{-\lambda
P_0}}{\lambda}\hat{\epsilon}^A\hat{P}_A\Phi\right)+\nonumber\\
&+& \hat{P}^i\left(\Phi e^{-\lambda
P_0}\hat{P}_i\hat{\epsilon}^A\hat{P}_A\Phi-\hat{P}_i\Phi\hat{\epsilon}^A\hat{P}_A\Phi\right)\bigg]=\nonumber\\
&=& \frac{1}{2} \int d^4x\Bigg\{e^{\frac{\lambda
P_0}{2}}\tilde{P}^0\bigg\{\left(\frac{2}{\lambda}+\lambda
m^2-\frac{e^{\lambda P_0}}{\lambda}\right)\bigg[[(
 \lambda \hat{P}_0+  e^{ -\lambda P_0} ) \hat{\epsilon}^0 + \lambda
\hat{P_i} \hat{\epsilon}^i+\nonumber\\
 &+& (\lambda \hat{P}_4+1 - e^{ -\lambda P_0})
 \hat{\epsilon}^4]\Phi\hat{P}_0\Phi+ [\lambda e^{ -\lambda P_0} \hat{P}_i \hat{\epsilon}^0 +  \hat{\epsilon}^i - \lambda e^{ -\lambda P_0} \hat{P}_i
 \hat{\epsilon}^4]\Phi\hat{P_i}\Phi+\nonumber\\
 &+&[\lambda
\hat{P}_0  \hat{\epsilon}^0 + \lambda
 \hat{P}_i \hat{\epsilon}^i + (\lambda \hat{P}_4+1)
  \hat{\epsilon}^4]\Phi\hat{P}_4\Phi\bigg]+\nonumber\\
&-&\bigg[(\lambda \hat{P}_0+  e^{ -\lambda P_0} ) \hat{\epsilon}^0 +
\lambda
 \hat{P_i} \hat{\epsilon}^i
+(\lambda \hat{P}_4+1-  e^{ -\lambda P_0} )
 \hat{\epsilon}^4\bigg]\Phi\frac{e^{-\lambda P_0}}{\lambda}\hat{P}_0\Phi+\nonumber\\
 &-& \bigg[\lambda e^{ -\lambda P_0} \hat{P}_i \hat{\epsilon}^0 +  \hat{\epsilon}^i - \lambda e^{ -\lambda P_0} \hat{P}_i
 \hat{\epsilon}^4\bigg]\Phi\frac{e^{-\lambda P_0}}{\lambda}\hat{P_i}\Phi+\nonumber\\
 &-&\bigg[\lambda
\hat{P}_0  \hat{\epsilon}^0 + \lambda
 \hat{P}_i \hat{\epsilon}^i + (\lambda \hat{P}_4+1)
  \hat{\epsilon}^4\bigg]\Phi\frac{e^{-\lambda
  P_0}}{\lambda}\hat{P}_4\Phi\bigg\}+\nonumber\\
&+& \hat{P}_i\bigg\{\bigg[(
 \lambda \hat{P}_0+  e^{ -\lambda P_0} ) \hat{\epsilon}^0 + \lambda
 \hat{P_i} \hat{\epsilon}^i
+(\lambda \hat{P}_4+1-  e^{ -\lambda P_0} )
 \hat{\epsilon}^4\bigg]\Phi\hat{P}_i\hat{P}_0\Phi+\nonumber\\
 &+& \bigg[\lambda e^{ -\lambda P_0} \hat{P}_i \hat{\epsilon}^0 +  \hat{\epsilon}^i - \lambda e^{ -\lambda P_0} \hat{P}_i
 \hat{\epsilon}^4\bigg]\Phi\hat{P}_i\hat{P_i}\Phi+\nonumber\\
 &+&\bigg[\lambda
\hat{P}_0  \hat{\epsilon}^0 + \lambda
 \hat{P}_i \hat{\epsilon}^i + (\lambda \hat{P}_4+1)
  \hat{\epsilon}^4\bigg]\Phi\hat{P}_i\hat{P}_4\Phi+\nonumber\\
&-&\bigg[( \lambda \hat{P}_0+  e^{ -\lambda P_0} ) \hat{\epsilon}^0
+ \lambda
 \hat{P_i} \hat{\epsilon}^i
+(\lambda \hat{P}_4+1-  e^{ -\lambda P_0} )
 \hat{\epsilon}^4\bigg]\hat{P}_i\Phi\hat{P}_0\Phi+\nonumber\\
 &-& \bigg[\lambda e^{ -\lambda P_0} \hat{P}_i \hat{\epsilon}^0 +  \hat{\epsilon}^i - \lambda e^{ -\lambda P_0} \hat{P}_i
 \hat{\epsilon}^4\bigg]\hat{P}_i\Phi\hat{P_i}\Phi+\nonumber\\
 &-&\bigg[\lambda\hat{P}_0  \hat{\epsilon}^0 + \lambda
 \hat{P}_i \hat{\epsilon}^i + (\lambda \hat{P}_4+1)
  \hat{\epsilon}^4\bigg]\hat{P}_i\Phi\hat{P}_4\Phi\bigg\}\Bigg\}\,,
\end{eqnarray}\\
\noindent where we have specialized to the case of fields such that
$\tilde{P}^\mu \tilde{P}_\mu\Phi=m^2\Phi$, since of course we
perform the Noether analysis on fields that are solutions of the
equation of motion.

Thus, the variation of the Lagrangian density takes the form\\
\begin{equation}\label{eq:protodiv}
\hat{\epsilon}^A\left(\,e^{\frac{\lambda
P_0}{2}}\tilde{P}^0\,J_{0A}+\hat{P}^iJ_{iA}\right)=0\,,
\end{equation}\\
\noindent where\\
\begin{eqnarray}
J_{00}\,&=&\,\frac{1}{2}\bigg\{\left(\frac{2}{\lambda}+\lambda
m^2-\frac{e^{\lambda P_0}}{\lambda}\right)\left[(\lambda\hat{P}_0+
e^{ -\lambda P_0} )\Phi\hat{P}_0\Phi + \lambda
 P_i\Phi\hat{P_i}\Phi
+\lambda \hat{P}_0\Phi\hat{P}_4\Phi\right]+\nonumber\\
\!\!&-&\!\!(\lambda\hat{P}_0+ e^{ -\lambda P_0} )\Phi
\frac{e^{-\lambda P_0}}{\lambda}\hat{P}_0\Phi - \lambda
 P_i\Phi\frac{e^{-\lambda P_0}}{\lambda}\hat{P_i}\Phi-
\lambda \hat{P}_0\Phi\frac{e^{-\lambda P_0}}{\lambda}\hat{P}_4\Phi\bigg\}\,, \nonumber \\
\nonumber \\
J_{0i}\,&=&\,\frac{1}{2}\bigg\{\left(\frac{2}{\lambda}+\lambda
m^2-\frac{e^{\lambda
P_0}}{\lambda}\right)\bigg[\lambda\hat{P}_i\Phi\hat{P}_0\Phi+\Phi\hat{P}_i\Phi+\lambda\hat{P}_i\Phi\hat{P}_4\Phi\bigg]+\nonumber\\
&-&\,\lambda\hat{P}_i\frac{e^{-\lambda
P_0}}{\lambda}\Phi\hat{P}_0\Phi-\Phi\hat{P}_i\frac{e^{-\lambda
P_0}}{\lambda}\Phi-\lambda\hat{P}_i\Phi\frac{e^{-\lambda
P_0}}{\lambda}\hat{P}_4\Phi\bigg\}\,, \nonumber\\
\nonumber \\
J_{04}\,&=&\,\frac{1}{2}\bigg\{\bigg(\frac{2}{\lambda}+\lambda
m^2-\frac{e^{\lambda P_0}}{\lambda}\bigg)\bigg[(\lambda\hat{P}_4+1-
e^{ -\lambda P_0} )\Phi\hat{P}_0\Phi - \lambda
 P_i\Phi\hat{P_i}\Phi
+\nonumber\\
&+& (\lambda\hat{P}_4+1)\Phi\hat{P}_4\Phi\bigg]-(\lambda\hat{P}_4+1-
e^{-\lambda P_0} )\Phi \frac{e^{-\lambda P_0}}{\lambda}\hat{P}_0\Phi +\nonumber\\
&+&\lambda
 P_i\Phi\frac{e^{-\lambda P_0}}{\lambda}\hat{P_i}\Phi-
 (\lambda\hat{P}_4+1)\Phi\frac{e^{-\lambda
P_0}}{\lambda}\hat{P}_4\Phi\bigg\}\,. \label{eq:J}
\end{eqnarray}\\
We want now to outline that eq. (\ref{eq:protodiv}), produced by the
Noether analysis, guarantees the time-independence of the charges
obtained from the
currents (\ref{eq:J}). In fact, remembering that\\
\begin{equation}
e^{\frac{\lambda}{2}P_0}\tilde{P_0}=(\hat{P_0}+\hat{P_4})=\frac{e^{\lambda
P_0}-1}{\lambda}\,,
\end{equation}\\
\noindent eq. (\ref{eq:protodiv}) constitutes a ``conservation
equation'' in $\kappa$-Minkowski spacetime with the proper generator
associated to temporal translations, i.e. it is the combination
$\hat{P_0}+\hat{P_4}$ which vanishes on time-independent
fields and not $\hat{P_0}$ alone. In particular, defining\\
\begin{equation}\label{eq:HatD0}
 \hat{\mathcal{D}}_0\equiv e^{\frac{\lambda}{2}P_0}\tilde{P_0}\,,
\end{equation}\\
\noindent the action of the proper time-translation generator $\hat{\mathcal{D}}_0$ on the currents obtained from
the Noether analysis vanishes:\\
\begin{equation}\label{eq:HatP0_action}
\hat{\mathcal{D}}_0J_{0A}=0\,.
\end{equation}\\
 The fact that the
Lagrangian density variation occurs exactly in the form
(\ref{eq:protodiv}) is thus a relief if one is looking for an
analogous of the 4-divergence of the currents in the Noether
analysis of ordinary theories in classical Minkowski spacetime. To
make this concept clearer, let us introduce the following rule of
spatial integration in
$\kappa$-Minkowski:\\
\begin{equation}
\int d^3x e^{i{p}\cdot \hat{x}}e^{-ip_0\,\hat{x}_0}
=\delta(\vec{p})\,e^{-ip_0\,\hat{x}_0}\, ,  \label{eq:integratio}
\end{equation}\\
\noindent so that for a $\kappa$-Minkowski field $\Psi(\hat{x})=\int
d^4p\, \tilde{\Psi}(p_0,\vec{p})\, \exp(i{p}\cdot\hat{x})\,
\exp(-ip_0\,\hat{x}_0)$ one obtains\\
\begin{equation}
\int d^3x \Psi(\hat{x}_0,\hat{x})= \int dp_0
\tilde{\Psi}(p_0,\vec{0})\, e^{-i p_0\,\hat{x}_0} \,.
\end{equation}\\
\noindent This spatial integration rule together with the action of
the operator $\hat{\mathcal{D}}_0$
(\ref{eq:HatP0_action}) allows to
write\\
\begin{equation}\label{eq:4Divergence}
\hat{\mathcal{D}}_0 \int d^3x \,J_{0A} = \int d^3x
\,\hat{\mathcal{D}}_0\,J_{0A}=- \int d^3x \hat{P}^iJ_{iA} =0 ~,
\end{equation}
\noindent from which the time independence of the charges $\int d^3x
\,J_{0A}$ follows from a generalization of the classical
4-divergence. We want to stress how all this argument holds without
the introduction of the Weyl map and, thereby, without any reference
to the classical case. In particular, we do not apply the technique
(Gauss theorem) valid in classical Minkowski, that allows to
transform the last integral of (\ref{eq:4Divergence}) into an
integral over a surface where the fields vanish. The vanishing of
$\int d^3x \hat{P}^iJ_{iA}$ follows directly from the action of the
operator $\hat{P}^i$ on a product of functions of
$\kappa$-Minkowski, codified in the structure of the coproduct
(\ref{eq:coproduct2}),\\
\begin{eqnarray*}
&\hat{P}_i\,\left( e^{i\vec{k} \cdot \vec{x}}e^{-i  k_0 {x}_0}\cdot
e^{i\vec{p} \cdot \vec{x}}e^{- i p_0
{x}_0}\right)=\hat{P}_i\,\left(e^{i(\vec{k}+
e^{-\lambda k_0}\vec{p}) \cdot \vec{x}}e^{- i (k_0+p_0) {x}_0}\right)= \\
&=({k_i}+e^{-\lambda k_0}{p_i})\, e^{\lambda
(k_0+p_0)}\,\left(e^{i(\vec{k}+
e^{-\lambda k_0}\vec{p}) \cdot \vec{x}}e^{- i(k_0+p_0) {x}_0}\right)=      \\
&=({k_i}e^{\lambda(k_0+ p_0)}+e^{\lambda p_0}{p_i})\,\left(
e^{i(\vec{k}+
e^{-\lambda k_0}\vec{p}) \cdot \vec{x}}e^{-i (k_0+p_0) {x}_0}\right)=\\
&=\hat{P}_i\left( e^{i\vec{k} \cdot \vec{x}}e^{- ik_0 {x}_0}
\right)\cdot e^{\lambda P_0}\left(  e^{i\vec{p} \cdot \vec{x}}e^{-
ip_0 {x}_0}\right)+ \left( e^{i\vec{k} \cdot \vec{x}}e^{- i k_0
{x}_0} \right)\cdot \hat{P}_i \left(  e^{i\vec{p} \cdot \vec{x}}
e^{- i p_0 {x}_0}\right)\\
\end{eqnarray*}
\noindent and the appearance of $\delta({k_i}+e^{-\lambda
k_0}{p_i})$, when expanding the fields $\Phi(\hat{x})$ over the
time-to-the-right-ordered plane waves, as a consequence of the
integration rule (\ref{eq:integratio}).

However, the conservation of the charges $\int d^3x \,J_{0A}$ will
be directly verified in the next section, where we explicitly
compute them.

\subsection{Conserved charges}\label{par:5DCharges}

We are now ready to derive the charges, that must be evaluated on
the solutions of the equation of motion whose general form is given
in (\ref{eq:Solutionm=0}). To show that they are time-independent we
proceed analyzing separately $J_{00}$, $J_{0i}$ and $J_{04}$. For
$J_{00}$ we
have:\\
\begin{eqnarray}\label{eq:ProtoQ0}
\hat{Q}_0&=&\int d^3x J_{00}\,=\,\frac{1}{2}\int d^3 x \Bigg\{  \left(
\lambda\hat{P}_0+ e^{ -\lambda P_0} \right)\,\cdot\,\nonumber\\
&\cdot& \left[  \phi \left( \frac{1-e^{-\lambda P_0} }{\lambda}
\right)-
\left( \frac{e^{\lambda P_0} -1}{\lambda}\right)\, \phi+\lambda\, m^2 \phi    \right]\, \hat{P}_0\, \phi+\nonumber\\
&+& \lambda P_i \,  \left[  \phi \,\left( \frac{1-e^{-\lambda P_0} }{\lambda}  \right)
- \left( \frac{e^{\lambda P_0} -1}{\lambda}\right) \, \phi+\lambda\, m^2 \phi     \right] \, \hat{P}_i\, \phi+\nonumber\\
&+& \lambda \hat{P}_0\,  \left[    \phi \left( \frac{1-e^{-\lambda P_0} }{\lambda}  \right)-
 \left( \frac{e^{\lambda P_0} -1}{\lambda}\right)\, \phi+\lambda\, m^2 \phi         \right]\, \hat{P}_4\, \phi \Bigg\}=\nonumber\\
&=&\,\frac{1}{2}\int d^3 x\, d^4k\,  d^4p\, \tilde{\Phi}(k_0,k_i)\,
\tilde{\Phi}(p_0,p_i) \,e^{i( \vec{k}+e^{-\lambda k_0} \vec{p})\cdot
\vec{x}}\, e^{-i(k_0+p_0)\,x_0} \,  \cdot \nonumber\\
&\cdot&\,\delta(C_\lambda (k)-m^2) \, \delta(C_\lambda (p)-m^2)\,\cdot \nonumber\\
&\cdot&\,\Bigg\{ \left( \lambda\hat{k}_0+ e^{ -\lambda k_0}
\right)\,
\left( \frac{2-e^{-\lambda p_0} -e^{\lambda k_0}}{\lambda}  +\lambda\, m^2  \right)\, \hat{p}_0 +\nonumber\\
\!\!\!&+& \!\!\!\lambda k_i \left(\! \frac{2-e^{-\lambda p_0} -e^{\lambda k_0}}{\lambda}  +\lambda\, m^2  \!\right)
\hat{p}_i + \lambda\hat{k}_0  \left( \!\frac{2-e^{-\lambda p_0} -e^{\lambda
k_0}}{\lambda}  +\lambda m^2  \!\right)
 \hat{p}_4 \Bigg\}\!\!\!=\!\!\!\nonumber\\
 &=&\,\frac{1}{2}\int d^4k\,  d^4p\, \tilde{\Phi}(k_0,k_i)
\tilde{\Phi}(p_0,p_i) \,  e^{-i(k_0+p_0)\,x_0}\,e^{3\lambda k_0} \,  \cdot\nonumber\\
&\cdot& \delta(\vec{p} +\vec{k}\, e^{\lambda k_0})
\,\delta(C_\lambda
(k)-m^2) \, \delta(C_\lambda (p)-m^2)\,\cdot \nonumber\\
&\cdot&\!\!\! \Bigg\{ \left( \!\frac{2-e^{-\lambda p_0} -e^{\lambda
k_0}}{\lambda}  +\lambda\, m^2  \!\right)\, \Bigg[  \left(\!
\lambda\hat{k}_0+ e^{ -\lambda k_0} \! \right) \, \hat{p}_0
\,+\,\lambda k_i\, \hat{p}_i \,+\,   \lambda\hat{k}_0  \,
  \hat{p}_4\Bigg]
  \Bigg\}\!\!\!\nonumber\\
&=&\,\frac{1}{2}\int d^4k\,  dp_0\, \tilde{\Phi}(k_0,k_i)
\tilde{\Phi}(p_0,-k_i\,e^{\lambda k_0})\,e^{-i(k_0+p_0)\,x_0}
\,e^{3\lambda k_0} \,\cdot\nonumber\\
 &\cdot&\,\delta(\tilde{k_0}^2 -
e^{\lambda k_0}k^2-m^2) \,\delta(\tilde{p_0}^2 - e^{\lambda
(p_0+k_0)}\tilde{k_0}^2+m^2(e^{\lambda (p_0+k_0)}-1))\,  \cdot \nonumber\\
&\cdot& \Bigg\{ \left( \frac{2-e^{-\lambda p_0} -e^{\lambda
k_0}}{\lambda}  +\lambda\, m^2  \right)\, \Bigg[ \lambda\hat{k}_0\,
\left(\hat{p}_0+\hat{p}_4 \right)\,+\nonumber\\
&+&\, e^{ -\lambda k_0} \, \left(\frac{e^{\lambda p_0}-1}{\lambda}
\, - \frac{\lambda m^2}{2} \right)
 - \lambda k_i^2\, e^{\lambda(p_0+k_0)}\Bigg]
  \Bigg\}\,,\nonumber\\
\end{eqnarray}
\noindent where $\tilde{k}_\mu$ and $\hat{k}_A$  are functions of the
Fourier parameters $k_\alpha$ of the same form as, respectively,
$\tilde{P}_\mu$ and $\hat{P}_A$, explicitly\\
\begin{equation}
\{\tilde{k}_0,\vec{\tilde{k}}\}|_{k_0,\vec{k}}\equiv
\big\{\frac{2}{\lambda}\sinh(\frac{\lambda}{2} k_0),\,\vec{k} \,e^{\frac{\lambda}{2} k_0}\big\}
\end{equation}
\begin{equation}
\{\hat{k}_0,\,\vec{\hat{k}}
,\,\hat{k}_4\}|_{k_0,\vec{k}}\equiv\big\{\frac{1}{\lambda} ( \sinh {\lambda k_0} +
\frac{\lambda^2}{2} \vec{k}^2 e^{ \lambda k_0} )\,,\vec{k} e^{ \lambda k_0},\, \frac{1}{\lambda} ( \cosh {\lambda k_0} -1 -
\frac{\lambda^2}{2} \vec{k}^2 e^{ \lambda k_0} )\big\} \,,
\end{equation}\\
\noindent and we used the relation\\
\begin{equation}
\hat{P}_0=\frac{e^{\lambda P_0}-1}{\lambda}  \, - \frac{\lambda
m^2}{2}\,.
\end{equation}\\
\noindent Looking at the requirement enforced by the second delta
function\\
\begin{equation*}
\tilde{p_0}^2 - e^{\lambda (p_0+k_0)}\tilde{k_0}^2+m^2(e^{\lambda
(p_0+k_0)}-1)=0\,,
\end{equation*}\\
\noindent one notices that it leads to two possible solutions\\
\begin{equation*}
 p_0^{(1)}= -k_0 ~~~~~~~~~~~~ e^{-\lambda
p_0^{(2)}}= 2-e^{\lambda k_0}+\lambda^2 m^2\,.
\end{equation*}\\
\noindent On the first solution the $\hat{Q}_0$ functional result to be
time-independent, while on the second solution the time independence
appears because of the vanishing of the $\hat{Q}_0$ functional. In fact,
the presence of the term $\left( \frac{2-e^{-\lambda p_0}
-e^{\lambda k_0}}{\lambda} +\lambda\, m^2 \right)$ inside the
expression of $\hat{Q}_0$ gives straightforwardly a vanishing charge on
the second solution.

Substituting the first solution of the second delta function, $p_0=
-k_0$, the value of the time-independent $\hat{Q}_0$ functional will be
given
by\\
\begin{eqnarray*}
\hat{Q}_0&=&\frac{1}{2}\int  d^4k dp_0 \Phi(k)
\Phi(p_0,\dot{-}\vec{k})e^{3\lambda k_0}e^{-i(k_0 +
p_0)t}\delta(\tilde{k_0}^2 - e^{\lambda
k_0}k^2-m^2)\,\cdot\nonumber\\
&\cdot&\Bigg\{ \left( \frac{2-e^{-\lambda p_0} -e^{\lambda
k_0}}{\lambda}  +\lambda\, m^2  \right)\,\cdot \nonumber\\
&\cdot&\Bigg[ \lambda\hat{k}_0\, \left(\hat{p}_0+\hat{p}_4
\right)\,+ e^{ -\lambda k_0} \, \left(\frac{e^{\lambda
p_0}-1}{\lambda} \, - \frac{\lambda m^2}{2} \right)
 - \lambda k_i^2\, e^{\lambda(p_0+k_0)}\Bigg]
  \Bigg\}\cdot\nonumber\\
&\cdot&\frac{\delta(k_0+p_0)}{|\partial_{p_0}[\tilde{p_0}^2 -
e^{\lambda (p_0+k_0)}\tilde{k_0}^2+m^2(e^{\lambda
(p_0+k_0)}-1)]_{p_0=-k_0}|}\nonumber\\
\end{eqnarray*}
\begin{equation} \label{eq:Q_0}
 \Rightarrow\,\,\,\,\hat{Q}_0=-\frac{1}{2}\int d^4k\,  \tilde{\Phi}(k)
\tilde{\Phi}(\dot{-}k) \,e^{3\lambda k_0}\, \delta(C_\lambda
(k)-m^2) \, \frac{(-2\tilde{k_0}e^{\frac{\lambda}{2}k_0}+\lambda
m^2)}{|-2\tilde{k_0}e^{\frac{\lambda}{2}k_0}+\lambda m^2|}\,
\hat{k}_0\,,
\end{equation}\\
\noindent where we introduced the notations $k \equiv
(k_0,\vec{k})$, $\dot{-}k \equiv (-k_0, -\vec{k}e^{\lambda k_0})$.

The proof of the time independence of the $\hat{Q}_i$ and
$\hat{Q}_4$ functionals will be similar to the one has been shown
for the $\hat{Q}_0$ functional. In particular\\
\begin{eqnarray}\label{eq:ProtoQi}
\hat{Q}_i\!\!&=&\!\!\int\!\! d^3x J_{0i}\,=\,\frac{1}{2}\int d^3 x
\Bigg\{ \lambda\hat{P}_i\, \left[  \phi \left( \frac{1-e^{-\lambda
P_0} }{\lambda} \right)-
\left( \frac{e^{\lambda P_0} -1}{\lambda}\right)\, \phi+\lambda\, m^2 \phi    \right]\, \cdot\!\!\nonumber\\
&\cdot&\,\hat{P}_0\, \phi\,+ \,  \left[  \phi \,\left(
\frac{1-e^{-\lambda P_0} }{\lambda} \right)
- \left( \frac{e^{\lambda P_0} -1}{\lambda}\right) \, \phi+\lambda\, m^2 \phi     \right] \, \hat{P}_i\, \phi+\nonumber\\
&+& \lambda \hat{P}_i\,  \left[    \phi \left( \frac{1-e^{-\lambda
P_0} }{\lambda}  \right)-
 \left( \frac{e^{\lambda P_0} -1}{\lambda}\right)\, \phi+\lambda\, m^2 \phi         \right]\, \hat{P}_4\, \phi \Bigg\}=\nonumber\\
&=&\,\frac{1}{2}\int d^3 x\, d^4k\,  d^4p\, \tilde{\Phi}(k_0,k_i)\,
\tilde{\Phi}(p_0,p_i) \,e^{i( \vec{k}+e^{-\lambda k_0} \vec{p})\cdot
\vec{x}}\, e^{-i(k_0+p_0)\,x_0} \,  \cdot \nonumber\\
&\cdot&\,\delta(C_\lambda (k)-m^2) \, \delta(C_\lambda (p)-m^2)\,
\Bigg\{ \lambda\hat{k}_i\,
\left( \frac{2-e^{-\lambda p_0} -e^{\lambda k_0}}{\lambda}  +\lambda\, m^2  \right)\, \hat{p}_0 +\nonumber\\
&+&\!\!\! \left( \frac{2-e^{-\lambda p_0} -e^{\lambda k_0}}{\lambda}
+\lambda\, m^2  \right) \hat{p}_i \,+ \, \lambda\hat{k}_i \left(
\frac{2-e^{-\lambda p_0} -e^{\lambda k_0}}{\lambda}  +\lambda\, m^2
\right)
 \hat{p}_4 \Bigg\}=\!\!\!\nonumber\\
 &=&\,\frac{1}{2}\int d^4k\,  d^4p\, \tilde{\Phi}(k_0,k_i)
\tilde{\Phi}(p_0,p_i) \,  e^{-i(k_0+p_0)\,x_0}\,e^{3\lambda k_0} \,  \cdot\nonumber\\
&\cdot& \delta(\vec{p} +\vec{k}\, e^{\lambda k_0})
\,\delta(C_\lambda
(k)-m^2) \, \delta(C_\lambda (p)-m^2)\,\cdot \nonumber\\
&\cdot& \Bigg\{ \left( \frac{2-e^{-\lambda p_0} -e^{\lambda
k_0}}{\lambda}  +\lambda\, m^2  \right)\, \Bigg[ \lambda\hat{k}_i\,
\hat{p}_0 \,+\, \hat{p}_i \,+\, \lambda\hat{k}_i \, \hat{p}_4\Bigg]
  \Bigg\}\nonumber\\
&=&\,\frac{1}{2}\int d^4k\,  d^4p\, \tilde{\Phi}(k_0,k_i)
\tilde{\Phi}(p_0,-k_i\,e^{\lambda k_0})\,e^{-i(k_0+p_0)\,x_0}
\,e^{3\lambda k_0} \,\cdot\nonumber\\
 &\cdot&\,\delta(\tilde{k_0}^2 -
e^{\lambda k_0}k^2-m^2) \,\delta(\tilde{p_0}^2 - e^{\lambda
(p_0+k_0)}\tilde{k_0}^2+m^2(e^{\lambda (p_0+k_0)}-1))\,  \cdot \nonumber\\
&\cdot& \Bigg\{ \left( \frac{2-e^{-\lambda p_0} -e^{\lambda
k_0}}{\lambda}  +\lambda\, m^2  \right)\, \Bigg[ \lambda\hat{k}_i
\left(\hat{p}_0+\hat{p}_4
\right)\,-\,k_i\,e^{\lambda(k_0+p_0)}\Bigg]
  \Bigg\}\nonumber\\
\end{eqnarray}
\noindent and\\
\begin{eqnarray}\label{eq:ProtoQ4}
\hat{Q}_4&=&\int d^3x J_{04}\,=\,\frac{1}{2}\int d^3 x \Bigg\{
\left(
\lambda\hat{P}_4+1- e^{ -\lambda P_0} \right)\,\cdot\,\nonumber\\
&\cdot& \left[  \phi \left( \frac{1-e^{-\lambda P_0} }{\lambda}
\right)-
\left( \frac{e^{\lambda P_0} -1}{\lambda}\right)\, \phi+\lambda\, m^2 \phi \right]\, \hat{P}_0\, \phi+\nonumber\\
&-& \lambda P_i \,  \left[  \phi \,\left( \frac{1-e^{-\lambda P_0}
}{\lambda}  \right)
- \left( \frac{e^{\lambda P_0} -1}{\lambda}\right) \, \phi+\lambda\, m^2 \phi \right] \, \hat{P}_i\, \phi+\nonumber\\
&+& \left(\lambda \hat{P}_4+1\right)\,  \left[    \phi \left(
\frac{1-e^{-\lambda P_0} }{\lambda}  \right)-
 \left( \frac{e^{\lambda P_0} -1}{\lambda}\right)\, \phi+\lambda\, m^2 \phi \right]\, \hat{P}_4\, \phi \Bigg\}=\nonumber\\
&=&\,\frac{1}{2}\int d^3 x\, d^4k\,  d^4p\, \tilde{\Phi}(k_0,k_i)\,
\tilde{\Phi}(p_0,p_i) \,e^{i( \vec{k}+e^{-\lambda k_0} \vec{p})\cdot
\vec{x}}\, e^{-i(k_0+p_0)\,x_0} \,  \cdot \nonumber\\
&\cdot&\,\delta(C_\lambda (k)-m^2) \, \delta(C_\lambda (p)-m^2)\,\cdot \nonumber\\
&\cdot&\,\Bigg\{ \left( \lambda\hat{k}_4+1- e^{ -\lambda k_0}
\right)\,
\left( \frac{2-e^{-\lambda p_0} -e^{\lambda k_0}}{\lambda}  +\lambda\, m^2  \right)\, \hat{p}_0 +\nonumber\\
&-&\, \lambda k_i \,\left( \frac{2-e^{-\lambda p_0} -e^{\lambda
k_0}}{\lambda}  +\lambda\, m^2  \right)
\hat{p}_i \,+ \nonumber\\
&+&\, \left(\lambda \hat{k}_4+1\right) \left( \frac{2-e^{-\lambda
p_0} -e^{\lambda k_0}}{\lambda}  +\lambda\, m^2  \right)
 \hat{p}_4 \Bigg\}=\nonumber\\
 &=&\,\frac{1}{2}\int d^4k\,  d^4p\, \tilde{\Phi}(k_0,k_i)
\tilde{\Phi}(p_0,p_i) \,  e^{-i(k_0+p_0)\,x_0}\,e^{3\lambda k_0} \,  \cdot\nonumber\\
&\cdot& \delta(\vec{p} +\vec{k}\, e^{\lambda k_0})
\,\delta(C_\lambda
(k)-m^2) \, \delta(C_\lambda (p)-m^2)\,\cdot \nonumber\\
&\cdot& \Bigg\{ \left( \frac{2-e^{-\lambda p_0} -e^{\lambda
k_0}}{\lambda}  +\lambda\, m^2  \right)\,\cdot \nonumber\\
&\cdot&\Bigg[  \left( \lambda\hat{k}_4+1- e^{ -\lambda k_0}  \right)
\, \hat{p}_0 \,-\,\lambda k_i\, \hat{p}_i \,+\,   \left(\lambda
\hat{k}_4+1\right)  \,
  \hat{p}_4\Bigg]
  \Bigg\}\nonumber\\
&=&\,\frac{1}{2}\int d^4k\,  d^4p\, \tilde{\Phi}(k_0,k_i)
\tilde{\Phi}(p_0,-k_i\,e^{\lambda k_0})\,e^{-i(k_0+p_0)\,x_0}
\,e^{3\lambda k_0} \,\cdot\nonumber\\
 &\cdot&\,\delta(\tilde{k_0}^2 -
e^{\lambda k_0}k^2-m^2) \,\delta(\tilde{p_0}^2 - e^{\lambda
(p_0+k_0)}\tilde{k_0}^2+m^2(e^{\lambda (p_0+k_0)}-1))\,  \cdot \nonumber\\
&\cdot& \Bigg\{ \left( \frac{2-e^{-\lambda p_0} -e^{\lambda
k_0}}{\lambda}  +\lambda\, m^2  \right)\,\cdot \nonumber\\
&\cdot&\!\! \Bigg[\left(\lambda \hat{k}_4+1\right)\,
\left(\hat{p}_0+\hat{p}_4 \right)\,-\, e^{ -\lambda k_0} \,
\left(\frac{e^{\lambda p_0}-1}{\lambda} \, - \frac{\lambda m^2}{2}
\right)
 + \lambda k_i^2\, e^{\lambda(p_0+k_0)}\Bigg]
  \Bigg\}\,.\!\!\nonumber\\
\end{eqnarray}
\noindent It is now clear that both the relations (\ref{eq:ProtoQi})
and (\ref{eq:ProtoQ4}) vanish on the solution $e^{-\lambda k_0}=
2-e^{\lambda p_0}+\lambda^2 m^2$, while for the $p_0=-k_0$ solution
the values of the time independent functionals are recovered:\\
\begin{equation}\label{eq:Q_i}
\hat{Q}_i\,=\,-\frac{1}{2}\int d^4k\,  \tilde{\Phi}(k)
\tilde{\Phi}(\dot{-}k) \,e^{3\lambda k_0}\, \delta(C_\lambda
(k)-m^2) \, \frac{(-2\tilde{k_0}e^{\frac{\lambda}{2}k_0}+\lambda
m^2)}{|-2\tilde{k_0}e^{\frac{\lambda}{2}k_0}+\lambda m^2|}\,
\hat{k}_i\,
\end{equation}\\
\noindent and\\
\begin{equation}\label{eq:Q_4}
\!\hat{Q}_4\,=\,-\frac{1}{2}\int d^4k\,  \tilde{\Phi}(k)
\tilde{\Phi}(\dot{-}k) \,e^{3\lambda k_0}\, \delta(C_\lambda
(k)-m^2) \, \frac{(-2\tilde{k_0}e^{\frac{\lambda}{2}k_0}+\lambda
m^2)}{|-2\tilde{k_0}e^{\frac{\lambda}{2}k_0}+\lambda m^2|}\,
\hat{k}_4\,.\!\!\!\!\!
\end{equation}\\
\noindent  Thus we can rewrite the charges in a more compact form\\
\begin{equation}\label{eq:QA}
\!\!\!\Bigg(\begin{array}{c}
   \hat{Q}_0 \\
   \hat{Q}_i \\
   \hat{Q}_4
 \end{array}\Bigg)\!\!=\!-\frac{1}{2}\int  d^4k \Phi(k) \Phi(\dot{-}k)e^{3\lambda k_0}\,
\frac{(-2\tilde{k_0}e^{\frac{\lambda}{2}k_0}+\lambda
m^2)}{|-2\tilde{k_0}e^{\frac{\lambda}{2}k_0}+\lambda
m^2|}\Bigg(\begin{array}{c}
   \hat{k}_0 \\
   \hat{k}_i \\
   \hat{k}_4
 \end{array}\Bigg)\!\delta(C_{\lambda}(k)-m^2)\,.\!\!\!\!\!\!
\end{equation}\\
Thereby, our analysis leads to 5 translation-symmetry conserved
charges from the 5D-calculus setup and, both in the massless and in
the massive case, the
 charges we obtain are not classical: they are
functional of the fields with a non-linear dependence on the
Plank-scale $\lambda$ and a delta of the deformed casimir. Just in
the limit $\lambda\rightarrow 0$ we reobtain the classical charges.
Besides, our Noether analysis constructively led us to a
``conservation equation" of the form
${\mathcal{\hat{D}}}_{0}\,J^{0}_A+\hat{P}_{i}\,J^{i}_A=0$.

\subsection{On a possible different choice of the 5D differential
calculus basis}\label{par:DifferentBasis}

We have seen in section \ref{par:5DCurrents} that the proper time
derivative operator is given by
${\mathcal{\hat{D}}}_0=\hat{P}_0+\hat{P}_4$. One may now look for a
change of the differential calculus basis which enables us to
rewrite the differential $df$ in terms of the operator
${\mathcal{\hat{D}}}_0$, i.e. performing a change of basis for the
transformation parameters such that the external derivative operator
$d$ still satisfies the Leibnitz rule. It is easy to see that the
following change of basis (rotation) for the one-form
generators\footnote{This change of basis was introduced by Sitarz
\cite{Sitarz} just for a reason of convenience in the presentation
of the 5D differential
calculus, without any physical intent.}\\
\begin{equation}\label{eq:dxChange}
\bar{d}\hat{x}_0=(d\hat{x}_0+
d\hat{x}_4)/\sqrt{2}~~~~~~~~~\bar{d}\hat{x}_i=d\hat{x}_i~~~~~~~~~\bar{d}\hat{x}_4=(d\hat{x}_0-d\hat{x}_4)/\sqrt{2}\,,
\end{equation}\\
\noindent endowed with the commutation relations\\
\begin{equation*}
[\hat{x}_0, \bar{d}\hat{x}_0] = i \lambda \bar{d}\hat{x}_0\,,
~~~~~~~~~ [\hat{x}_0, \bar{d}\hat{x}_4] =- i \lambda
\bar{d}\hat{x}_4\,, ~~~~~~~~~ [\hat{x}_0, \bar{d}\hat{x}_j] = 0\,,
\end{equation*}
\begin{equation}\label{eq:BasisChange}
[\hat{x}_j, \bar{d}\hat{x}_4] =0\,,~~~~~~~~~   [\hat{x}_j,
\bar{d}\hat{x}_0] = -\sqrt{2}\,i \lambda \bar{d}\hat{x}_j\,,
~~~~~~~~~ [\hat{x}_j, \bar{d}\hat{x}_k] = -\sqrt{2}\,i \lambda
\delta_{jk} \bar{d}\hat{x}_4\,,
\end{equation}\\
\noindent suggest a natural way to write the differential
$df$, in particular\\
\begin{equation}\label{eq:df2}
df\,=\,(\bar{d}\hat{x}^0\bar{\mathcal{D}}_0+\bar{d}\hat{x}^i\bar{\mathcal{D}}_i+\bar{d}\hat{x}^4\bar{\mathcal{D}}_4)\,f\,\equiv\,\bar{d}f\,,
\end{equation}\\
\noindent where\\
\begin{equation}
\bar{\mathcal{D}}_0=(\hat{P}_0+\hat{P}_4)/\sqrt{2}=\mathcal{\hat{D}}_0/\sqrt{2}~~~~~~~~~\bar{\mathcal{D}}_i\equiv\hat{P}_i~~~~~~~~~
\bar{\mathcal{D}}_4=(\hat{P}_0-\hat{P}_4)/\sqrt{2}\,.
\end{equation}\\
The reason here to perform all the previous Noether analysis with
this new basis for the transformation parameters (and hence for the
translation generators), is to look for a more constraining
characterization of the energy observable. The introduction of the
proper time derivative operator from the beginning of the analysis,
i.e. inside the definition of the differential of a generic
$\kappa$-Minkowski element, might lead to a stronger intuition to
identify a plausible energy charge\footnote{Even though we started
from the request that the time derivative operator
$\mathcal{\hat{D}}_0$ enter the expression of the differential $df$,
arriving in this way to the one-form generators change of basis
(\ref{eq:dxChange}) satisfying this request, it can be shown that
working within this new basis for the differential calculus one
straightforwardly obtains the new basis for the translation
generators
$\{\bar{\mathcal{D}}_0\,,\bar{\mathcal{D}}_i\,,\bar{\mathcal{D}}_4\}$.
A less rigorous but more physically intuitive procedure might
consist of a direct manipulation of the $\delta \mathcal{L}$
expression (\ref{eq:protodiv}). For example, led by the intuition
that the charge associated to the current $J_{00}+J_{04}$ could
represent a good candidate for the energy observable, one could
notice that $\hat{\epsilon}^A\,{\mathcal{\hat{D}}}^0\,J_{0A}=
{\mathcal{\hat{D}}}^0\big(d\hat{x}^0(J_{00}+J_{04})+d\hat{x}^iJ_{0i}+(d\hat{x}^4-d\hat{x}^0)J_{04}\big)$.}.

Following all the steps of section \ref{par:5DCurrents}, one arrives
to the expression for the Lagrangian density variation
\begin{equation}
\delta\mathcal{L}=\bar{d}\hat{x}^A\left(\bar{\mathcal{D}}^0
\bar{J}_{0A}+ \bar{\mathcal{D}}^i \bar{J}_{iA}\right) \,,
\end{equation}
\noindent where
\begin{equation}
\bar{J}_{00}\equiv J_{00}+J_{04}\,,~~~~~~\bar{J}_{0 j}\equiv
\,\sqrt{2}\,J_{0 j}\,, ~~~~~~\bar{J}_{0 4}\equiv J_{00}-J_{04}\,,
\end{equation}
\begin{equation}
\bar{J}_{i0}\equiv (J_{i0}+J_{i4})/\sqrt{2}\,,~~~~~~\bar{J}_{i
j}\equiv J_{ij}\,, ~~~~~~\bar{J}_{i 4}\equiv
(J_{i0}-J_{i4})/\sqrt{2}\,.
\end{equation}\\
\noindent Thereby, the translation symmetry charges obtained from
the rotation of the $d\hat{x}_A$ basis (\ref{eq:dxChange}) are\\
\begin{eqnarray}\label{Qbarra}
\!\!\left(\begin{array}{c}
\bar{Q}_0\\
\bar{Q}_i\\
\bar{Q}_4\\
\end{array}{}\right)
\!\!\!&=&\!\!\!-\frac{1}{2}\int \! d^4k
\left|\tilde{\Phi}(k)\right|^2 \left(\begin{array}{c}
\hat{k}_0+\hat{k}_4\\
\sqrt{2}\,\hat{k}_i\\
\hat{k}_0-\hat{k}_4\\
\end{array}{}\right)  \frac{(-2\tilde{k_0}e^{\frac{\lambda}{2}k_0}+\lambda
m^2)}{|-2\tilde{k_0}e^{\frac{\lambda}{2}k_0}+\lambda
m^2|}\delta(C_{\lambda}(k)-m^2)\!=\!\!\nonumber\\
&=& \left(\begin{array}{c}
\hat{Q}_0+\hat{Q}_4\\
\sqrt{2}\,\hat{Q}_i\\
\hat{Q}_0-\hat{Q}_4\\
\end{array}{}\right)\,.
\end{eqnarray}\\
\noindent We will see in the next Chapter how $\bar{Q}_0$ might turn
out to be a valuable tool, since it is the conserved charge
associated with the transformation parameter $\bar{d}\hat{x}_0$, and
therefore (in light of the fact that in $\bar{d}f$ we have
$\bar{d}\hat{x}_0$ multiplying $\bar{\mathcal{D}}_0$, which is a
plausible time-translation generator) is a plausible candidate for
the energy charge.

\clearpage

\chapter{Energy-momentum dispersion relation}
\label{ch:DispersionRelation}

The interpretation that Quantum Group language gives to ``momenta"
as generators of translations (i.e. the real physical particle
momenta) is based on the notion of quantum group symmetry. Chapters
\ref{ch:Analysis4D} and \ref{ch:Analysis5D} provide two example of
the freedom there exist in the description of $\kappa$-Minkowski
symmetries by anyone of a large number of basis of the
$\kappa$-Poincar\'{e} Hopf algebra. The nature of this
symmetry-description degeneracy remains obscure from a physics
perspective, in particular we are used to associate energy-momentum
with the translation generators and it is not conceivable that a
given operative definition of energy-momentum could be equivalently
described in terms of different translation generators. The
difference would be easily established by testing, for example, the
different dispersion relations (a meaningful physical property) that
the different momenta satisfy.

In this context the claim for a Plank-scale modification of the
energy-momentum relation is a crucial key-point. We saw in
(\ref{eq:4DDispersion}) how, using a four-dimensional differential
calculus and the Majid-Ruegg $\kappa$-Poincar\'{e} basis for the
translation generators, a non-linear Plank-scale modification of the
dispersion relation for a free massless scalar field can be
obtained. In this chapter we will look for a possible modification
of the energy-momentum relation using the charges we obtained with
the five-dimensional bicovariant differential calculus
(\ref{eq:5Dcommutators}).

Since the 5D differential calculus is bicovariant under the action
of the full $\kappa$-Poincar\'{e} algebra and the basis generators
$\hat{P}_0, \hat{P}_i$ of translations transform under
$\kappa$-Poincar\'{e} action in the same way as the operators
$P_\mu$ in the commutative case transform under the standard
Poincar\'{e} action, it could be expected that the energy-momentum
relation remains classical. But this aspect of the 5D differential
calculus should not mislead the analysis of $\kappa$-Minkowski
translational symmetry, in fact the linearity of the
$\kappa$-Poincar\'{e} action on the commutation relation
(\ref{eq:5Dcommutators}) induces a highly non-trivial structure in
the coalgerba sector of the generators $\hat{P}_A$ and thereby a
non-trivial modification of the quantum symmetry.

Recovering a special-relativistic dispersion relation at the end of
the analysis would seem less likely than expected and we will see
that this indeed does not happen when considering a massive scalar
field.\\

\section{Dispersion relation for regularized plane-wave field $\Phi \in \mathbb{C}$}\label{par:PlaneWave}

The expressions for the charges obtained in (\ref{eq:QA}) can now be
used to investigate if there is any Plank-scale modification of the
energy-momentum relation with respect to the special-relativistic
(Poincar\'{e}-Lie-algebra) limit. We intend to probe the structure
of the dispersion relation by using a ``regularized plane-wave''
field. In preparation for that we first rewrite the charges
(\ref{eq:QA}) in a more compact form.

For a real scalar classic field $\Phi$, solution of
$C_{\lambda}(k)\Phi=m^2\Phi$ on $\kappa$-Minkowski\\
\begin{equation*}
\Phi(x)=\int
d^4k\tilde{\Phi}(k)\delta(C_{\lambda}(k)-m^2)e^{i\vec{k}\cdot\vec{x}}e^{-ik_0x_0}
\end{equation*}\\
\noindent holds the reality condition\\
\begin{equation}\label{eq:reality}
\tilde{\Phi}(k_0,\vec{k})=\left(\tilde{\Phi}(-k_0,-\vec{k}e^{\lambda
k_0})\right)^*e^{3\lambda k_0}
\end{equation}\\
\noindent that allows us to rewrite the charges as\\
\begin{equation}\label{eq:QAcompact}
\Bigg(\begin{array}{c}
   \hat{Q}_0 \\
   \hat{Q}_i \\
   \hat{Q}_4
 \end{array}\Bigg)=-\frac{1}{2}\int  d^4k |\tilde{\Phi}(k_0,\vec{k})|^2\,
\frac{(-2\tilde{k_0}e^{\frac{\lambda}{2}k_0}+\lambda
m^2)}{|-2\tilde{k_0}e^{\frac{\lambda}{2}k_0}+\lambda
m^2|}\,\Bigg(\begin{array}{c}
   \hat{k}_0 \\
   \hat{k}_i \\
   \hat{k}_4
 \end{array}\Bigg)\,\delta(C_{\lambda}(k)-m^2)\,.
\end{equation}\\
We now want to demonstrate that eq. (\ref{eq:reality}) holds for
complex fields too, i.e.\\
\begin{equation}\label{eq:Complexity}
\tilde{\Phi}(k_0,\vec{k})=\left(\tilde{\Phi}^*(-k_0,-\vec{k}e^{\lambda
k_0})\right)^*e^{3\lambda k_0}
\end{equation}\\
\noindent In fact\\
\begin{eqnarray}
\Phi^*(x)\!\!\!&=&\!\!\!\!\!\int\!\!\!
d^4k\tilde{\Phi}^*(k)\delta(C_{\lambda}(k)-m^2)e^{i\vec{k}\cdot\vec{x}}e^{-ik_0x_0}=
\nonumber\\
\!\!\!&=&\!\!\!\!\!\int \!\!d^4k'e^{-3\lambda
k_0'}\tilde{\Phi}^*(k_0',\vec{k}'e^{-\lambda
k_0'})\delta(C_{\lambda}(k'_0,\vec{k}'e^{-\lambda
k_0'})-m^2)e^{i\vec{k}'e^{-\lambda k_0'}\cdot\vec{x}}e^{-ik'_0x_0}=
\nonumber\\
\!\!\!&=&\!\!\!\!\!\int\!\! d^4ke^{3\lambda
k_0}\tilde{\Phi}^*(-k_0,-\vec{k}e^{\lambda
k_0})\delta(C_{\lambda}(-k_0,-\vec{k}e^{\lambda
k_0})-m^2)e^{-i\vec{k}e^{\lambda k_0}\cdot\vec{x}}e^{+ik_0x_0}=
\nonumber\\
\!\!\!&=&\!\!\!\!\!\int \!\!d^4ke^{3\lambda
k_0}\tilde{\Phi}^*(-k_0,-\vec{k}e^{\lambda
k_0})\delta(C_{\lambda}(k_0,\vec{k})-m^2)\left(e^{i\vec{k}\cdot\vec{x}}e^{-ik_0x_0}\right)^*;\!\!\!\!\!\!\!\!
\end{eqnarray}\label{eq:Complexity1}\!\!\!\!\\
\noindent conjugating now the last term of (\ref{eq:Complexity1})\\
\begin{equation}\label{eq:Complexity2}
\Phi(x)=\int d^4k\left(e^{3\lambda
k_0}\tilde{\Phi}^*(-k_0,-\vec{k}e^{\lambda
k_0})\right)^*\delta(C_{\lambda}(k_0,\vec{k})-m^2)e^{i\vec{k}\cdot\vec{x}}e^{-ik_0x_0}
\end{equation}\\
\noindent and comparing (\ref{eq:Complexity2}) with\\
\begin{equation*}
\Phi(x)=\int
d^4k\tilde{\Phi}(k)\delta(C_{\lambda}(k)-m^2)e^{i\vec{k}\cdot\vec{x}}e^{-ik_0x_0}\,,
\end{equation*}\\
\noindent one immediately has eq. (\ref{eq:Complexity}), which can
be rewritten as\\
\begin{equation}\label{eq:Complexity3}
e^{-3\lambda
k_0}\left(\tilde{\Phi}(k_0,\vec{k})\right)^*=\tilde{\Phi}^*(-k_0,-\vec{k}e^{\lambda
k_0})\,.
\end{equation}\\
We want to compute the translation-symmetry charges for a complex
scalar field in order to compare our results with those of
\cite{Friedel, Freidel2}. In the previous chapter we considered real
scalar fields, but actually the steps of the analysis are very
similar for the case of complex fields. Essentially it reduces to
the fact that in appropriate places one must consider the complex
conjugate $\Phi^*(x)$ of the field $\Phi(x)$. The action to use for
a complex
field is\\
\begin{eqnarray}
S[\Phi]&=&\int d^4x \mathcal{L}[\Phi(x)]\nonumber\\
\mathcal{L}[\Phi(x)]&=& \left(\Phi^*(x) \,C_\lambda\,\Phi(x) -
m^2\Phi^*(x)\Phi(x)\right)\,,
\end{eqnarray}\\
\noindent and proceeding exactly in the same way as in the previous
chapter one then easily arrives once again to the equation
$d\hat{x}^A\left(\,e^{\frac{\lambda
P_0}{2}}\tilde{P}^0\,J_{0A}+\hat{P}^iJ_{iA}\right)=0$, with $J$'s of
the same form as in the previous section but involving
$\Phi^*(x)$ in appropriate places. In particular, one finds\\
\begin{eqnarray}
J_{00}\!\!\!\!&=&\!\!\!\!\bigg\{\left(\frac{2}{\lambda}+\lambda
m^2-\frac{e^{\lambda P_0}}{\lambda}\right)\left[(\lambda\hat{P}_0+
e^{ -\lambda P_0} )\Phi^*\hat{P}_0\Phi + \lambda
 P_i\Phi^*\hat{P_i}\Phi
+\lambda \hat{P}_0\Phi^*\hat{P}_4\Phi\right]+\nonumber\\
&-&\!\!\!\!(\lambda\hat{P}_0+ e^{ -\lambda P_0} )\Phi^*
\frac{e^{-\lambda P_0}}{\lambda}\hat{P}_0\Phi - \lambda
 P_i\Phi^*\frac{e^{-\lambda P_0}}{\lambda}\hat{P_i}\Phi-
\lambda \hat{P}_0\Phi^*\frac{e^{-\lambda P_0}}{\lambda}\hat{P}_4\Phi\bigg\}\,, \nonumber \\
\nonumber \\
J_{0i}\!\!\!\!&=&\!\!\!\!\bigg\{\left(\frac{2}{\lambda}+\lambda
m^2-\frac{e^{\lambda
P_0}}{\lambda}\right)\bigg[\lambda\hat{P}_i\Phi^*\hat{P}_0\Phi+\Phi^*\hat{P}_i\Phi+\lambda\hat{P}_i\Phi^*\hat{P}_4\Phi\bigg]+\nonumber\\
&-&\!\!\!\!\,\lambda\hat{P}_i\frac{e^{-\lambda
P_0}}{\lambda}\Phi^*\hat{P}_0\Phi-\Phi^*\hat{P}_i\frac{e^{-\lambda
P_0}}{\lambda}\Phi-\lambda\hat{P}_i\Phi^*\frac{e^{-\lambda
P_0}}{\lambda}\hat{P}_4\Phi\bigg\}\,, \nonumber\\
\nonumber \\
J_{04}\!\!\!\!&=&\!\!\!\!\bigg\{\bigg(\frac{2}{\lambda}+\lambda
m^2-\frac{e^{\lambda P_0}}{\lambda}\bigg)\bigg[(\lambda\hat{P}_4+1-
e^{ -\lambda P_0} )\Phi^*\hat{P}_0\Phi - \lambda
 P_i\Phi^*\hat{P_i}\Phi
+\nonumber\\
&+&\!\!\!\!
(\lambda\hat{P}_4+1)\Phi^*\hat{P}_4\Phi\bigg]-(\lambda\hat{P}_4+1-
e^{-\lambda P_0} )\Phi^* \frac{e^{-\lambda P_0}}{\lambda}\hat{P}_0\Phi +\nonumber\\
\!\!\!\!&+&\!\!\!\!\lambda
 P_i\Phi^*\frac{e^{-\lambda P_0}}{\lambda}\hat{P_i}\Phi-
 (\lambda\hat{P}_4+1)\Phi^*\frac{e^{-\lambda
P_0}}{\lambda}\hat{P}_4\Phi\bigg\}\,.\!\!\!\!
\end{eqnarray}\label{eq:Jcomplex}\\
Hence the Noether analysis reported in Chapter \ref{ch:Analysis5D}
straightforwardly gives, for a complex scalar field, the following
expression for the charges:\\
\begin{equation}\label{eq:QAcomplex}
\!\!\!\!\Bigg(\begin{array}{c}
   \hat{Q}_0 \\
   \hat{Q}_i \\
   \hat{Q}_4
 \end{array}\Bigg)=-\int  d^4k \Phi(k) \Phi^*(\dot{-}k)e^{3\lambda k_0}\,
\frac{(-2\tilde{k_0}e^{\frac{\lambda}{2}k_0}+\lambda
m^2)}{|-2\tilde{k_0}e^{\frac{\lambda}{2}k_0}+\lambda
m^2|}\,\Bigg(\begin{array}{c}
   \hat{k}_0 \\
   \hat{k}_i \\
   \hat{k}_4
 \end{array}\Bigg)\,\delta(C_{\lambda}(k)-m^2)\,.\!\!\!\!
\end{equation}\\
\noindent Eq. (\ref{eq:Complexity3}) enables us to rewrite the
charges for a complex scalar field in the more compact form\\
\begin{equation}\label{eq:QAcomplexcompact}
\!\!\!\!\Bigg(\begin{array}{c}
   \hat{Q}_0 \\
   \hat{Q}_i \\
   \hat{Q}_4
 \end{array}\Bigg)=-\int  d^4k |\tilde{\Phi}(k_0,\vec{k})|^2\,
\frac{(-2\tilde{k_0}e^{\frac{\lambda}{2}k_0}+\lambda
m^2)}{|-2\tilde{k_0}e^{\frac{\lambda}{2}k_0}+\lambda
m^2|}\,\Bigg(\begin{array}{c}
   \hat{k}_0 \\
   \hat{k}_i \\
   \hat{k}_4
 \end{array}\Bigg)\,\delta(C_{\lambda}(k)-m^2)\,.\!\!\!\!
\end{equation}\\
\noindent And therefore it is clear that also for complex fields the
translation-symmetry charges are real.

To write the plane-wave field $\Phi_0^{p.w.}(x)$ solution of the
deformed Klein-Gordon equation we need to calculate first the
solutions of
$\delta(C_{\lambda}(k)-m^2)$:\\
\begin{equation*}
\delta(C_{\lambda}(k)-m^2)=\delta\left((\frac{2}{\lambda}\sinh{\frac{\lambda
k_0}{2}})^2-|\vec{k}|^2e^{\lambda k_0}-m^2\right)=
\end{equation*}\\
\begin{equation}
=\frac{1}{2\sqrt{m^2+|\vec{k}|^2+\lambda^2m^4/4}}(\delta(k_0
-k^+_0)+\delta(k_0 -k^-_0))\,,
\end{equation}\\
\noindent where\\
\begin{equation*}
k^+_0=\frac{1}{\lambda}\ln\left(\frac{1+(\lambda m)^2/2+\lambda
\sqrt{m^2+|\vec{k}|^2+\lambda^2m^4/4}}{1-(\lambda
|\vec{k}|)^2}\right)
\end{equation*}\\
\begin{equation}
k^-_0=\frac{1}{\lambda}\ln\left(\frac{1+(\lambda m)^2/2-\lambda
\sqrt{m^2+|\vec{k}|^2+\lambda^2m^4/4}}{1-(\lambda
|\vec{k}|)^2}\right)\,;
\end{equation}\\
\noindent from the signs analysis it's easy to see that $k^+_0$ is
positive and $k^-_0$ is negative in the definition dominion
$|\vec{k}|<\frac{1}{\lambda}$. It can also be seen that $k^+_0$ is
real only in the dominion $|\vec{k}|<\frac{1}{\lambda}$. Besides,
it's obvious how in the ``classic limit'' $\lambda\rightarrow0$,
$k^+_0$ and $k^-_0$ go respectively to the positive,
$\sqrt{|\vec{k}|^2+m^2}$, and negative, $-\sqrt{|\vec{k}|^2+m^2}$,
``classic'' frequencies.

Setting $N=2\sqrt{m^2+|\vec{k}|^2+\lambda^2m^4/4}$, the particular
regularized plane-wave $\Phi_0^{p.w.}(x)$, solution of the equation of motion, can be written as:\\
\begin{eqnarray}
\Phi_0^{p.w.}(x)&=&\int d^4k
\frac{\sqrt{N}\theta(k_0)\delta(\vec{k}-\vec{p})
}{\sqrt{V}}e^{i\vec{k}\cdot\vec{x}}e^{-ik_0x_0}\delta(C_{\lambda}(k_0,\vec{k})-m^2)=
\nonumber\\
&=&\int \frac{d^4k}{N} \frac{\sqrt{N}\delta(\vec{k}-\vec{p})
}{\sqrt{V}}e^{i\vec{k}\cdot\vec{x}}e^{-ik_0x_0}\delta(k_0 -k^+_0)=
\nonumber\\
&=&\frac{1}{(2V\sqrt{m^2+|\vec{p}|^2+\lambda^2m^4/4})^{\frac{1}{2}}}e^{i\vec{p}\cdot\vec{x}}e^{-ip^+_0x_0}\,,
\end{eqnarray}\\
\noindent where $V$ is a normalization spatial volume of the
plane-wave.

Noting that $(-2\tilde{k_0}e^{\frac{\lambda}{2}k_0}+\lambda
m^2)$ is negative for $|\vec{k}|<\frac{1}{\lambda}$, we are now ready to compute the charges:\\
\begin{eqnarray}\label{eq:QAplanewave}
\!\!\!\!\Bigg(\begin{array}{c}
   \hat{Q}_0^{p.w.} \\
   \hat{Q}_i^{p.w.} \\
   \hat{Q}_4^{p.w.}
 \end{array}\Bigg)\!\!\!\!&=&\!\!\!\!-\!\!\int\!\!  d^4k\Bigg(\!\begin{array}{c}
   \hat{k}_0 \\
   \hat{k}_i \\
   \hat{k}_4
 \end{array}\!\Bigg)
|\frac{\sqrt{N}\theta(k_0)\delta(\vec{k}-\vec{p}) }{\sqrt{V}}|^2
\frac{(-2\tilde{k_0}e^{\frac{\lambda}{2}k_0}+\lambda
m^2)}{|-2\tilde{k_0}e^{\frac{\lambda}{2}k_0}+\lambda
m^2|}\delta(C_{\lambda}(k)-m^2)=\nonumber\\
\!\!\!\!&=&\!\!\!\!-\!\!\int\!\! d^4k \Bigg(\!\begin{array}{c}
   \hat{k}_0 \\
   \hat{k}_i \\
   \hat{k}_4
 \end{array}\!\Bigg)\!\left|\sqrt{N}\right|^2\!\frac{(-2\tilde{k_0}e^{\frac{\lambda}{2}k_0}+\lambda
m^2)}{|-2\tilde{k_0}e^{\frac{\lambda}{2}k_0}+\lambda
m^2|}\theta(k_0)\delta(\vec{k}-\vec{p})\delta(C_{\lambda}(k)-m^2)=\!\!\!\!\nonumber\\
&=&\,\int d^4k \,\Bigg(\begin{array}{c}
   \hat{k}_0 \\
   \hat{k}_i \\
   \hat{k}_4
 \end{array}\Bigg)\,\frac{\left|\sqrt{N}\right|^2}{N}\delta(k_0
-k^+_0)\delta(\vec{k}-\vec{p})=\nonumber\\
&=&\,\int d^3k\,\Bigg(\begin{array}{c}
   \hat{k}_0(k^+_0,\vec{k}) \\
   \hat{k}_i (k^+_0,\vec{k})\\
   \hat{k}_4(k^+_0,\vec{k})
 \end{array}\Bigg)\,
\delta(\vec{k}-\vec{p})=\nonumber\\
\!\!&=&\,\Bigg(\begin{array}{c}
   \hat{k}_0\big|_{k_0=p^+_0, \vec{k}=\vec{p}} \\
   \hat{k}_i \big|_{k_0=p^+_0, \vec{k}=\vec{p}}\\
   \hat{k}_4\big|_{k_0=p^+_0, \vec{k}=\vec{p}})
 \end{array}\Bigg)\!\!
\end{eqnarray}\\
\noindent``on shell'' with respect to the deformed Casimir of the
bicrossproduct basis $C_\lambda(p_0^+,\vec{p})$.

In light of the hypothesis of a modified dispersion relation for
particles in the context of Quantum Groups approach to the problem
of Quantum Gravity, we may now investigate this possibility. A key
role in this investigation is played by the energy observable, i.e.
the identification of a plausible candidate for the energy
observable is a fundamental step for any claim.

It is perhaps intriguing that, from the equation of motion $C_\lambda(p_0^+,\vec{p})=m^2$,\\
\begin{eqnarray}
(\hat{Q}^{p.w.}_0)^2-(\hat{Q}^{p.w.}_i)^2 &=&\left(\frac{e^{\lambda
p^+_0}-1}{\lambda}-\frac{\lambda
m^2}{2}\right)^2-\left(\vec{p}\,e^{\lambda p^+_0}\right)^2=
\nonumber\\
&=&
m^2\left(1+\frac{\lambda^2m^2}{4}\right)=m^2+(\hat{Q}^{p.w.}_4)^2\,,
\end{eqnarray}\\
\noindent but this should be analyzed taking into consideration the
fact that, in light of the observations we reported in section
\ref{par:5DCurrents} on time translations, $\hat{Q}^{p.w.}_0$
clearly cannot be the energy carried by our regularized plane wave.

The Noether analysis reported in section \ref{par:DifferentBasis}
for a rotated basis of the transformation parameters, might now be
taken into consideration to contemplate a role for the combination
$\hat{Q}^{p.w.}_0+\hat{Q}^{p.w.}_4$, which has emerged as the
conserved charge associated with a transformation parameter
($\bar{d}\hat{x}_0$) that can be meaningfully described as a
time-translation parameter. Thus, taking
$\hat{Q}^{p.w.}_0+\hat{Q}^{p.w.}_4$ as a candidate for the energy
observable, we find that\\
\begin{eqnarray}\label{eq:PlaneWaveDispersion}
(\hat{Q}^{p.w.}_0+\hat{Q}^{p.w.}_4)^2-(\hat{Q}^{p.w.}_i)^2
&=&\left(\frac{e^{\lambda
p^+_0}-1}{\lambda}\right)^2-\left(\vec{p}\,e^{\lambda
p^+_0}\right)^2=
\nonumber\\
&=& \left(\lambda (\hat{Q}^{p.w.}_0+  \hat{Q}^{p.w.}_4)
+1\right)m^2\,.
\end{eqnarray}\\
Eq. (\ref{eq:PlaneWaveDispersion}) shows how in the massless case
there is no Plank-scale modification of the energy-momentum
relation, the dispersion relation is classical, as in \cite{Friedel,
Freidel2}, even though the charges are not. While, in the massive
case, there is a $\lambda$ deformation and the energy-momentum
relation is no more special-relativistic, in fact the right-side
term of (\ref{eq:PlaneWaveDispersion}) is not a relativistic
invariant.

However, it is interesting to notice that, as in the analysis with a
four-dimensional differential calculus reported in Chapter
\ref{ch:Analysis4D}, this modification vanishes if one increases
arbitrary the intensity of the fields. In fact, if we rescale the
fields $\Phi_0$ by a factor A, we have\\
\begin{equation}
\Phi^R_0=A\Phi_0~~~\Rightarrow~~~(\hat{Q}^{p.w.\,R}_0+\hat{Q}^{p.w.\,R}_4,\hat{Q}_i^{p.w.\,R})=
A^2(\hat{Q}^{p.w.}_0+\hat{Q}^{p.w.}_4,{\hat{Q}}_i^{p.w.})
\end{equation}\\
\noindent and, rewriting eq. (\ref{eq:PlaneWaveDispersion}),\\
\begin{equation*}
\frac{(\hat{Q}^{p.w.\,R}_0+\hat{Q}^{p.w.\,R}_4)^2}{A^4}
-\frac{(\hat{Q}_i^{p.w.\,R})^2}{A^4}=m^2\left(1+\lambda
\frac{(\hat{Q}^{p.w.\,R}_0+\hat{Q}^{p.w.\,R}_4)}{A^2}\right)
\end{equation*}
\begin{equation}
\!\!\Rightarrow~~~(\hat{Q}^{p.w.\,R}_0+\hat{Q}^{p.w.\,R}_4)^2-(\hat{Q}_i^{p.w.\,R})^2=(A^2m)^2\left(1+\lambda
\frac{(\hat{Q}^{p.w.\,R}_0+\hat{Q}^{p.w.\,R}_4)}{A^2}\right).\!\!\!
\end{equation}\\
\noindent So, in the limit $A\rightarrow\infty$, the
special-relativistic relation is reestablished with mass $m^R=A^2m$.

\clearpage

\chapter*{Conclusions}\label{par:Conclusions}
\addcontentsline{toc}{chapter}{Conclusions}

In this thesis work we have investigated the connection between
Hopf-algebra-type symmetry (Quantum symmetry) and five-dimensional
bicovariant differential calculus, both concepts already present in
the literature but never combined together within a rigorous and
comprehensive analysis. The concept of Quantum symmetry in the
context of $\kappa$-Minkowski noncommutative spacetime has revealed
many interesting aspects but also some ambiguities. The formulation
in terms of Hopf algebra (quantum) version of the classical
Poincar\'{e} group for the symmetries of a free scalar field theory
in $\kappa$-Minkowski provided a solid background for a Noether
analysis of these symmetries. But the classical interpretation of
real physical particle momenta as the conserved charges associated
to the translation generators seems to be puzzling in the Quantum
Group language. In fact, in Chapter \ref{ch:NoncommutativeGeometry}
we illustrated the freedom there exist in the description of
$\kappa$-Minkowski symmetries by anyone of a large number of
$\kappa$-Poincar\'{e} Hopf algebra basis. In particular, we are left
with a choice between different realization of the concept of
translations in the noncommutative spacetime. This
symmetry-description degeneracy  raises a puzzling question: which
translation generators basis gives the real physical energy-momentum
charges?

Besides, the discovery of the central role that the introduction of
a differential calculus has in order to be able to complete a
Noether analysis brings about further ambiguities. We have seen in
Chapters \ref{ch:Analysis4D} and \ref{ch:Analysis5D} that, while in
the commutative case there is only one natural differential
calculus, involving the conventional derivatives, in the
$\kappa$-Minkowski case the introduction of a differential calculus
is not unique. The construction of the $\kappa$-Minkowski spacetime
enables us to use the tools of noncommutative geometry to construct
$\kappa$ deformations of field theory. The differential calculus,
being the most important tool, is therefore a crucial point of these
efforts. The search of a differential calculus that is left
invariant under the action of the full $\kappa$-Poincar\'{e} algebra
seems to be a reasonable choice and might have motivated some
authors to think that the ``classical case" might be recovered. The
fact that the new translation generators basis $\hat{P}_A$
introduced by the 5D differential calculus transform under
$\kappa$-Poincar\'{e} action with the same commutation relations of
the commutative case, i.e as the classical operators $P_\mu$ on
Minkowski spacetime, is quite surprising and might motivate the
attribution to these generators of a special or privileged role. In
Chapter \ref{ch:Analysis5D} we have shown that this sort of
linearity of the operators $\hat{P}_A$ has as counterpart an high
non-trivial structure of the coalgebra sector and of the commutation
relations of the one-form elements of the 5D differential calculus
with the time-to-the-right-ordered plane wave basis of
$\kappa$-Minkowski. Thereby the overall structure of the quantum
symmetry is once again far from satisfactory. This thesis work shows
that the elegant, and, from a certain point of view, natural
requirement of bicovariance of the differential calculus adopted is
not strong enough to eliminate the peculiarities of the new type of
symmetry we have to deal with in the Quantum Groups scenario.

Nevertheless, some interesting properties of the 5D differential
calculus are revealed once we use it to perform a Noether analysis
of translation symmetries in $\kappa$-Minkowski. The fact that we
have five $d\hat{x}_A$ one-forms, and thus we expect five currents
in the analysis, could represent a challenge for the physical
interpretation, once we use these currents in order to look for
conserved charges. The results of Chapter \ref{ch:Analysis5D} show
that the 5D-calculus-based translation transformations can indeed be
implemented as symmetries of theories in $\kappa$-Minkowski. Our
analysis performed directly within the noncommutative theory also
allowed us to investigate explicitly the properties of the 5
``would-be currents", dissolving all the initial worries and
constructively leading us to current-conservation-like equations
written in terms of the operator ${\mathcal{\hat{D}}}_{0}$ which,
rather then $\hat{P}_0$, is a plausible candidate for the generator
of time translations.
 The real physical interpretation problem concerns
the possibility or not to properly call these charges the
\textit{energy-momentum charges}. In fact, even though the change of
basis for the 5D differential calculus introduced in section
\ref{par:DifferentBasis} led us to identify the new parameter
$\bar{d}\hat{x}_0$ as a time-translation parameter and the charge
$\bar{Q}_0$ as a plausible candidate of time-translation-symmetry
charge, several logical-consistency checks should be performed
before any definite claim. More on this point can be found in
\cite{k-Noether5D}, which also compares the analysis here reported
in Chapter \ref{ch:Analysis5D} with \cite{Friedel, Freidel2}.

A possible way out of this physical description ambiguity would be
provided by testing experimentally the different dispersion
relations (a meaningful physical property) that the different
momenta satisfy. In Section \ref{par:4DNoetherAnalysism=0} and
Chapter \ref{ch:DispersionRelation} we used the general formulas for
the translation-symmetry charges obtained with the 4D and 5D
differential calculi to derive their form for complex plane wave
fields and their dispersion relations. In both cases we saw a
$\lambda$ deformation of the special-relativistic form. However,
when we rescale the fields by a factor $A$ the $\lambda$-dependent
correction becomes less and less important as $A$ is increased, and
actually in the $A\rightarrow\infty$ limit the new effect disappears
and the dispersion relation regain its special-relativistic form.
Thereby, from a phenomenological perspective, the possibility to
discriminate between the different choices of generators basis and
differential calculi seems very unlikely, since for all practical
purposes (all realistically-large classical-field configurations)
the associated new effects are quantitatively irrelevant.

In this context the construction of a Quantum Field Theory in
$\kappa$-Minkowski might have an important role in clarifying the
status of energy-momentum observables and in dissolving the
ambiguity concerning the description of translations. It seems
plausible that, while classical fields are essentially unaffected by
the symmetry deformation, quantum particles in $\kappa$-Minkowski
spacetime be affected by a significant modification of the
dispersion relation. Perhaps the theory we considered does not have
enough structure to give proper physical significance to
energy-momentum, while a Quantum Field Theory analysis, following
the approach here advocated, might lead to further constraints to
the connotation of energy-momentum observables.

\appendix

\chapter{Bicrossproduct Hopf algebras} \label{par:AppendixBicrossproduct}

\noindent In order to provide the definition of ``Bicrossproduct"
Hopf algebra we need first to introduce the concepts of action and
coaction of an algebra. An algebra can act on other structures. A
left action of an algebra $H$ over an algebra $A$ is a linear map
$\alpha:H\otimes A\rightarrow A$ such that:

\begin{equation*}
\alpha((h\cdot g)\otimes a)=\alpha(h\otimes\alpha(g\otimes
a))~~~~h,g\in H, a\in A
\end{equation*}
\begin{equation} \label{eq:Action}
\alpha(h\otimes a)=\epsilon(h)a
\end{equation}

\noindent We can use the short notation $\alpha(h\otimes
a)=h\triangleright a$, so the (\ref{eq:Action}) can be written as

\begin{equation}
(hg)\triangleright(a)=h\triangleright(g\triangleright a)
\end{equation}
\begin{equation}
h\triangleright1=\epsilon(h)1.
\end{equation}

Usually in physical applications, the request of \textit{covariant
action} is made in order that the action of a Hopf algebra preserves
the structure of the object on which it acts. We say that an Hopf
algebra $H$ acts \textit{covariantly} (from the left) over an
algebra $A$ (or equivalently that $A$ is a left $H$-module algebra)
if $\forall h\in H$:

\begin{equation*}
h\triangleright(a\cdot b)=(h_{(1)}\triangleright
a)(h_{(2)}\triangleright b)~~~~a,b\in A.
\end{equation*}

The action of $H$ over a coalgebra $C$ ($C$ is a left $H$-module
coalgebra) states that:

\begin{equation*}
\Delta(h\triangleright c)=(h_{(1)}\triangleright
c_{(1)})\otimes(h_{(2)}\triangleright c_{(2)})=(\Delta
h)\triangleright \Delta c
\end{equation*}
\begin{equation}
\epsilon(h\triangleright c)=\epsilon(h)\epsilon(c),~~~~c\in C.
\end{equation}

In the same way it can be defined a right action of a Hopf algebra
$H$ on $A$ (algebra, coalgebra or Hopf algebra):

\begin{equation*}
\triangleleft: A\otimes H\rightarrow A
\end{equation*}

\noindent and a \textit{covariant} right-action should satisfy

\begin{equation*}
(a\cdot b)\triangleleft h=(a\triangleleft h_{(1)})(b\triangleleft
h_{(2)}),~~~~h\in H,~~a,b\in A
\end{equation*}
\noindent The duality relations connect the left action over an
algebra $A$ and the corresponding right dual action over the dual
algebra $A^*$ in the following way:

\begin{equation}\label{eq:Right-Left}
<a,h\triangleright b>=<a\triangleleft^*h,b>,~~~~b\in A, ~~a\in
A^*,~~h\in H.
\end{equation}

Let us make two examples of action, the adjoint action and the
canonical action, and show that the latter reduces to the physical
notion of translation in the case it is applied to the standard
generators of the Poincar\'{e}-translations acting on its dual space
(i.e. the commutative Minkowski space).

The left and right \textit{adjoint actions} of a Hopf algebra $H$ on
itself are linear maps $H\otimes H\rightarrow H$ such that

\begin{equation*}
a\triangleright^{ad}b=a_{(1)}bS(a_{(2)}),
\end{equation*}
\begin{equation}\label{eq:Adjoint}
b\triangleleft^{ad}a=S(a_{(1)})ba_{(2)},~~~~a,b\in H
\end{equation}

\noindent These actions are covariant.

The left and right \textit{canonical actions} of an algebra $A$ over
the dual coalgebra $C\equiv A^*$ are defined as:

\begin{equation*}
a\triangleright^{can}c=c_{(1)}<a,c_{(2)}>,
\end{equation*}
\begin{equation}\label{eq:Canonical}
c\triangleleft^{can}a=<c_{(1)},a>c_{(2)},~~~~a\in A,~~c\in C
\end{equation}

\noindent These actions are covariant as well.

The translation sector $T$ of the Poincar\'{e} algebra is a Lie
algebra (trivial Hopf algebra) generated by the operators $P_\mu$
that has the following Hopf Algebra structure:

\begin{equation}
[P_\mu,P_\nu]=0,~~~\Delta P_\mu=P_\mu\otimes1+1\otimes
P_\mu,~~~\epsilon(P_\mu)=0,~~~S(P_\mu)=0.
\end{equation}

\noindent the duality axioms (\ref{eq:axiom1}-\ref{eq:axiom5}) allow
us to reconstruct the Hopf-algebra structure of the Minkowski space
$M$ from the Hopf-algebra structure of its dual space $T$. In fact,
assuming that the duality relations between the generators $P_\mu$
of $T$ and the generators $x_\mu$ of its dual space $M=T^*$
be\footnote{We use a $(+,-,-,-)$ signature.}

\begin{equation}
<P_\mu,x_\nu>=-i\eta_{\mu\nu},
\end{equation}

\noindent one can easily find also $M$ has a Lie-algebra structure

\begin{equation}
[x_\mu,x_\nu]=0,~~~\Delta(x_\mu)=x_\mu\otimes1+1\otimes
x_\mu,~~~S(x_\mu)=-x.
\end{equation}
\noindent Using these relations, the canonical action of $P_\mu\in
T$ on $x_\mu\in T^*$ can be obtained:

\begin{equation}
P^\mu\triangleright^{can}x^\nu=x^\nu_{(1)}<P^\mu,x^\nu_{(2)}>=<P^\mu,x^\nu>=-i\eta_{\mu\nu}
\end{equation}

\noindent This is just the usual definition of the Poincar\'{e}
translations in the case of commutative Minkowski spacetime in which
the translation generators take the differential form
$P_\mu=-i\partial_\mu$ and its action over the coordinates is just
$P_\mu x_\nu=-i\eta_{\mu\nu}$.

The dual concept to the action of an algebra is the
\textit{coaction} of a coalgebra. the left coaction of a coalgebra
$C$ over an algebra $A$ is defined as a linear application
$\beta_L:A\rightarrow C\otimes A$. The map $\beta_L$ satisfies:

\begin{equation}
(id\otimes\Delta)\circ\beta_L=(\Delta\otimes id)\circ\beta_L
\end{equation}
\begin{equation}
(\epsilon\otimes id)\circ\beta_L=id
\end{equation}

\noindent The coaction gives a corepresentation of a coalgebra. A
\textit{covariant coaction} is required to respect the algebra
structure on which it (co)acts. Thus:

\begin{equation}
\beta_L(ab)=\beta_L(a)\beta_L(b),~~~\beta_L(1)=1\otimes 1,~~~a,b\in
A.
\end{equation}

\noindent We will adopt the following notation for the coaction:

\begin{equation}
\beta(a)=\sum_ia_i^{(\bar{1})}\otimes
a_i^{(\bar{2})}=a^{(\bar{1})}\otimes
a^{(\bar{2})}~~~a,a^{(\bar{2})}\in A,~~a^{(\bar{1})}\in C.
\end{equation}

The notions of action and coaction allow us to define a special
class of algebras that can be constructed by the composition of two
Hopf algebras. These algebras are called \textit{bicrossproduct
algebras} and take an important role in our study since the
$\kappa$-Poincar\'{e} algebra has been shown to be of this type
\cite{Majid2}. Consider two Hopf algebras, $A$ and $X$. Suppose that
we know a right action $\triangleleft$ of the algebra $X$ over the
algebra $A$ and a left coaction of $A$ on $X$:

\begin{equation}
\triangleleft:~~~A\otimes X\rightarrow A,
\end{equation}
\begin{equation}
\beta_L:~~~X\rightarrow A\otimes X.
\end{equation}

\noindent A bicrossproduct algebra (usually indicated with the
symbol $X\triangleright\triangleleft A$)is the tensor product
algebra $X\otimes A$ with the maps:

\begin{equation}
(x\otimes a)\cdot(y\otimes b)=xy_{(1)}\otimes(a\triangleleft
y_{(2)})b~~~~~~~~~~~~~~~~~(product)
\end{equation}
\begin{equation}
1_{X\triangleright\triangleleft
A}=1_X\otimes1_A~~~~~~~~~~~~~~~~~~~~~~~~~~~~~~~~~~~~(unity)
\end{equation}
\begin{equation}
\Delta(x\otimes a)=[x_{(1)}\otimes
x_{(2)}^{(\bar{1})}a_{(1)}]\otimes[x_{(2)}^{(\bar{2})}\otimes
a_{(2)}]~~~~~~~~(coproduct)
\end{equation}
\begin{equation}
\epsilon(x\otimes
a)=\epsilon(x)\epsilon(a)~~~~~~~~~~~~~~~~~~~~~~~~~~~~~~~~~~~~~~~~~(counit)
\end{equation}
\begin{equation}
S(x\otimes a)=(1_X\otimes
S(x^{(\bar{1})}a))\cdot(S(x^{(\bar{2})}\otimes
1_A)~~~~~~~~(antipode)
\end{equation}

\noindent where $a\in A,~~x,y\in X$.

One can consider as generators of the bicrossproduct algebra
$X\triangleright\triangleleft A$ the elements of the type
$A=1\otimes a$ and $X=x\otimes1$. In fact, following the definition
above, the single element $x\otimes a\in
X\triangleright\triangleleft A$ is given by the product $XA$:

\begin{equation}
XA=(x\otimes1)(1\otimes a)=x\otimes(1\triangleleft1)a=x\otimes a,
\end{equation}

\noindent while the other product $AX$ is:

\begin{equation}
AX=(1\otimes a)(x\otimes1)=x_{(1)}\otimes(a\triangleleft x_{(2)}).
\end{equation}

\noindent Thus, the bicrossproduct algebra
$X\triangleright\triangleleft A$ can be viewed as the enveloping
algebra generated by $X$ and $A$, modulo the commutation relations:

\begin{equation}\label{eq:BicrossCommutator}
[X,A]=x\otimes a-x_{(1)}\otimes(a\triangleleft x_{(2)}).
\end{equation}

The bicrossproduct $\kappa$-Poincar\'{e} algebra
$U(so(1,3))\triangleright\triangleleft T$ is constructed in this
way, choosing $X=U(so(1,3))$, the Lie algebra of Lorentz rotations,
and $A=T$, the algebra of the Poincar\'{e} translations with a
deformed coalgebra sector ($T$ is a non-trivial Hopf algebra). In
particular, in the Majid-Ruegg construction \cite{Majid2}, the
(deformed) coalgebra of $T$ is chosen such that $T$ is a dual space
to $\kappa$-Minkowski, i.e. to the Lie algebra generated by the
elements $\hat{x}_\mu$ which satisfy the commutation relations

\begin{equation}
[\hat{x}_j,\hat{x}_0]=i\lambda\hat{x}_j,~[\hat{x}_j,\hat{x}_k]=0.
\end{equation}

Quantum groups of bicrossproduct type were first proposed by S.
Majid \cite{Majid2} as key ingredient for the unification of Gravity
and Quantum Mechanics. This point of view is alternative to the idea
of quantization of a theory as the result of a process applied to
the underlying classical space (as for example in the case of
deformation quantization, that is based on the classical notion of
Poisson brackets or the case of String Theory description of quantum
Gravity where one quantizes strings moving in a classical
spacetime). Models should instead be built guided by the intrinsic
Noncommutative Geometry at the level of noncommutative algebras.
Only at the end one can consider classical geometry (with Poisson
brackets) as classical limits and not as a starting point, like in
\textit{deformation quantization}. In a quantum world, in fact,
phase-space and probably spacetime should be ``fuzzy" and only
approximately described by classical geometry.

The idea at the basis of the introduction of Quantum Groups in a
Quantum Gravity approach is a principle of \textit{self-duality},
that is peculiar in the Hopf algebras and that should allow to put
\textit{quantum mechanics} and \textit{gravity} on equal (but
mutually dual) footing.

In order to explain this concept we make the example of the
classical phase-space $(q_j, p_j)\in R^{2n}, j=1,...,n$. In this
case we can consider the group $G$ of elements $W_k=e^{ik_jq_j}$
labeled by the parameter $k_j\in R^n$. This group has an abelian
composition law $W_{k_1}W_{k_2}=W_{k_1+k_2}$. The algebra of the
functions of positions is given by the enveloping algebra $U(g)$
(where $g$ is the Lie algebra of $G$). The algebra of the momenta
can be viewed as the algebra $C[G]$ dual to $U(g)$, and the
generators $p_j$ of this algebra can be introduced via the
relations:

\begin{equation}
p_j(W_k)\equiv <p_j,W_k>=k_j
\end{equation}

\noindent that follow from the duality relation
$<p_j,q_l>=-i\delta_{jl}$. Thus:

\begin{equation}
p_j(W_{k_1}W_{k_2})=p_j(W_{k_1+k_2})=(k_1+k_2)_j.
\end{equation}

\noindent On the other hand (by duality),

\begin{equation*}
p_j(W_{k_1}W_{k_2})=<p_j,W_{k_1}W_{k_2}>=<\Delta(p_j),W_{k_1}\otimes
W_{k_2}>=
\end{equation*}
\begin{equation}
=\Delta(p_j)(W_{k_1}\otimes W_{k_2}).
\end{equation}

\noindent Thus the coproduct turns out to be:
$\Delta(p_j)=p_j\otimes1+1\otimes p_j$. From this example one can
see the relation between the coproduct and the composition law in
the momentum sector. The commutativity of the positions is connected
with the cocommutivity of coproduct in the space of momenta (then
the momentum space is flat, \textit{it has an abelian composition
law}). If the space of positions is noncommutative, the group law of
$G$ will be in general non-abelian $W_{k_1}W_{k_2}\neq W_{k_1+k_2}$
and the coproduct of momenta will be noncocommutative. Thus a
noncommutative position space corresponds to a curved momentum
space. The existence of \textit{self-duality} however states also
the opposite: a curved space in positions corresponds to a
noncommutative momentum space.

In these terms, $\kappa$-Poincar\'{e} can have two different
physical interpretations: one as quantum group symmetry, the other
as quantized phase-space. In fact we have

\begin{equation}\label{eq:duality}
P_k=U(so(1,3))\triangleright\triangleleft
T=U(so(1,3))\triangleright\triangleleft \mathbb{C}(P),
\end{equation}

\noindent where on one side we have $T$ as the deformed enveloping
algebra of the Poincar\'{e} momenta sector. (\ref{eq:duality})
underlines that the same construction can be described in terms of
$\mathbb{C}(P)$, the algebra of functions on the ``classical" but
curved momenta space. Thus, our new phase-space is constructed by
$\kappa$-Minkowski noncommutative spacetime and a curved momenta
space. The two sectors are connected by a particular Fourier
transform that we introduced in section (\ref{par:Weylmaps})

The search for a quantum algebra of observable which is a Hopf
algebra translates in a search of a simple model in which quantum
and gravitational effects are unified and in which they are dual to
each other.

Let us analyze more deeply the bicrossproduct structure of the
Majid-Ruegg basis and let us show that it acts \textit{covariantly}
on $\kappa$-Minkowski. In the construction of a bicrossproduct
algebra there are no prescriptions on the action of the elements of
the bicrossproduct algebra itself. The choice of this action can be
dictated by the generalization of the classical action of the
standard Poincar\'{e} generators. In the case of the Poincar\'{e}
algebra the action of the Lorentz rotations over the translation
generators is represented by the commutators, and this action
coincides with the adjoint action. Taking into account that the
Poincar\'{e} algebra is generated by elements $h$ that satisfy
$\Delta(h)=h\otimes1+1\otimes h$, $S(h)=-h$, one easily finds

\begin{equation}
p^j\triangleleft^{Ad} h=S(h_{(1)})p^jh_{(2)}=[h,p^j].
\end{equation}

\noindent This suggest to consider the adjoint action as a good
generalization of the action of the Lorentz rotations and boosts
over the translation generators, also in the deformed case. It is
surprising that in the case of $\kappa$-Poincar\'{e} bicrossproduct
basis the adjoint action is still given by the commutators. Taking
into account the definition (\ref{eq:Adjoint}) and the
$\kappa$-Poincar\'{e} structures in the bicrossproduct basis
(\ref{eq:MajidCoproducts})-(\ref{eq:antipodes}), one finds:

\begin{equation}\label{eq:Naction}
P^\mu\triangleleft^{Ad} N^j=S(N^j_{(1)})P^\mu N^j_{(2)}=[P^\mu,N^j]
\end{equation}
\begin{equation}\label{eq:Maction}
P^\mu\triangleleft^{Ad} M^j=S(M^j_{(1)})P^\mu M^j_{(2)}=[P^\mu,M^j].
\end{equation}

\noindent Assuming that $M_j,N_j$ act on $\mathcal{T}$ via the
adjoint action, we can determine their action on $\kappa$-Minkowski
generators through the duality structure of bicrossproduct Hopf
algebras.

When we previously defined the canonical action, we have seen that
the action of the ordinary translations on the commutative spacetime
coordinates is described by this type of action. So, it appears
natural to consider the canonical action as the generalization of
the action of the translation generators $P_\mu$ on
$\kappa$-Minkowski as well.

Under these assumptions we can show that $\kappa$-Minkowski
transforms covariantly under the action of the $\kappa$-Poincar\'{e}
generators.

$P_\mu$ acts on its dual space $\mathcal{T}^*$ (i.e.
$\kappa$-Minkowski) by canonical action

\begin{equation*}
t\triangleright
x=<x_{(1)},t>x_{(2)},~~~~t\in\mathcal{T},x\in\mathcal{T}^*,
\end{equation*}

\noindent from which it follows that $P_\mu$ acts as a derivation on
the $\kappa$-Minkowski generators:

\begin{equation*}
P_\mu\triangleright x_\nu=-i\eta_{\mu\nu}.
\end{equation*}

\noindent Their extension to products of the spacetime coordinates
is via the covariance condition $t\triangleright
xy=(t_{(1)}\triangleright x)(t_{(2)}\triangleright y)$, in
particular,

\begin{equation}
P_0\triangleright x_0x_k=-ix_k~~~~P_0\triangleright x_kx_0=-ix_k
\end{equation}
\begin{equation}
P_j\triangleright x_0x_k=(1\triangleright x_0)(P_j\triangleright
x_k)=i\delta_{jk}x_0
\end{equation}
\begin{equation}
P_j\triangleright x_kx_0=(P_j\triangleright x_k)(e^{-\lambda
P_0}\triangleright x_0)=i(x_0+i\lambda)\delta_{jk}\,;
\end{equation}

\noindent therefore, the $\kappa$-Minkowski commutation relations
are invariant under the $\kappa$-Poincar\'{e} translations.

To derive also the action of $(M_j,N_j)$ on $\mathcal{T}^*$ one can
use the fact that their right action on $\mathcal{T}$ dualizes to an
action on the left on $\mathcal{T}^*$ (\ref{eq:Right-Left}):

\begin{equation}
<P_\mu\triangleleft^{ad}M_k,x>=-<P_\mu,M_k\triangleright^{ad}x>
\end{equation}
\begin{equation}
<P_\mu\triangleleft^{ad}N_k,x>=-<P_\mu,N_k\triangleright^{ad}x>,~~~~x\in\mathcal{T}^*
\end{equation}

\noindent so, in our case, using (\ref{eq:Maction},
\ref{eq:Naction}) and the commutators (\ref{eq:MajidAlgebra}), one
finds:

\begin{equation}
M_j\triangleright
\hat{x}_k=i\epsilon_{jkl}\hat{x}_l,~~~~M_j\triangleright
\hat{x}_0=0,~~~~N_j\triangleright
\hat{x}_k=i\delta_{jk}\hat{x}_0,~~~~N_j\triangleright
\hat{x}_0=i\hat{x}_i
\end{equation}

\noindent and, extending these actions via the covariance property
of the adjoint action $h\triangleright xy=(h_{(1)}\triangleright
x)(h_{(2)}\triangleright y)$, $h=M_j,N_j$, $x,y\in\mathcal{T}^*$,

\begin{equation*}
M_j\triangleright
\hat{x}_0\hat{x}_k=i\epsilon_{jkl}\hat{x}_0\hat{x}_l~~~~M_j\triangleright
\hat{x}_k\hat{x}_0=i\epsilon_{jkl}\hat{x}_l\hat{x}_0
\end{equation*}\\
\begin{equation*}
N_j\triangleright \hat{x}_0\hat{x}_k=(N_j\triangleright
\hat{x}_0)(e^{-\lambda P_0}\triangleright\hat{x}_k)+(1\triangleright
\hat{x}_0)(N_j\triangleright \hat{x}_k)-\lambda
\epsilon_{jrl}(M_r\triangleright\hat{x}_0)(P_l\triangleright\hat{x}_k)=
\end{equation*}
\begin{equation*}
=i\hat{x}_j\hat{x}_k+i\delta_{jk}\hat{x}^2_0
\end{equation*}\\
\begin{equation*}
N_j\triangleright \hat{x}_k\hat{x}_0=(N_j\triangleright
\hat{x}_k)(e^{-\lambda P_0}\triangleright\hat{x}_0)+(1\triangleright
\hat{x}_k)(N_j\triangleright \hat{x}_0)-\lambda
\epsilon_{jrl}(M_r\triangleright\hat{x}_k)(P_l\triangleright\hat{x}_0)=
\end{equation*}
\begin{equation*}
=i\hat{x}_k\hat{x}_j+i\delta_{jk}\hat{x}^2_0-\lambda\delta_{jk}\hat{x}_0
\end{equation*}\\
\begin{equation*}
N_j\triangleright \hat{x}_k\hat{x}_l=(N_j\triangleright
\hat{x}_k)(e^{-\lambda P_0}\triangleright\hat{x}_l)+(1\triangleright
\hat{x}_k)(N_j\triangleright \hat{x}_l)-\lambda
\epsilon_{jrs}(M_r\triangleright\hat{x}_k)(P_s\triangleright\hat{x}_l)=
\end{equation*}
\begin{equation*}
=i\delta_{jk}\hat{x}_0\hat{x}_l+
i\delta_{jl}\hat{x}_k\hat{x}_0+\lambda(\delta_{lk}\hat{x}_j-\delta_{jk}\hat{x}_l)
\end{equation*}\\
\begin{equation*}
N_j\triangleright \hat{x}_l\hat{x}_k=(N_j\triangleright
\hat{x}_l)(e^{-\lambda P_0}\triangleright\hat{x}_k)+(1\triangleright
\hat{x}_l)(N_j\triangleright \hat{x}_k)-\lambda
\epsilon_{jrs}(M_r\triangleright\hat{x}_l)(P_s\triangleright\hat{x}_k)=
\end{equation*}
\begin{equation*}
=i\delta_{jl}\hat{x}_0\hat{x}_k+
i\delta_{jk}\hat{x}_l\hat{x}_0+\lambda(\delta_{kl}\hat{x}_j-\delta_{jl}\hat{x}_k)
\end{equation*}\\

\begin{equation}
\Rightarrow M_j\triangleright
[\hat{x}_k,\hat{x}_0]=i\epsilon_{jkl}[\hat{x}_l,\hat{x}_0]=-\epsilon_{jkl}\lambda\hat{x}_l=M_j\triangleright
(i\lambda\hat{x}_k)
\end{equation}\\
\begin{equation}
\Rightarrow N_j\triangleright
[\hat{x}_k,\hat{x}_0]=-\lambda\delta_{jk}\hat{x}_0=N_j\triangleright(i\lambda\hat{x}_k)
\end{equation}\\
\begin{equation}
\Rightarrow N_j\triangleright
[\hat{x}_k,\hat{x}_l]=i\delta_{jk}[\hat{x}_0,\hat{x}_l]+
i\delta_{jl}[\hat{x}_k,\hat{x}_0]+\lambda\delta_{jl}\hat{x}_k-\delta_{jk}\hat{x}_l=0=N_j\triangleright0\,.
\end{equation}\\

\noindent Thereby the $\kappa$-Minkowski commutation relations
remain unmodified also under the action of the boost-rotation
generators of $\kappa$-Poincar\'{e} Hopf algebra.

\clearpage

\chapter{Five-dimensional differential calculus}\label{par:AppendixCalculus}

 \noindent In a noncommutative
spacetime it is a highly non-trivial exercise to establish the
differential calculus. On $\kappa$-Minkowski spacetime one can
construct two distinct differential calculi. In chapter
\ref{ch:Analysis4D} we presented the four-dimensional translational
invariant calculus proposed by Majid and Oeckl \cite{Majid3}, which
is however not covariant under the action of the full
$\kappa$-Poincar\'{e} algebra. In this appendix we want to show how
the calculus (\ref{eq:5Dcommutators}) bicovariant under the action
of the full $\kappa$-Poincar\'{e} algebra is obtained and why it is
necessarily five dimensional.

Let us recall that the $\kappa$-Poincar\'{e} algebra in the
Majid-Ruegg bicrossproduct basis is defined by the commutation
relation between the Lorentz and translational sectors:

\begin{equation*}
[P_\mu, P_\nu]=0
\end{equation*}
\begin{equation*}
[M_j, P_0]=0
\end{equation*}
\begin{equation*}
[M_j, P_k]=i\epsilon_{jkl}P_l
\end{equation*}
\begin{equation*}
[N_j, P_0]=iP_j
\end{equation*}
\begin{equation}\label{eq:MajidAlgebra2}
[N_j,P_k]=i\delta_{jk}\left(\frac{1}{2\lambda}(1-e^{-2\lambda
P_0})+\frac{\lambda}{2}P^2\right)-i\lambda P_jP_k
\end{equation}

\noindent and the following deformed coproducts:

\begin{equation*}
\Delta (P_0) = P_0 \otimes 1 + 1 \otimes P_0
\end{equation*}
\begin{equation*}
\Delta (P_j)= P_j \otimes e^{-\lambda P_0} +1 \otimes P_j
\end{equation*}
\begin{equation*}
\Delta (M_j) = M_j\otimes 1 + 1 \otimes M_j
\end{equation*}
\begin{equation}
\Delta (N_j)= N_j \otimes e^{-\lambda P_0} +1 \otimes N_j
-\lambda\epsilon_{jkl}M_k\otimes P_l\,.
\end{equation}

We have seen in the first chapter that as the $\kappa$ deformation
of Minkowski space we take the dual Hopf algebra of the translation
algebra $\mathcal{T}$ and we denote its generators by $\hat{x}_\mu$.
Thus $\kappa$-Minkowski is defined by:

\begin{equation} \label{eq:Minkowski}
[\hat{x}_i,\hat{x}_j]=0~~~~[\hat{x}_i,\hat{x}_0]=i\lambda\hat{x}_i,
\end{equation}
\begin{equation}
\Delta\hat{x}_\mu=\hat{x}_\mu\otimes1+1\otimes\hat{x}_\mu.
\end{equation}

\noindent We have shown in Appendix A that the canonical action of
translations on $\kappa$-Minkowski spacetime is:

\begin{equation*}
t\triangleright x=<x_{(1)},t>x_{(2)},~~~~\forall x\in
\mathcal{T^*},\forall t\in \mathcal{T},
\end{equation*}

\noindent with the shorthand notation $\Delta x=\sum x_{(1)}\otimes
x_{(2)}$.

From the bicrossproduct structure of $\kappa$-Poincar\'{e} we have
the action of $U(so(1,3))$ on translations $\mathcal{T}$, which, by
duality, can be translated into action on the generators of
$\kappa$-Minkowski:

\begin{equation}\label{eq:LorentzAction}
M_i\triangleright\hat{x}_j=i\epsilon_{ijk}\hat{x}_k,~~~~
M_i\triangleright\hat{x}_0=0,~~~~
N_i\triangleright\hat{x}_j=i\delta_{ij}\hat{x}_0,~~~~
N_i\triangleright\hat{x}_0=i\hat{x}_i,
\end{equation}

\noindent which generalizes to the whole algebra by the covariance
condition:

\begin{equation}\label{eq:Covariance}
h\triangleright xy=(h_{(1)}\triangleright x)(h_{(2)}\triangleright
y),~~~~\forall h\in U(so(1,3)), x,y\in \mathcal{T^*}.
\end{equation}

The problem we are to solve here is the following. In commutative
spacetime positions commute with differentials (one forms). However
here we are working with noncommutative spacetime, and thus we
cannot assume a priori that positions commute with one forms.
Instead, let us take a basis of one forms, which should include
differentials $d\hat{x}_\mu$, and denote the elements of this basis
by $\chi_a, a=0,...,N, N\geq 4$. We assume that the commutator
$[\hat{x}_\mu,\chi_a]$ must have the following expansion:

\begin{equation} \label{eq:DiffCalc}
[\hat{x}_\mu,\chi_a]=\sum_{\mu,a,b}A^b_{\mu a}\chi_b.
\end{equation}

\noindent There are, of course, some consistency conditions for the
above relations, which come from the mixed Jacoby identity:

\begin{equation}\label{eq:Jacoby}
[[\hat{x}_\mu,\hat{x}_\nu],\chi_a]+[[\hat{x}_\nu,\chi_a],\hat{x}_\mu]+
[[\chi_a,\hat{x}_\mu],\hat{x}_\nu]=0.
\end{equation}

\noindent If we rewrite, for simplicity of notation, the commutation
relations (\ref{eq:Minkowski}) in a more general form:

\begin{equation}\label{eq:FormalKMinkowski}
[\hat{x}_\mu,\hat{x}_\nu]=B^\rho_{\mu\nu}\hat{x}_\rho,
\end{equation}

\noindent the relation (\ref{eq:Jacoby}) takes the form:

\begin{equation}\label{eq:Jacoby2}
A^a_{\nu c}A^c_{\mu b}-A^a_{\mu c}A^c_{\nu
b}=B^\rho_{\mu\nu}A^a_{\rho b}.
\end{equation}

\noindent Next, expressing $d\hat{x}_\mu$ as a linear combination of
$\chi_a$:

\begin{equation}
d\hat{x}_\mu=D_\mu^a\chi_a,
\end{equation}

\noindent if we apply the exterior derivative to both sides of
(\ref{eq:FormalKMinkowski}) and we impose the Leibnitz rule, we
obtain another restriction:

\begin{equation}\label{eq:Leibnitz2}
D^b_\nu A^a_{\mu b}-D^b_\mu A^a_{\nu b}=B^\rho_{\mu\nu}D^a_\rho.
\end{equation}

\noindent Both relations (\ref{eq:Jacoby2}) and (\ref{eq:Leibnitz2})
are necessary consistency conditions to determine a bicovariant
differential calculus on $\kappa$-Minkowski spacetime.

We need now to append these conditions with the covariance
requirement, i.e. the condition that both sides of
(\ref{eq:DiffCalc}) transform in the same way under the actions of
rotations and boosts. We shall postulate that the action of the
Lorentz algebra (\ref{eq:LorentzAction})-(\ref{eq:Covariance})
extends to the differential algebra in a natural covariant way,
i.e.:

\begin{equation}
h\triangleright(ydx)=(h_{(1)}\triangleright
y)(d(h_{(2)}\triangleright
x)),~~~~h\triangleright(dxy)=(d(h_{(1)}\triangleright x))
(h_{(2)}\triangleright y).
\end{equation}

\noindent From the above definition and the action
(\ref{eq:LorentzAction}) we obtain for the left side of
(\ref{eq:DiffCalc}) the following identities:

\begin{equation}\label{eq:Identity1}
N_k\triangleright[\hat{x}_i,d\hat{x}_j]=i\delta_{ki}[\hat{x}_0,d\hat{x}_j]+i\delta_{kj}[\hat{x}_i,d\hat{x}_0]+
\lambda(\delta_{kj}d\hat{x}_i-\delta_{ij}d\hat{x}_k),
\end{equation}\\
\begin{equation}\label{eq:Identity2}
N_k\triangleright[\hat{x}_0,d\hat{x}_i]=i[\hat{x}_k,d\hat{x}_i]+i\delta_{ki}[\hat{x}_0,d\hat{x}_0]+\lambda\delta_{ki}d\hat{x}_0,
\end{equation}\\
\begin{equation}\label{eq:Identity3}
N_k\triangleright[\hat{x}_i,d\hat{x}_0]=i[\hat{x}_i,d\hat{x}_k]+i\delta_{ki}[\hat{x}_0,d\hat{x}_0],
\end{equation}\\
\begin{equation}\label{eq:Identity4}
N_k\triangleright[\hat{x}_0,d\hat{x}_0]=i[\hat{x}_k,d\hat{x}_0]+i[\hat{x}_0,d\hat{x}_k]+\lambda
d\hat{x}_k,
\end{equation}\\
\begin{equation}\label{eq:Identity5}
M_k\triangleright[\hat{x}_i,d\hat{x}_j]=i\epsilon_{kis}[\hat{x}_s,d\hat{x}_j]+i\epsilon_{kjs}[\hat{x}_i,d\hat{x}_s],
\end{equation}\\
\begin{equation}\label{eq:Identity6}
M_k\triangleright[\hat{x}_0,d\hat{x}_i]=i\epsilon_{kis}[\hat{x}_0,d\hat{x}_s],
\end{equation}\\
\begin{equation}\label{eq:Identity7}
M_k\triangleright[\hat{x}_i,d\hat{x}_0]=i\epsilon_{kis}[\hat{x}_s,d\hat{x}_0],
\end{equation}\\
\begin{equation}\label{eq:Identity8}
M_k\triangleright[\hat{x}_0,d\hat{x}_0]=0;
\end{equation}\\

\noindent we will demonstrate just the first of the identities
above, the others follows analogously:

\begin{equation*}
N_k\triangleright[\hat{x}_i,d\hat{x}_j]=N_k\triangleright(\hat{x}_id\hat{x}_j-d\hat{x}_j\hat{x}_i)=
\end{equation*}
\begin{equation*}
=(N_k\triangleright\hat{x}_i)d(e^{-\lambda
P_0}\triangleright\hat{x}_j)+1\triangleright\hat{x}_i(d(N_k\triangleright\hat{x}_j))-\lambda\epsilon_{klm}M_l\triangleright\hat{x}_i(d(P_m\triangleright\hat{x}_j))+
\end{equation*}
\begin{equation*}
- d(N_k\triangleright\hat{x}_j)e^{-\lambda
P_0}\triangleright\hat{x}_i-d(1\triangleright\hat{x}_j)(N_k\triangleright\hat{x}_i)+d(\lambda\epsilon_{klm}M_l\triangleright\hat{x}_j)P_m\triangleright\hat{x}_i=
\end{equation*}
\begin{equation*}
=i\delta_{ki}\hat{x}_0d\hat{x}_j+i\delta_{kj}\hat{x}_id\hat{x}_0-\lambda\epsilon_{kli}\epsilon_{ljr}d\hat{x}_r
-i\delta_{ki}d\hat{x}_0\hat{x}_i-i\delta_{ki}d\hat{x}_j\hat{x}_0=
\end{equation*}
\begin{equation*}
=i\delta_{ki}[\hat{x}_0,d\hat{x}_j]+i\delta_{kj}[\hat{x}_i,d\hat{x}_0]+
\lambda(\delta_{kj}\delta_{ir}-\delta_{ij}\delta_{kr})d\hat{x}_r=
\end{equation*}
\begin{equation*}
=i\delta_{ki}[\hat{x}_0,d\hat{x}_j]+i\delta_{kj}[\hat{x}_i,d\hat{x}_0]+
\lambda(\delta_{kj}d\hat{x}_i-\delta_{ij}d\hat{x}_k).
\end{equation*}
\\
Transforming also the right side of (\ref{eq:DiffCalc}) under the
action of rotations and boosts (with $\chi_\mu=d\hat{x}_\mu$) and
imposing the equivalence with the identities
(\ref{eq:Identity1})-(\ref{eq:Identity8}) we obtain a system of
linear equations for the coefficients $A^\rho_{\mu\nu}$ which we can
solve, reminding that the consistency conditions (\ref{eq:Jacoby2})
and (\ref{eq:Leibnitz2}) must be satisfied.

Now, if we consider only 4D bicovariant calculi, it appears that the
solution is unique and gives us the following relations:

\begin{equation}
[\hat{x}_i,d\hat{x}_j]=i\delta_{ij}\lambda
d\hat{x}_0~~~~[\hat{x}_i,d\hat{x}_0]=i\lambda d\hat{x}_i
\end{equation}
\begin{equation}
[\hat{x}_0,d\hat{x}_j]=0 ~~~~[\hat{x}_0,d\hat{x}_0]=0,
\end{equation}

\noindent which, however, do not define a differential calculus as
they fail to obey the condition (\ref{eq:Jacoby2}). Therefore we
conclude that there not exist a four-dimensional bicovariant
differential calculus on $\kappa$-Minkowski spacetime.

Thus we see that the basis of one-forms of the bicovariant
differential calculus is indeed $\chi_a=(d\hat{x}_\mu,d\hat{x}_4)$.
Since $d\hat{x}_4$ does not carry the spacetime index, it must be
invariant under the action of the Lorentz generators

\begin{equation}
N_i\triangleright d\hat{x}_4=0~~~~M_i\triangleright d\hat{x}_4=0.
\end{equation}

\noindent Now, solving the system of linear equations
(\ref{eq:Identity1})-(\ref{eq:Identity8}) and imposing the
consistency conditions (\ref{eq:Jacoby2}) and (\ref{eq:Leibnitz2}),
one finds that the commutation relations between the coordinates
$\hat{x}_\mu$ and all the generating one-forms
$d\hat{x}_\mu,d\hat{x}_4$ are the following:

\begin{equation}\label{eq:Calculus1}
[\hat{x}_0, d\hat{x}_4] = i \lambda d\hat{x}_0 ~~~ [\hat{x}_0,
d\hat{x}_0] = i \lambda d\hat{x}_4 ~~~ [\hat{x}_0, d\hat{x}_i] = 0
\end{equation}

\begin{equation}\label{eq:Calculus2}
[\hat{x}_i, d\hat{x}_4] = [\hat{x}_i, d\hat{x}_0] = -i \lambda
d\hat{x}_i ~~~ [\hat{x}_i, d\hat{x}_j] = i \lambda \delta_{ij}
(d\hat{x}_4 - d\hat{x}_0).
\end{equation}\\

\noindent These relations define a five-dimensional bicovariant
differential calculus on $\kappa$-Minkowski spacetime.


\addcontentsline{toc}{chapter}{Bibliography}

\begin{thebibliography}{99}

\bibitem{Stachel} J. Stachel, \textit{Early History of Quantum
Gravity}, in ``Black Holes, Gravitational Radiation and the
Universe'', ed. by B.R. Iyer, B. Bhawal eds. (Kluwer Academic
Publisher, Netherlands, 1999).

\bibitem{Witten} M.B. Green, J.H. Schwarz and E. Witten, \textit{Superstring
theory}, Cambridge University Press, Cambridge, 1987.

\bibitem{Polchinski} J. Polchinski, \textit{Superstring Theory and
Beyond}, Cambridge University Press, Cambridge, 1998.

\bibitem{Rovelli} C. Rovelli, \textit{Loop Quantum Gravity}, Living Rev. Rel. 1, 1 (1998).

\bibitem{Ashtekar} A. Ashtekar, \textit{Quantum Geometry and Gravity: Recent
Advances}, gr-qc/0112038.

\bibitem{Smolin} L. Smolin, \textit{How far are we from the quantum theory of
gravity?}, hep-th/0303185.

\bibitem{Thiemann} T. Thiemann, \textit{Lectures on Loop Quantum
Gravity}, Lect. Notes Phys. 631, 41 (2003).

\bibitem{mead} C.A.~Mead,
\textit{Possible connection between gravitation and fundamental
length}, Phys.~Rev.~135 (1964) B849.

\bibitem{padma} T.~Padmanabhan,
\textit{Limitations On The Operational Definition Of Space-Time
Events And Quantum Gravity}, Class.~Quant.~Grav.~4 (1987)  L107.

\bibitem{ng1994} Y.J.~Ng and H.~Van Dam, \textit{Limit to space-time
measurement}, Mod.~Phys.~Lett.~A9 (1994) 335.

\bibitem{gacmpla} G.~Amelino-Camelia,
\textit{General Relativity and Quantum Cosmology},
Mod.~Phys.~Lett.~A9 (1994) 3415.

\bibitem{garay} See, {\it e.g.}, L.J.~Garay,
 \textit{Quantum gravity and minimum length},
Int.~J.~Mod.~Phys.~A10 (1995) 145, and references therein.

 \bibitem{Amelino1} G. Amelino-Camelia, \textit{Relativity in space-times
 with short-distances structure governed by an
 observer-independent (Planckin) length scale}, Int. J. Mod. Phys.
 D11 (2002) 35; G. Amelino-Camelia,
 \textit{Testable secnario for relativity with minimum-length}, Phys. Lett.
 B510 (2001) 255; G. Amelino-Camelia,
 \textit{Doubly Special Relativity}, Nature 418, 34 (2002).

 \bibitem{Majid1} S. Majid, \textit{Hopf algebras for physics at the
 Planck  scale}, Class. Quantum Grav. 5 (1988) 1587.

\bibitem{Majid2} S. Majid and H. Ruegg, \textit{Bicrossproduct structure
of Kappa Poincar\'{e} group and noncommutative geometry}, Phys.
Lett. B 334 (1994) 348.

\bibitem{Amelino2} G. Amelino-Camelia, J. Lukierski and A.
Nowicki, \textit{Kappa-deformed covariant phase space and
quantum-gravity uncertainty relations}, Phys. Atom. Nucl. 61 (1998)
1811. \textit{Distance measurement and Kappa-deformed propagation of
light and heavy probes}, Int. J. Mod. Phys. A 14 (1999) 4575.

\bibitem{Amelino3} A. Agostini, G. Amelino-Camelia, M. Arzano, A.
MArciano and R. Tacchi, \textit{Generalizing the Noether theorem for
Hopf-algebra spacetime symmetries}, Mod.~Phys.~Lett.~A22 (2007)
1779.

\bibitem{Lukierski} J.~Lukierski, H.~Ruegg and W.J.~Zakrzewski
\textit{\it Classical and quantum-mechanics of free
$\kappa$-relativistic systems}, Ann.~Phys.~243 (1995) 90.

\bibitem{Kowalski} J. Kowalski-Glikman and S. Nowak,
\textit{Non-commutative space-time of doubly special relativity
theories}, Int. J. Mod. Phys. D 12 (2003) 299.

\bibitem{Amelino4} A. Agostini, G. Amelino-Camelia and F.
D'andrea, \textit{Hopf algebra description of
noncommutative-spacetime symmetries}, Int.~J.~Mod.~Phys.~A19 (2004)
5187.

\bibitem{Majid3} R. Oeckl, \textit{Classification of differential calculi
on $U_q(b_+)$, classical limit and duality}, J. Math. Phys. 40
(1999) 3588-3604; S. Majid, \textit{Foundation of Quantum Group
Theory}, Cambridge University Press, 1995.

\bibitem{Sitarz} A. Sitarz, \textit{Noncommutative Differential Calculus
on the Kappa-Minkowski Space}, Phys. Lett. B349 (1995) 42-48.

\bibitem{Weyl} H. Weyl, \textit{The theory of groups and Quantum
Mechanics}, Dover, 1931.

\bibitem{Amelino5} G. Amelino-Camelia, J.R. Ellis, N.E.
Mavromatos, D.V. Nanopoulus and S. Sarkar, \textit{Potential
Sensitivity of Gamma-Ray Burster Observations to Wave Dispersion in
Vacuo}, Nature 393, 763 (1998).

\bibitem{Amelino6} G. Amelino-Camelia and T. Piran, \textit{Planck-scale
deformation of Lorentz symmetry as a solution to the UHECR and the
TeV-gamma paradoxes}, Phys. Rev. D 64, 036005 (2001).

\bibitem{Kowalski2} J. Kowalski-Glikman, \textit{Testing dispersion
relations of quantum kappa-Poincare algebra on cosmological
background}, Phys. Lett. B499 (2001) 1.

\bibitem{Friedel} L. Friedel, J. Kowalski-Glikman and S. Nowak,
\textit{From noncommutative $\kappa$-Minkowski to Minkowski
space-time}, Phys. Lett. B648 (2007) 70-75.

\bibitem{Freidel2} L. Freidel, J. Kowalski-Glikman and S. Nowak,
\textit{Field theory on $\kappa$-Minkowski space revisited: Noether
charges and breaking of Lorentz symmetry}, arXiv:hep-th/0706.3658v1.

\bibitem{Snyder} H.S. Snyder, \textit{Quantized Space-Time}, Phys. Rev. 71
(1947) 38.

\bibitem{Connes} A. Connes, \textit{Noncommutative Geometry}, Academic
Press, (1994); G. Landi, \textit{An introduction to noncommutative
spaces and their geometry}, Springer (1998).

\bibitem{Woronowicz} S.L. Woronowicz, \textit{Differential Calculus On
Compact Matrix Pseudogroups (Quantum Groups)}, Commun. Math. Phys.
122, 125 (1989).

\bibitem{Kowalski3} J. Kowalski-Glikman, \textit{Obsrver-independent quanta
of mass and length}, Phys. Lett. A286 (2001) 391-394.

\bibitem{Majid4} S. Majid, \textit{Non-commutative-geometric Groups by
Bicrossproduct Construction}, (PhD thesis, Harvard mathematical
physics, 1988).

\bibitem{Celeghini} E. Celeghini, R. Giacchetti, E. Sorace and M.
Tarlini, \textit{Three dimensional Quantum Groups from Contraction
of $SU(2)_Q$}, J. Math. Phys. 31 (1990)2548.

\bibitem{Lukierski2} J. Lukierski, A. Nowicki and H. Ruegg, \textit{New
quantum Poincar\'{e} algebra and $\kappa$ deformed field theory},
Phys. Lett. B293 (1992) 344; J. Lukierski, A. Nowicki, H. Ruegg and
V.N. Tolstoi, \textit{Q-deformation of Poincar\'{e} algebra}, Phys.
Lett. B264 (1991) 331.

\bibitem{Amelino7} G. Amelino-Camelia and S. Majid, \textit{Waves on
noncommutative spacetime and gamma-ray bursts}, Int. J. Mod. Phys. A
15 (2000) 4301.

\bibitem{Majid5} S. Majid and R. Oeckl, \textit{Twisting of quantum differentials and the Planck scale Hopf algebra},
Commun. Math. Phys. 205 (1999) 617-655.

\bibitem{Gonera} C. Gonera, P. Kosinski and P. Maslanka,
\textit{Differential calculi on quantum Minkowski space},
q-alg/9602007.

\bibitem{k-Noether5D}
G. Amelino-Camelia, A. Marcian\`o and D. Pranzetti, \textit{On the
5D differential calculus and translation transformations
 in 4D $\kappa$-Minkowski noncommutative spacetime},
hep-th/0709.2063.




\end{thebibliography}
\end{document}